\documentclass[3p, onecolumn, fleqn, preprint, authoryear, times]{elsarticle} % add authoryear for more pleasant looking citations

\usepackage{graphicx}
\usepackage{amsmath}
\usepackage{amssymb}
\usepackage{aas_macros}
\usepackage{multirow}
\usepackage{adjustbox}

\usepackage{hyperref}
\hypersetup{colorlinks=true, urlcolor=blue, citecolor=blue, linkcolor=black}
\begin{document}

\title{Constrained hyperbolic divergence cleaning in smoothed particle magnetohydrodynamics with variable cleaning speeds}

\author[cita,exeter]{Terrence S. Tricco\corref{cor1}}
\ead{ttricco@cita.utoronto.ca}
\author[monash]{Daniel J. Price}
\ead{daniel.price@monash.edu}
\author[exeter]{Matthew R. Bate}
\ead{mbate@astro.ex.ac.uk}

\cortext[cor1]{Corresponding author.}

\address[cita]{Canadian Institute for Theoretical Astrophysics, University of Toronto, 60 St. George Street, Toronto, ON M5S 3H8, Canada}
\address[exeter]{School of Physics, University of Exeter, Stocker Road, Exeter, EX4 4QL, United Kingdom}
\address[monash]{Monash Centre for Astrophysics, School of Physics and Astronomy, Monash University, Clayton, VIC, 3800, Australia}

\begin{abstract}
We present an updated constrained hyperbolic/parabolic divergence cleaning algorithm for smoothed particle magnetohydrodynamics (SPMHD) that remains conservative with wave cleaning speeds which vary in space and time. This is accomplished by evolving the quantity $\psi / c_{\rm h}$ instead of $\psi$. Doing so allows each particle to carry an individual wave cleaning speed, $c_{\rm h}$, that can evolve in time without needing an explicit prescription for how it should evolve, preventing circumstances which we demonstrate could lead to runaway energy growth related to variable wave cleaning speeds. This modification requires only a minor adjustment to the cleaning equations and is trivial to adopt in existing codes. Finally, we demonstrate that our constrained hyperbolic/parabolic divergence cleaning algorithm, run for a large number of iterations, can reduce the divergence of the field to an arbitrarily small value, achieving $\nabla \cdot {\bf B}=0$ to machine precision. 
\end{abstract}


\maketitle

\section{Introduction}

Accurately evolving the equations of magnetohydrodynamics (MHD) in numerical simulations is crucial in astrophysical fluid dynamics. In smoothed particle magnetohydrodynamics (SPMHD) \citep{gm77, pm85, pm04a, pm04b, pm05, price12}, upholding the divergence-free constraint of the magnetic field has been the main technical difficulty. The usual approach is to evolve the magnetic field directly by the induction equation (as in \citealt{pm85}), but this preserves a divergence-free magnetic field only to truncation error. These errors cause more harm than just yielding an unphysical field. They introduce spurious monopole accelerations, which have to be carefully handled in SPMHD in order to ensure numerical stability, at the price of no longer exactly conserving momentum \citep{pm85, morris96, bot01}. Handling the divergence-free constraint on the magnetic field is therefore one of the most important aspects of accurate SPMHD simulations.

One option is to define the magnetic field in a way that manifestly enforces the divergence-free constraint. Use of the Euler potentials, ${\bf B} = \nabla \alpha \times \nabla \beta$ where $\alpha$ and $\beta$ are passive scalars, was proposed as early as \citet{pm85}, and recently the potentials have been used in simulations of protostar formation \citep{pb07}, star cluster formation \citep{pb08, pb09} and magnetised galaxies \citep{dp08, kotarbaetal09}. However, the Euler potentials cannot represent winding motions, prevent dynamo processes by construction \citep{brandenburg10}, and it is not clear how to incorporate non-ideal dissipation. A vector potential implementation, ${\bf B} = \nabla \times {\bf A}$, was tested for SPMHD by \citet{price10}, but was found to be numerically unstable. \citet{se15} recently proposed that the vector potential could be used, if one added numerical diffusion to the potential, enforced the Coulomb gauge condition on the vector potential ($\nabla \cdot {\bf A} = 0$) and smoothed the resulting magnetic field, though it is not clear how robust this approach is in practice.

The second option to handle the divergence-free constraint in SPMHD is to directly evolve the magnetic field with the induction equation, but then `clean' errors out of the field. For example, parabolic diffusion terms can be used to smooth the magnetic field at the resolution scale \citep{morris96}. The artificial resistivity formulation of \citet{pm04a, pm05} has been used for this purpose (e.g., \citealt{burzleetal11b}), however, artificial resistivity is intended for shock capturing and dissipates physical as well as unphysical components of the field. A similar idea is to periodically smooth the magnetic field to remove fluctuations below the resolution limit \citep{bot01}, but this adds computational expense, is time resolution dependent, and reduces the spatial resolution of the magnetic field.

At present, the best option for divergence cleaning in SPMHD is the `constrained' hyperbolic/parabolic divergence cleaning method of \citet{tp12}, an improved version of the method by \citet{dedneretal02}. The original idea from \citet{dedneretal02} was to couple an additional scalar field, $\psi$, to the induction equation according to
\begin{align}
\frac{\partial {\bf B}}{\partial t} & = \nabla \times ({\bf v} \times {\bf B}) - \nabla \psi, \label{eq:dbdt-gradpsi-dedner} \\
\frac{\partial \psi}{\partial t} & = - c_{\rm h}^2 (\nabla \cdot {\bf B}) - \frac{\psi}{\tau}, \label{eq:psievolution-dedner}
\end{align}
where ${\bf B}$ is the magnetic field and ${\bf v}$ is the velocity. These may be combined to produce a damped wave equation for the divergence of the magnetic field, 
\begin{equation}
\frac{{\partial^2} (\nabla \cdot {\bf B})}{\partial t^2} - c_{\rm h}^2 \nabla^2 (\nabla \cdot {\bf B}) + \frac{1}{\tau} \frac{\partial (\nabla \cdot {\bf B})}{\partial t} = 0.
\label{eq:divbwave}
\end{equation}
 From Equation~(\ref{eq:divbwave}), we see that Equation~(\ref{eq:dbdt-gradpsi-dedner}) and the first term on the right hand side of Equation~(\ref{eq:psievolution-dedner}) represent hyperbolic transport of divergence errors at a characteristic speed, $c_{\rm h}$, which we refer to as the `wave cleaning speed'. This is typically chosen to be the fast MHD wave speed so that it obeys the local Courant condition and does not impose any additional timestep constraint. The second term on the right hand side of Equation~(\ref{eq:psievolution-dedner}) produces parabolic diffusion on a timescale defined according to
\begin{equation}
\tau \equiv \frac{h}{\sigma c_{\rm h}},
\end{equation}
where $h$ is the smoothing length (resolution scale) and $\sigma$ is a dimensionless constant with empirically determined optimal values of $0.3$ and $1.0$ in 2D and 3D, respectively \citep{tp12}. The combination of hyperbolic and parabolic terms in Equations~(\ref{eq:dbdt-gradpsi-dedner})--(\ref{eq:psievolution-dedner}) spreads the divergence of the magnetic field over a larger area, reducing the impact of any single large source of error, while also allowing the diffusion to be more effective.

In \citet{tp12}, we showed that the original \citet{dedneretal02} approach could be unstable at density jumps and free surfaces, leading to exponential growth of magnetic energy. To remedy this, we derived a version of the cleaning equations under the constraint that the hyperbolic transport should conserve energy. Though $\psi$ is not a physical variable, conservation of energy for the hyperbolic term between the magnetic and $\psi$ fields ensures that, when the parabolic term is included, magnetic energy can only ever be removed by divergence cleaning, never added, guaranteeing numerical stability. The `constrained' or `conservative' cleaning equations we derived in \citet{tp12} are given by
\begin{align}
\frac{{\rm d}{\bf B}}{{\rm d}t} & = ({\bf B}\cdot\nabla) {\bf v} - {\bf B}(\nabla\cdot{\bf v}) - \nabla \psi, \label{eq:dbdt-gradpsi-original} \\
\frac{{\rm d}\psi}{{\rm d}t} & = - c_{\rm h}^2 (\nabla \cdot {\bf B}) - \frac{\psi}{\tau} - \frac{1}{2} \psi (\nabla \cdot {\bf v}) , \label{eq:psievolution-original}
\end{align}
where ${\rm d}/{\rm d}t \equiv \partial/\partial t + {\bf v}\cdot\nabla$ is the Lagrangian time derivative. The formulation of the induction equation (Equation~(\ref{eq:dbdt-gradpsi-original})) in the absence of the $\nabla\psi$ term follows the `divergence preserving scheme' of \citet{powelletal99} (see also \citealt{janhunen00,dellar01}), meaning that divergence errors are preserved by the flow in the absence of cleaning. The third term in Equation~(\ref{eq:psievolution-original}) was introduced by \citet{tp12} to account for changes in $\psi$ from compression or rarefaction of the gas, and is necessary to ensure total energy conservation in the absence of damping.  The practical advantage of this algorithm for SPMHD is that it adds no additional timestep constraint, is simple to implement, computationally efficient, and has been successfully used to enforce the divergence-free constraint in simulations of jets and outflows during protostar formation \citep{ptb12, btp14, lbp15, wpb16}. However, our original method was derived assuming that the cleaning speed, $c_{\rm h}$, is constant in both space and time, but this is not true in practice and presents a source of non-conservation of energy. Furthermore, source terms are added to the right hand side of Equation~(\ref{eq:divbwave}) when $c_{\rm h}$ or $\tau$ are time or spatially variable, by the addition of the $\tfrac{1}{2} \psi (\nabla \cdot {\bf v})$ term, and by solving the cleaning equations in the Lagrangian frame of motion. How these source terms change the propagation of divergence errors is not properly understood, but will be addressed in this work.
 
In this paper, we derive an improvement to constrained hyperbolic/parabolic divergence cleaning such that the hyperbolic evolution equations remain conservative even in the presence of a variable cleaning speed (Section~\ref{sec:cleaningv2}). We demonstrate that these equations create a generalised wave equation which naturally incorporates the source terms (Section~\ref{sec:sourceterms}). Aspects of the method are tested in Section~\ref{sec:idealised-tests} using a series of test problems. In particular, we will show that, if the time variability of the cleaning wave speed is not properly accounted for, the non-conservation of energy introduced may reduce the effectiveness of the divergence cleaning, and, worst case scenario, lead to runaway energy growth and numerical instability. In Section~\ref{sec:practical-tests}, the original and updated versions of the method are compared using standard MHD tests to quantify how much of an improvement the new scheme confers. Finally, in Section~\ref{sec:divbzero}, we demonstrate that, by iterating the divergence cleaning equations, it is possible to clean the magnetic field until $\nabla \cdot {\bf B}=0$ to machine precision in the chosen divergence operator. We summarise in Section~\ref{sec:summary}.

While our focus in this paper is on improved divergence cleaning methods for SPMHD, our analysis and in particular our reformulation of the cleaning equations should apply equally to implementations of hyperbolic/parabolic cleaning in grid-based MHD codes, particularly in the context of adaptive mesh refinement (AMR) where jumps in the cleaning speed may occur at refinement boundaries. Application to Eulerian MHD codes is beyond the scope of this paper but would be an interesting and worthwhile extension to our work.

\section{Constrained hyperbolic divergence cleaning with variable wave speeds}
\label{sec:cleaningv2}

The issue with variable wave cleaning speeds can be seen by considering the energy conservation of the cleaning equations. Equations~(\ref{eq:dbdt-gradpsi-original})--(\ref{eq:psievolution-original}) transfer energy back and forth between the ${\bf B}$ and $\psi$ fields, and, in the absence of damping, this transfer should conserve energy. If it does not, then the method may inject spurious energy into the magnetic field which can act against the cleaning efforts. 

\subsection{Constraints from energy conservation}
\label{sec:energ}
To derive the conservative cleaning equations, the energy content of the $\psi$ field needs to be known. The specific energy of the $\psi$ field was determined by \citet{tp12} to be
\begin{equation}
e_\psi = \frac{\psi^2}{2 \mu_0 \rho c_{\rm h}^2}.
\end{equation}
The total energy is given by
\begin{equation}
E = \int \left[ \frac12 {\bf v}^{2} + u + \frac12\frac{{\bf B}^{2}}{\mu_{0} \rho} + e_{\psi} \right] \rho {\rm d}V,
\end{equation}
where $u$ is the specific thermal energy and $\rho$ is the density, such that $\rho {\rm d}V$ is equivalent to the mass element ${\rm d}m$. The total energy must be conserved, that is, ${\rm d}E/{\rm d}t = 0$. Since we are concerned only with the cleaning terms added to the usual MHD equations (which conserve energy in the absence of divergence cleaning) we need only consider the additional term involving $\psi$ in Equation~(\ref{eq:dbdt-gradpsi-original}). This means that the time derivative of magnetic energy should balance the time derivative of $e_\psi$ according to
\begin{equation}
\frac{{\rm d}E}{{\rm d}t} = \int \left[ \frac{{\bf B}}{\mu_0 \rho} \cdot \left( \frac{{\rm d}{\bf B}}{{\rm d}t} \right)_\psi + \frac{{\rm d}}{{\rm d}t} \left( \frac{\psi^2}{2 \mu_0 \rho c_{\rm h}^2} \right) \right] \rho {\rm d}V = 0 , \label{eq:energy-conserv1}
\end{equation}
where the Lagrangian time derivative of the mass element $\rho {\rm d}V$ is zero. The ${\rm d} e_\psi /{\rm d}t$ term, when expanded, produces terms related to the time change of $\psi$, $\rho$ and $c_{\rm h}$ according to
\begin{equation}
\int \bigg[ \frac{{\bf B}}{\mu_0 \rho} \cdot \left( \frac{{\rm d}{\bf B}}{{\rm d}t} \right)_\psi + \frac{\psi}{\mu_0 \rho c_{\rm h}^2} \frac{{\rm d}\psi}{{\rm d}t} - \frac{\psi^2}{2 \mu_0 \rho^2 c_{\rm h}^2} \frac{{\rm d}\rho}{{\rm d}t} - \frac{\psi^2}{\mu_0 \rho c_{\rm h}^3} \frac{{\rm d}c_{\rm h}}{{\rm d}t} \bigg] \rho {\rm d}V = 0 .
\label{eq:dedt}
\end{equation}
We note that ${\rm d}\rho / {\rm d}t$ terms arising from the magnetic energy are balanced as part of the MHD equations, so do not need to considered here. The ${\rm d}\rho / {\rm d}t$ term resulting from the $e_\psi$ term was accounted for in \citet{tp12} by the addition of a $\tfrac{1}{2} \psi (\nabla \cdot {\bf v})$ term to the evolution equation for $\psi$ (Equation~(\ref{eq:psievolution-original})). The question is how to handle the ${\rm d}c_{\rm h} / {\rm d}t$ term.

Our approach is to use $\psi / c_{\rm h}$ as the evolved quantity instead of $\psi$. In this case, Equation~(\ref{eq:energy-conserv1}) when expanded yields
\begin{equation}
\int \left[ \frac{{\bf B}}{\mu_0 \rho} \cdot \left( \frac{{\rm d}{\bf B}}{{\rm d}t} \right)_\psi + \frac{\psi}{\mu_0 \rho c_{\rm h}} \frac{{\rm d}}{{\rm d}t} \left( \frac{\psi}{c_{\rm h}} \right) - \frac{\psi^2}{2 \mu_0 \rho^2 c_{\rm h}^2} \frac{{\rm d}\rho}{{\rm d}t} \right] \rho {\rm d}V = 0 ,\label{eq:energy-conserv2}
\end{equation}
such that the ${\rm d}c_{\rm h} / {\rm d}t$ term is included within the ${\rm d} / {\rm d}t (\psi / c_{\rm h})$ term. By evolving $\psi / c_{\rm h}$ instead of $\psi$, we avoid the need to explicitly prescribe ${\rm d} c_{\rm h} / {\rm d}t$. 

\subsection{Hyperbolic transport}

The evolution equation for $\psi / c_{\rm h}$ can be obtained in the following manner. By the chain rule, 
\begin{equation}
\frac{{\rm d}}{{\rm d}t} \left( \frac{\psi}{c_{\rm h}} \right) = \frac{1}{c_{\rm h}} \frac{{\rm d}\psi}{{\rm d}t} - \frac{\psi}{c_{\rm h}^2} \frac{{\rm d}c_{\rm h}}{{\rm d}t} .
\end{equation}
For the case where the co-moving time derivative ${\rm d} c_{\rm h} / {\rm d}t = 0$, it becomes clear that the hyperbolic term in the evolution equation of $\psi / c_{\rm h}$ must be
\begin{equation}
\frac{{\rm d}}{{\rm d}t} \left( \frac{\psi}{c_{\rm h}} \right) = - c_{\rm h} (\nabla \cdot {\bf B}),  \label{eq:new-psich-evolution}
\end{equation}
in order to be consistent with the existing formulation.

\subsection{Hyperbolic transport in SPMHD}
Equation~(\ref{eq:new-psich-evolution}) is discretised, as in \citet{tp12}, using the SPH difference operator for $\nabla\cdot{\bf B}$, giving
\begin{equation}
\frac{{\rm d}}{{\rm d}t} \left( \frac{\psi_a}{c_{{\rm h},a}} \right) = \frac{c_{{\rm h},a}}{\Omega_a \rho_a} \sum_b m_b ({\bf B}_a - {\bf B}_b) \cdot \nabla_a W_{ab}(h_a) ,
\end{equation}
where $m$ is the particle mass, $W_{ab}(h_a) \equiv W(\vert {\bf r}_a - {\bf r}_b \vert, h_a)$ is the smoothing kernel, and $\Omega$ is a factor to account for gradients in the smoothing length \citep{monaghan02, sh02}. The summations are over neighbouring particles, with subscripts $a$ and $b$ referring to the particle index.

The discretised version of $\nabla \psi$ in the induction equation is derived by ensuring that energy is conserved. The discrete equivalent of Equation~(\ref{eq:energy-conserv2}) is
\begin{equation}
\frac{{\rm d}E}{{\rm d}t} = \sum_a m_a \left[ \frac{{\bf B}_a}{\mu_0 \rho_a} \cdot \left( \frac{{\rm d}{\bf B}_a}{{\rm d}t} \right)_\psi + \frac{\psi_a}{\mu_0 \rho_a c_{{\rm h},a}} \frac{{\rm d}}{{\rm d}t} \left( \frac{\psi_a}{c_{{\rm h},a}} \right) \right] = 0 ,
\end{equation}
where for the moment we have neglected the ${\rm d}\rho / {\rm d}t$ term, considered later in Section~\ref{sec:drhodt}. Also ignoring the damping term for the moment (see Section~\ref{sec:damping}), this implies that
\begin{equation}
\sum_a \frac{m_a}{\mu_0 \rho_a} {\bf B}_a \cdot \left( \frac{{\rm d}{\bf B}_a}{{\rm d}t} \right)_\psi = - \sum_a \frac{m_a}{\mu_0 \rho_a} \frac{\psi_a}{\Omega_a \rho_a} \sum_b m_b ({\bf B}_a - {\bf B}_b) \cdot \nabla_a W_{ab}(h_a) .
\end{equation}
From here, the procedure is the same as that in \citet{tp12}, with the symmetric estimate being obtained for the $\nabla \psi$ term in ${\rm d}{\bf B} / {\rm d}t$, yielding
\begin{equation}
\left( \frac{{\rm d}{\bf B}_a}{{\rm d}t} \right)_\psi = - \rho_a \sum_b m_b \left[ \frac{\psi_a}{\Omega_a \rho_a^2} \nabla_a W_{ab}(h_a) + \frac{\psi_b}{\Omega_b \rho_b^2} \nabla_a W_{ab}(h_b) \right] .
\end{equation}
The key to the conservative properties of these equations is that the derivative estimates for $\nabla \cdot {\bf B}$ and $\nabla \psi$ form a conjugate pair (difference and symmetric operators, respectively; see \citet{price12} for discussion on derivative estimates in SPH). The occurrence of the pairing of these operators in SPH has been noted elsewhere, i.e., \citet{cr99, price10, tii13, wpa14}, and is discussed further in \citet{tp12}.

There is freedom to choose the discretisation of the divergence of the magnetic field. We investigated this in \citet{tp12}, since we thought it might make sense to use the symmetric operator for $\nabla \cdot {\bf B}$ since that is the operator that appears in the momentum equation. By conservation of energy, this was shown to require use of the SPH difference operator for $\nabla \psi$, again forming a conjugate pair. However, in \citet{tp12} we concluded that it is not advisable to use the symmetric operator of $\nabla \cdot {\bf B}$ for divergence cleaning, since the low order errors in the symmetric operator were found to produce artefacts in the physical components of the magnetic field and also over-dissipate the magnetic energy.

\subsection{Parabolic damping}
\label{sec:damping}

The parabolic damping term for the modified $\psi / c_{\rm h}$ evolution equation is obtained through a similar procedure to the hyperbolic term. It should reduce to the previous formalism for constant $c_{\rm h}$. Therefore, the parabolic damping term is
\begin{equation}
\frac{{\rm d}}{{\rm d}t} \left( \frac{\psi_a}{c_{{\rm h},a}} \right)_{\rm damp} = - \frac{1}{\tau} \frac{\psi_a}{c_{{\rm h},a}} . \label{eq:psich-damping}
\end{equation}
By similar arguments, the empirical values of $\sigma$ obtained by \citet{tp12} should be not affected by evolving $\psi / c_{\rm h}$ instead of $\psi$. It is straightforward to show that this term provides a negative definite contribution to the total energy. The rate of change of total energy from the damping term in the discrete system is given by
\begin{equation}
\frac{{\rm d}E}{{\rm d}t} = \sum_a m_a \frac{\psi_a}{\mu_0 \rho_a c_{{\rm h},a}} \frac{{\rm d}}{{\rm d}t} \left( \frac{\psi_a}{c_{{\rm h},a}} \right)_{\rm damp}.
\end{equation}
Inserting Equation~(\ref{eq:psich-damping}), we have
\begin{equation}
\frac{{\rm d}E}{{\rm d}t} = - \sum_a m_a \frac{\psi_a^2}{\mu_0 \rho_a c_{\rm h}^2 \tau},
\label{eq:psiheat}
\end{equation}
showing that the $\psi / c_{\rm h}$ damping term is guaranteed to remove energy from the system. This energy removal may be balanced by an equivalent deposit into thermal energy so that total energy is conserved, however there is no requirement to do this for stability reasons. As discussed in \citet{tp12}, the removal of magnetic energy and subsequent generation of thermal energy would be non-local due to the coupling of parabolic diffusion with hyperbolic transport. Therefore, we do not add the removed energy as heat.

\subsection{Compression and rarefaction of $\psi / c_{\rm h}$}
\label{sec:drhodt}

The ${\rm d} \rho / {\rm d}t$ term in Equation~(\ref{eq:energy-conserv2}) may be balanced by adding the following term to the evolution equation for $\psi / c_{\rm h}$,
\begin{equation}
\frac{{\rm d}}{{\rm d}t} \left( \frac{\psi}{c_{\rm h}} \right)_{{\rm d}\rho/{\rm d}t} = - \frac{\psi}{2 c_{\rm h}} (\nabla \cdot {\bf v}),
\label{eq:divv1}
\end{equation}
making use of the continuity equation [${\rm d}\rho/{\rm d}t = -\rho (\nabla\cdot{\bf v})$]. The SPMHD equivalent is
\begin{equation}
\frac{{\rm d}}{{\rm d}t} \left( \frac{\psi_a}{c_{{\rm h},a}} \right)_{{\rm d}\rho/{\rm d}t} = \frac{\psi_a}{2 c_{{\rm h},a}} \sum_b m_b ({\bf v}_a - {\bf v}_b) \cdot \nabla_a W_{ab}(h_a) ,
\label{eq:divv2}
\end{equation}
where we use the difference derivative operator for $\nabla \cdot {\bf v}$ to match the discretised continuity equation in SPH \citep{monaghan05}. 

An alternative approach to handle compression and rarefaction, as suggested by one referee of this paper, would be to evolve the variable $\psi / (c_{\rm h} \sqrt{\rho})$ instead of $\psi / c_{\rm h}$ (see also Section~\ref{sec:sourceterms}). Incorporating $\rho$ into the choice of variable removes the need to explicitly prescribe the ${\rm d} \rho / {\rm d}t$ term in Equation~(\ref{eq:energy-conserv2}), just as folding $c_{\rm h}$ into the evolved variable did for  ${\rm d} c_{\rm h} / {\rm d}t$. We prefer the approach using Equations~(\ref{eq:divv1})--(\ref{eq:divv2}) for practical reasons --- evolving $\psi / (c_{\rm h} \sqrt{\rho})$ introduces factors of $\sqrt{\rho}$ into the cleaning equations which are expensive to compute, particularly compared to $\nabla \cdot {\bf v}$ which is typically calculated already in SPMHD codes. Furthermore, evolving $\psi/c_{\rm h}$ or $\psi/(c_{\rm h} \sqrt{\rho})$ is analogous to evolving ${\bf B}/\rho$ instead of ${\bf B}$, both of which are commonly used in SPMHD, and neither of which have been found to confer any advantage over the other. 

We do note that our previous tests of the $\nabla \cdot {\bf v}$ term found that it provided no real benefit in terms of divergence error reduction \citep{tp12, tricco15}. The importance of this term is tested further in Section~\ref{sec:tests-divv}.

\subsection{Summary of modified cleaning equations}

The cleaning equations, modified to evolve $\psi / c_{\rm h}$ so that energy is conserved by the hyperbolic terms even in the presence of time-varying cleaning wave speeds, are given by
\begin{align}
 \frac{{\rm d}{\bf B}}{{\rm d}t} & = ({\bf B}\cdot\nabla) {\bf v} - {\bf B} (\nabla\cdot{\bf v}) - \nabla \psi , \label{eq:c1} \\
\frac{{\rm d}}{{\rm d}t} \left( \frac{\psi}{c_{\rm h}} \right) & = - c_{\rm h} (\nabla \cdot {\bf B}) - \frac{1}{\tau} \left(\frac{\psi}{c_{\rm h}}\right) - \frac12 \left(\frac{\psi}{c_{\rm h}}\right) (\nabla \cdot {\bf v}). \label{eq:c2}
\end{align}
The corresponding discrete set of conservative SPMHD cleaning equations are given by
\begin{align}
\left( \frac{{\rm d}{\bf B}_a}{{\rm d}t} \right)_\psi &=  - \rho_a \sum_b m_b \left[ \frac{\psi_a}{\Omega_a \rho_a^2} \nabla_a W_{ab}(h_a) + \frac{\psi_b}{\Omega_b \rho_b^2} \nabla_a W_{ab}(h_b) \right] , \label{eq:cleaning-spmhd1} \\
\frac{{\rm d}}{{\rm d}t} \left( \frac{\psi}{c_{{\rm h}}} \right)_{a} &= \frac{c_{{\rm h},a}}{\Omega_a \rho_a} \sum_b m_b ({\bf B}_a - {\bf B}_b) \cdot \nabla_a W_{ab}(h_a) - \frac{1}{\tau} \left(\frac{\psi}{c_{{\rm h}}}\right)_{a} + \frac12 \left(\frac{\psi}{c_{{\rm h}}}\right)_{a} \sum_b m_b ({\bf v}_a - {\bf v}_b) \cdot \nabla_a W_{ab}(h_a) . \label{eq:cleaning-spmhd2}
\end{align}
In an existing code which evolves $\psi$, the modifications needed to implement the new cleaning scheme evolving $\psi/c_{\rm h}$ are minor. Both $\psi$ and $\psi/c_{\rm h}$ are zero initially. In the code we typically set 
\begin{equation}
c_{{\rm h}, a} = \sqrt{v_{{\rm A}, a}^{2} + c_{{\rm s}, a}^{2}},
\end{equation}
where $v_{\rm A}$ is the Alfv\'en speed and $c_{\rm s}$ is the sound speed. This is used in the first term on the right hand side of Equation~(\ref{eq:cleaning-spmhd2}), and to construct $\psi$ from the evolved quantity $\psi / c_{\rm h}$ when evaluating the right hand side of Equation~(\ref{eq:cleaning-spmhd1}). Since it is easy to evaluate $c_{\rm h}$ both for particle $a$ and for the neighbouring particle $b$, it does not require extra storage in the code. Importantly, our cleaning equations are now guaranteed to conserve or dissipate energy even though this speed changes with time.

\subsection{Cleaning wave equation and source terms}
\label{sec:sourceterms}

One of the unanswered questions from our previous paper \citep{tp12} is whether the character of the wave equation (Equation~(\ref{eq:divbwave})) is changed by enforcing energy conservation in the cleaning equations. If one naively takes our new set of cleaning equations (\ref{eq:c1}--\ref{eq:c2}) and expands the terms using Eulerian time derivatives to match Equation~(\ref{eq:divbwave}), source terms appear on the right hand side related to derivatives of $c_{\rm h}$ in either time or space, if one derives the propagation equation for $\psi$ or $\nabla\cdot{\bf B}$, respectively (see e.g.\ \citealt{hr16} for details). Source terms also appear from use of the Lagrangian time derivative and from the addition of the $\frac12 (\psi / c_{\rm h}) (\nabla\cdot{\bf v})$ term in Equation~(\ref{eq:c2}). Nevertheless, these terms are necessary for the hyperbolic terms to conserve energy, as demonstrated in Sections~\ref{sec:energ}--\ref{sec:drhodt}. 

The propagation of divergence errors in our new formulation can be understood by writing Equations~(\ref{eq:c1})--(\ref{eq:c2}) in the form
\begin{align}
 \frac{{\rm d}{\bf B}}{{\rm d}t} & = ({\bf B}\cdot\nabla) {\bf v} - {\bf B} (\nabla\cdot{\bf v}) - \nabla \psi , \label{eq:d1} \\
\frac{1}{\sqrt{\rho} c_{\rm h}}\frac{{\rm d}}{{\rm d}t} \left( \frac{\psi}{\sqrt{\rho} c_{\rm h}} \right) & = - \frac{\nabla \cdot {\bf B}}{\rho} - \frac{1}{\rho c_{\rm h}} \left(\frac{\psi}{c_{\rm h}\tau}\right) , \label{eq:d2}
\end{align}
where Equation~(\ref{eq:d2}) has been written in terms of the variable $\psi / \sqrt{\rho} c_{\rm h}$ (see Section~\ref{sec:drhodt}). Taking the Lagrangian time derivative of Equation~(\ref{eq:d2}) gives
\begin{equation}
\frac{{\rm d}}{{\rm d}t}  \left[ \frac{1}{\sqrt{\rho} c_{\rm h}}\frac{{\rm d}}{{\rm d}t} \left( \frac{\psi}{\sqrt{\rho} c_{\rm h}} \right) \right] = - \frac{{\rm d}}{{\rm d}t} \left( \frac{\nabla \cdot {\bf B}}{\rho} \right) - \frac{{\rm d}}{{\rm d}t} \left[ \frac{1}{\sqrt{\rho} c_{\rm h}} \left(\frac{\psi}{\sqrt{\rho}c_{\rm h}\tau}\right) \right]. \label{eq:d3}
\end{equation}
Expanding the first term on the right hand side and using ${\rm d}\rho/{\rm d}t = -\rho (\nabla\cdot{\bf v})$, we have
\begin{equation}
\frac{{\rm d}}{{\rm d}t} \left( \frac{\nabla \cdot {\bf B}}{\rho} \right) = \frac{1}{\rho} \frac{{\rm d} }{{\rm d}t} (\nabla\cdot{\bf B}) + \frac{(\nabla\cdot{\bf B})(\nabla\cdot{\bf v})}{\rho}.
\end{equation}
Using the relation
\begin{equation}
\frac{{\rm d} }{{\rm d}t} (\nabla\cdot{\bf B})  = \nabla \cdot \left( \frac{{\rm d}{\bf B}}{{\rm d}t} \right) - \frac{\partial v^{i}}{\partial x^{j}} \frac{\partial B^{j}}{\partial x^{i}},
\end{equation}
and inserting Equation~(\ref{eq:d1}), we have
\begin{align}
\frac{{\rm d} }{{\rm d}t} (\nabla\cdot{\bf B}) & =\frac{\partial}{\partial x^{i}} \left ( B^{j} \frac{\partial v^{i}}{\partial x^{j}} \right)  - \frac{\partial}{\partial x^{i}} \left ( B^{i} \frac{\partial v^{j}}{\partial x^{j}} \right) - \nabla^{2} \psi - \frac{\partial v^{i}}{\partial x^{j}} \frac{\partial B^{j}}{\partial x^{i}}, \\
& = -(\nabla\cdot{\bf B})(\nabla\cdot{\bf v}) -  \nabla^{2} \psi,
\end{align}
giving
\begin{equation}
\frac{{\rm d}}{{\rm d}t} \left( \frac{\nabla \cdot {\bf B}}{\rho} \right) = -\frac{\nabla^{2}\psi}{\rho}.
\end{equation}
Finally, inserting this term in Equation~(\ref{eq:d2}), we obtain a generalised wave equation for $\psi$ in the form
\begin{equation}
\frac{{\rm d}}{{\rm d}t}  \left[ \frac{1}{\sqrt{\rho} c_{\rm h}}\frac{{\rm d}}{{\rm d}t} \left( \frac{\psi}{\sqrt{\rho} c_{\rm h}} \right) \right] - \frac{\nabla^{2} \psi}{\rho} + \frac{{\rm d}}{{\rm d}t} \left[ \frac{1}{\sqrt{\rho} c_{\rm h}} \left(\frac{\psi}{\sqrt{\rho}c_{\rm h}\tau}\right) \right] = 0. \label{eq:genwave}
\end{equation}
This shows that the propagation of divergence errors in our new method remains hyperbolic/parabolic in character, but that the wave propagation occurs in the co-moving frame and takes account of the time-variability of the density, wave speed and parabolic damping term with a rescaling of the time coordinate.
%
% We can simplify the expression above by defining a generalised time derivative operator
%\begin{equation}
%\frac{{\rm D} \psi }{{\rm D}\mathcal{T}} \equiv \frac{{\rm d} }{{\rm d}t} \left( \frac{\psi}{\sqrt{\rho} c_{\rm h}} \right),
%\end{equation}
%which simplifies Equation~\ref{eq:genwave} to
%\begin{equation}
%\frac{{\rm D}^{2}\psi }{{\rm D}\mathcal{T}^{2}}- \frac{\nabla^{2}\psi}{\rho} + \frac{{\rm D}}{{\rm D}\mathcal{T}} \left(\frac{\psi}{\lambda} \right) = 0.
%\end{equation}
If the velocity of the fluid is constant (implying ${\rm d}\rho/{\rm d}t = 0$) the time derivatives reduce to Eulerian derivatives, but still allow for a time dependent wave speed and damping term,
\begin{equation}
\frac{\partial}{\partial t} \left[ \frac{1}{c_{\rm h}} \frac{\partial}{\partial t} \left( \frac{\psi}{c_{\rm h}} \right) \right] - \nabla^{2}\psi + \frac{\partial}{\partial t} \left( \frac{\psi}{c_{\rm h}^{2} \tau} \right) = 0.
\end{equation}
If we further assume that $c_{\rm h}$ and $\tau$ are constant, this reduces to the usual damped wave equation
\begin{equation}
\frac{\partial^{2} \psi}{{\partial t}^{2}}  - c_{\rm h}^{2} \nabla^{2}\psi + \frac{1}{\tau} \frac{\partial \psi}{\partial t} = 0. 
\end{equation}
Importantly, the generalised wave equation does not imply that $\nabla\cdot{\bf B}$ locally always decreases, as one might naively expect. Rather, the amplitude of the divergence `wave' can both increase and decrease in response to changes in the wave speed or density --- corresponding physically to the refraction and reflection of waves in response to changes in the medium through which the wave travels. However, refraction and reflection occur in a way that conserves energy.

In the above, we have derived the propagation equation for $\psi$ rather than $\nabla\cdot{\bf B}$. With constant density and wave speed these two propagate in an identical manner (compare Equation~(\ref{eq:genwave}) above to Equation~(\ref{eq:divbwave})). Deriving the propagation equation for $\nabla\cdot{\bf B}$ in our generalised case is significantly more complicated, and as a result we have not proved in this paper that it propagates identically to $\psi$, but we expect the evolution of $\nabla\cdot{\bf B}$ to follow a similar equation. Figs.~\ref{fig:timevary-render} and \ref{fig:eulerian-render} demonstrate that this is indeed the case.

\section{Idealised tests}
\label{sec:idealised-tests}

Our first tests are designed to target specific aspects of the method. In particular, we highlight how time variations of the wave cleaning speed may lead to runaway energy growth. To fully explore this issue, spatial variations of the wave cleaning speed are also investigated, as are discontinuities in $\tau$. We take this opportunity to further test other elements of the method, specifically whether it is appropriate to use Lagrangian derivatives for the cleaning equations, as we have done, or to use Eulerian derivatives, as in the original \citet{dedneretal02} paper. Finally, we demonstrate that the $\tfrac12 (\psi / c_{\rm h}) (\nabla \cdot {\bf v})$ term to account for compression and rarefaction is indeed required to satisfy energy conservation.

\subsection{Fiducial model -- Divergence advection test}

All tests in this section are based on the divergence advection test used by \citet{dedneretal02}, \citet{pm05} and \citet{tp12}. It is a simple test of fluid flowing diagonally across a two-dimensional domain, with a uniform magnetic field that has divergence of the field introduced by adding a small perturbation. While idealised, its simplicity allows for targeted analysis on specific elements of the divergence cleaning method. 

The simulation is performed in the domain $x, y = [-0.5, 1.5]$ with periodic boundary conditions, using $50 \times 58$ particles arranged on a triangular lattice. The initial conditions are given by $\rho = 1$, $P = 6$ and $\gamma = 5/3$. The initial velocity field is ${\bf v} = [1,1]$. The magnetic field is $B_z = 1 / \sqrt{4 \pi}$, using $\mu_0 = 1$, with $B_x = B_y = 0$, except for a perturbation introduced into the $x$ component according to
\begin{equation}
B_x = \frac{1}{\sqrt{4 \pi}} \left[ \left( \frac{r}{r_0} \right)^8 - 2 \left( \frac{r}{r_0} \right)^4 + 1 \right], \hspace{8mm} \frac{r}{r_0} < 1,
\end{equation}
where $r = \sqrt{x^2 + y^2}$. The size of the perturbation is $r_0 = 1/\sqrt{8}$. This perturbation artificially introduces divergence into the magnetic field. 

These conditions yield a plasma beta, the ratio of thermal to magnetic pressure, of $\beta = 150$ in the region of strongest magnetic field strength. Since thermal pressure is dominant for these conditions, the simulations do not require the tensile instability correction term used in the magnetic force, which is necessary to prevent particle pairing when $\beta < 1$ \citep{pm85, morris96, bot01}. Since the correction term introduces a source of non-conservation of energy, we do not apply the correction term for these idealised simulations so that the energy conservation properties of the divergence cleaning method can be accurately measured. By running SPMHD in fully conservative form, these set of simulations will exactly conserve energy to the accuracy of the timestepping algorithm, and importantly, to the accuracy of the divergence cleaning method, which is our purpose. Furthermore, to isolate changes in divergence error as occurring due to the divergence cleaning method, these simulations are run without artificial resistivity.

%% color bar options
%% 8.0 x 8.0 inch page, character height 1.1

\begin{figure}
\centering
\begin{minipage}{0.87\textwidth}
\includegraphics[width=0.24\linewidth]{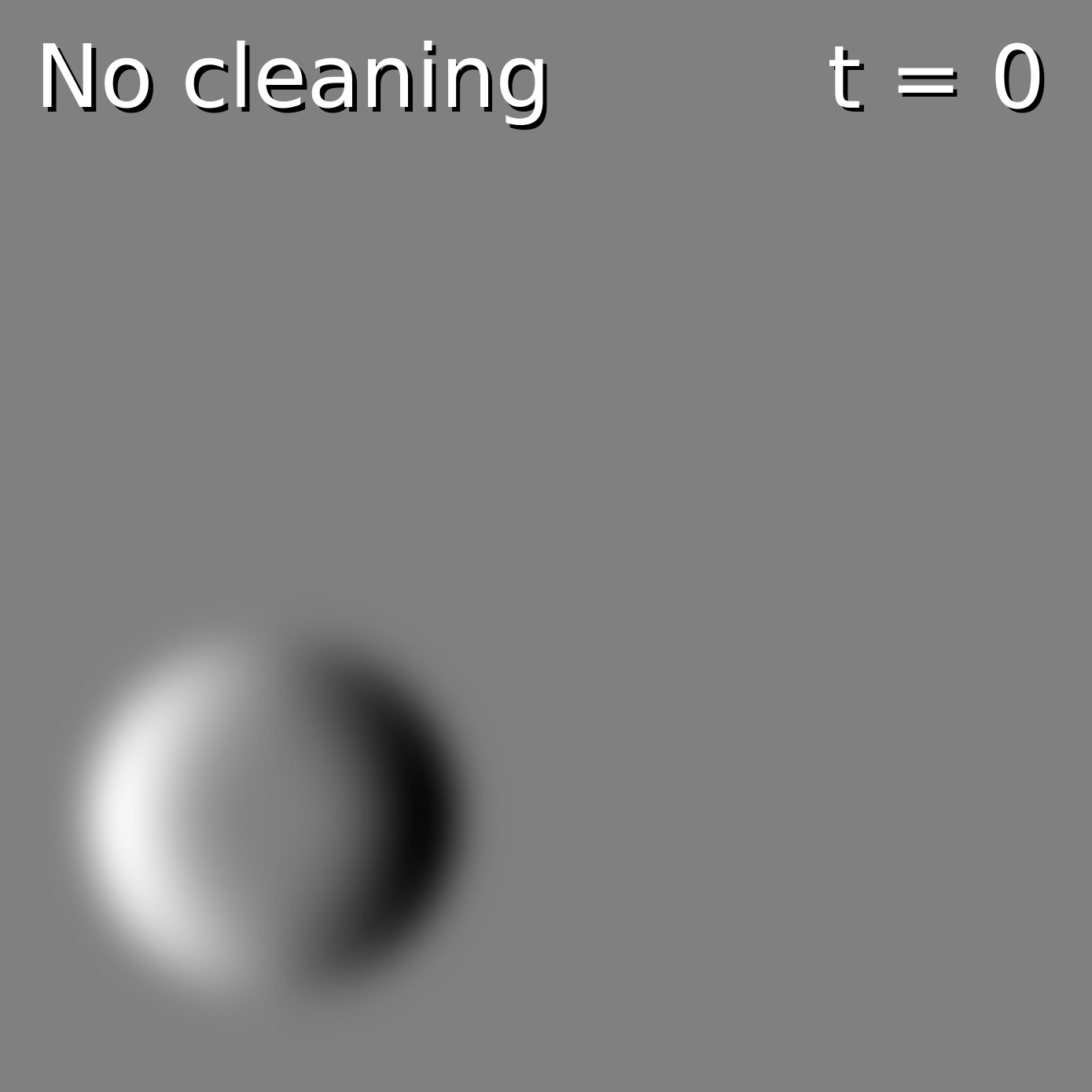}
\includegraphics[width=0.24\linewidth]{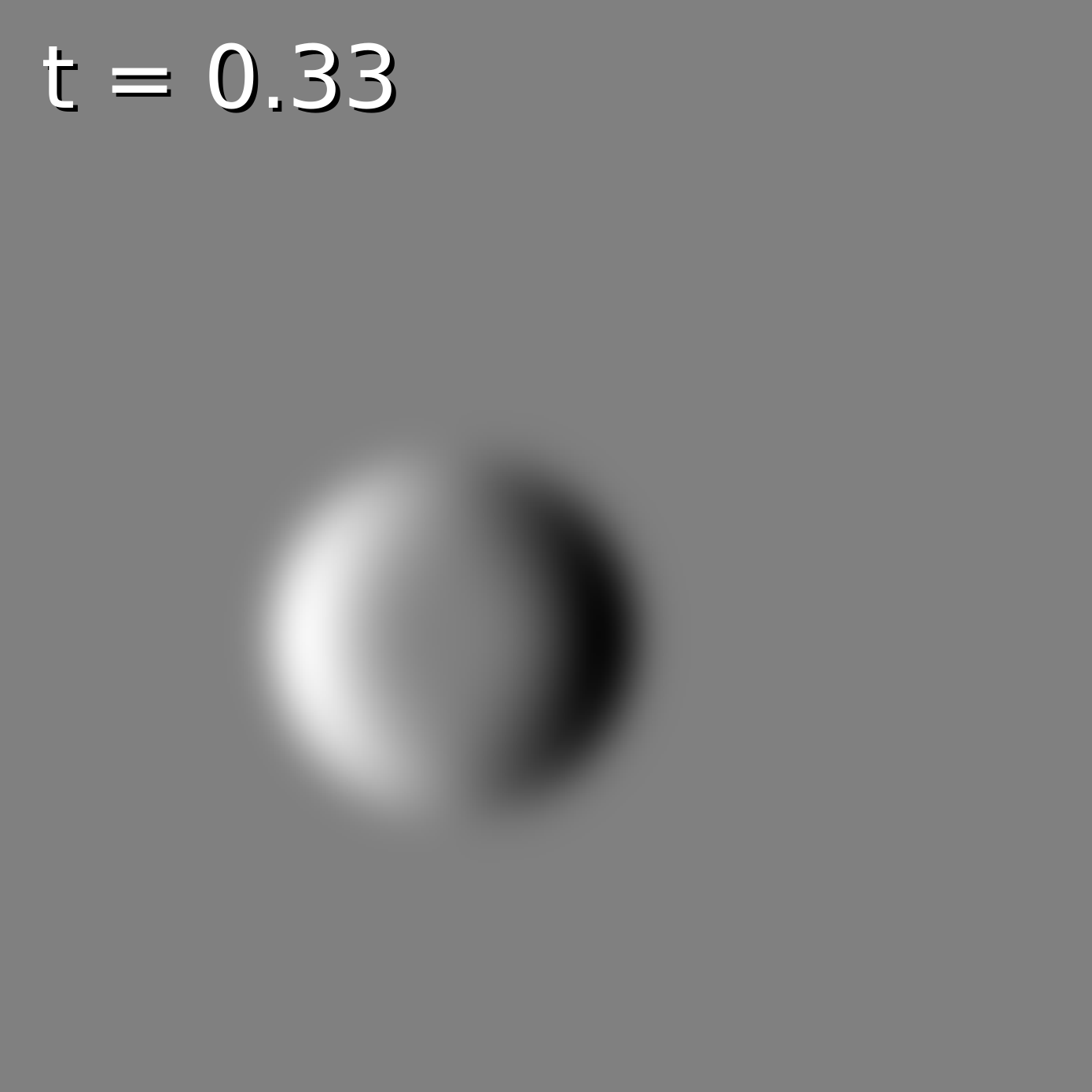}
\includegraphics[width=0.24\linewidth]{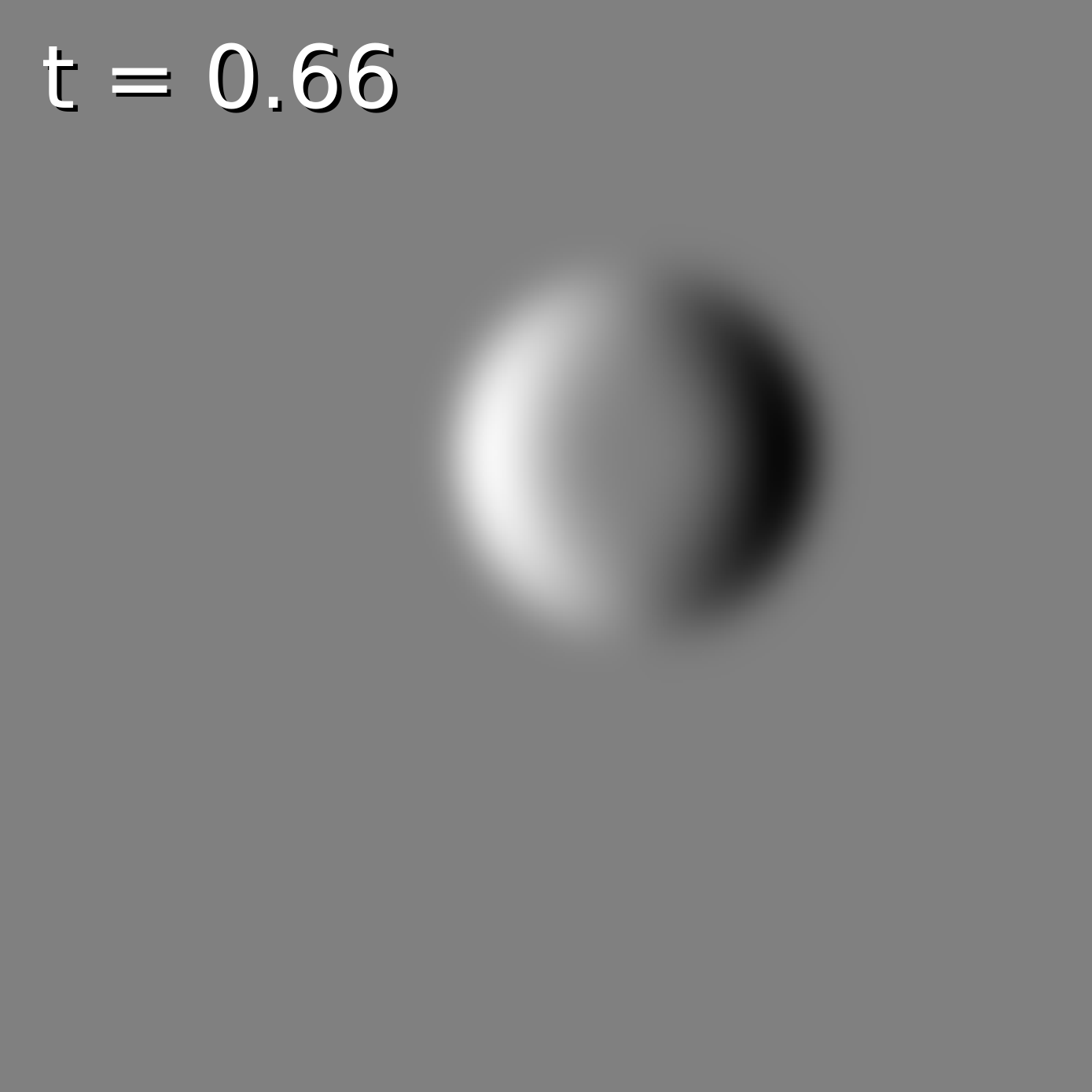}
\includegraphics[width=0.24\linewidth]{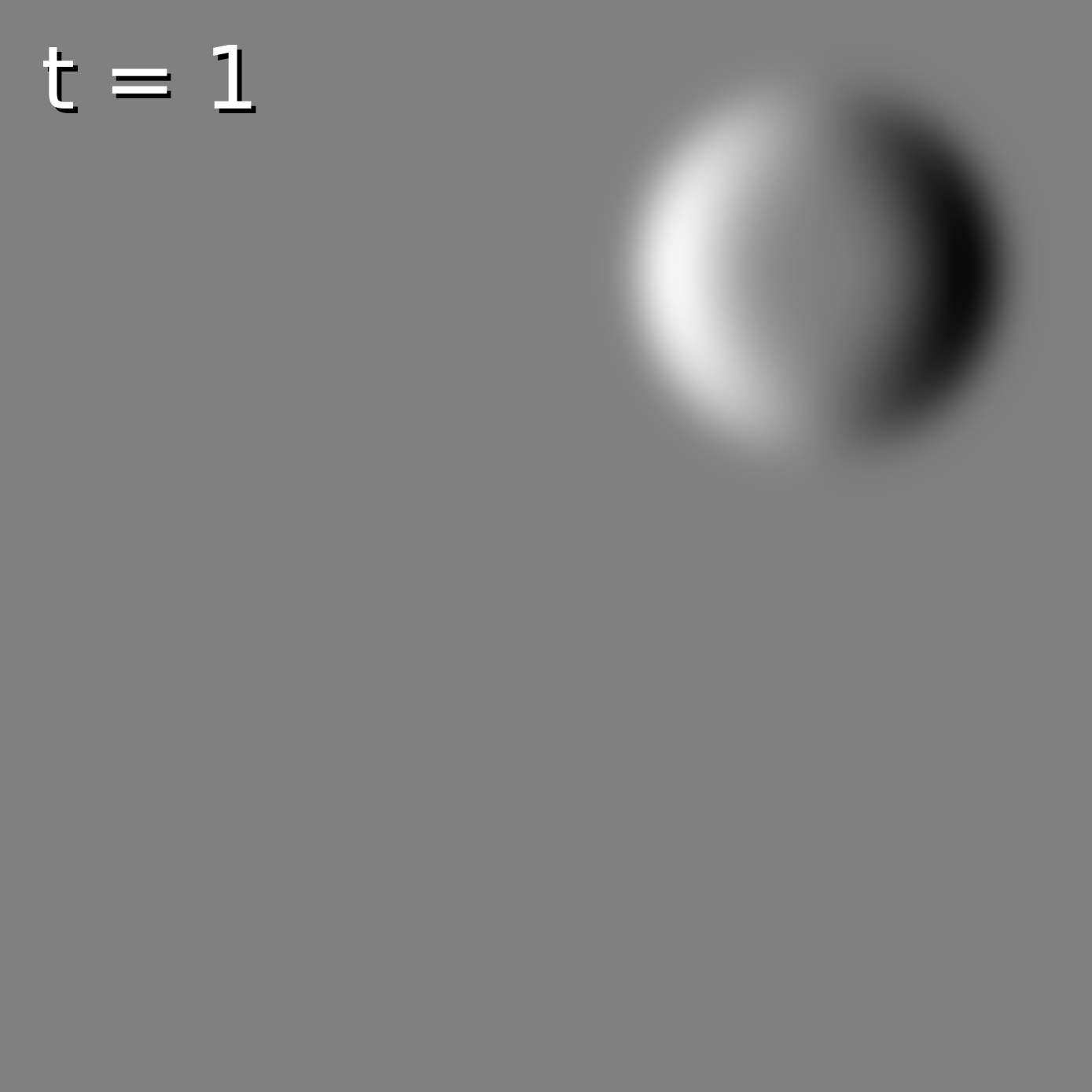} \\
\includegraphics[width=0.24\linewidth]{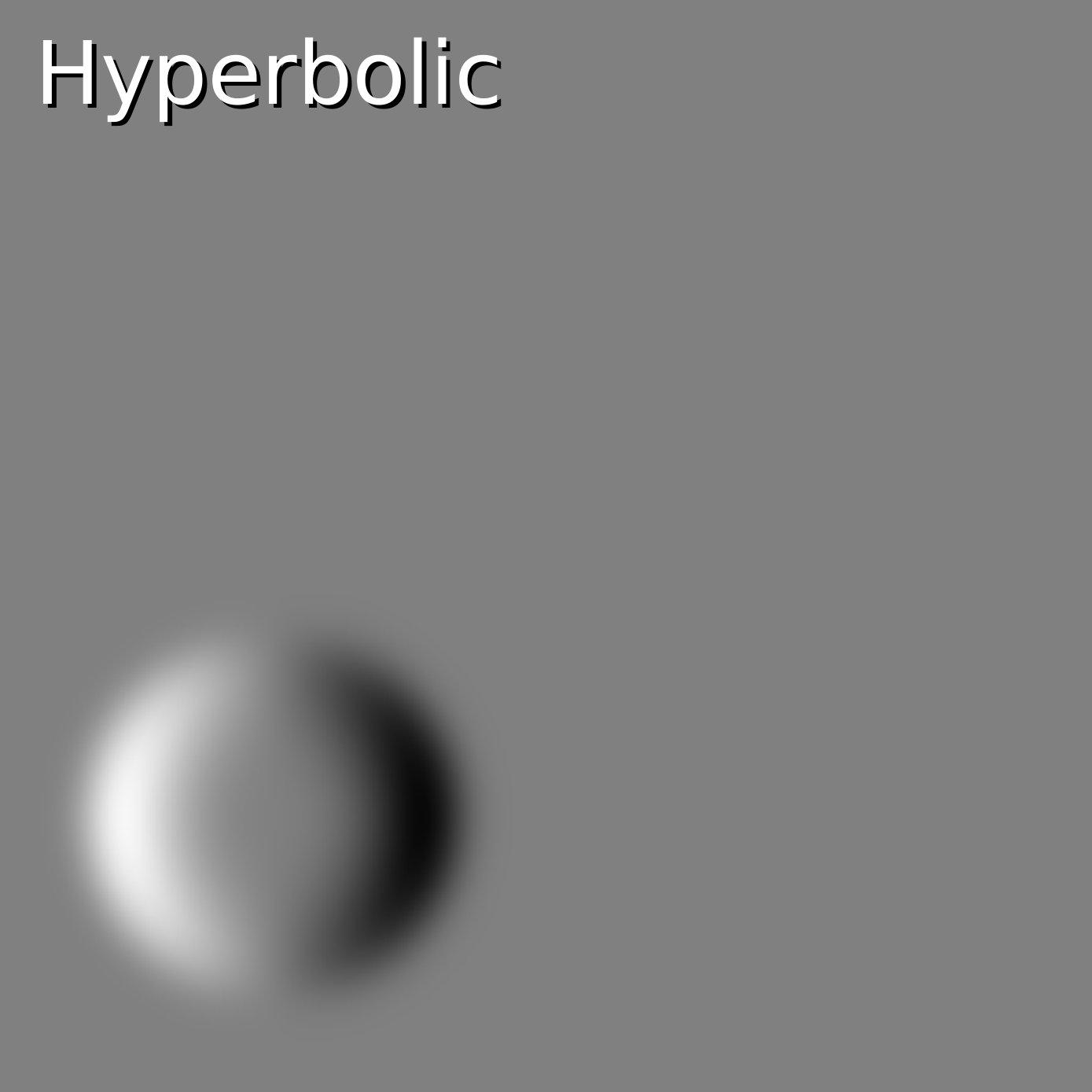}
\includegraphics[width=0.24\linewidth]{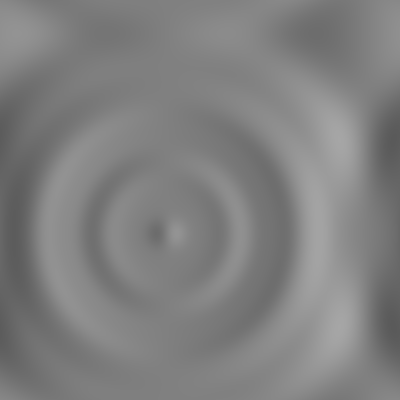}
\includegraphics[width=0.24\linewidth]{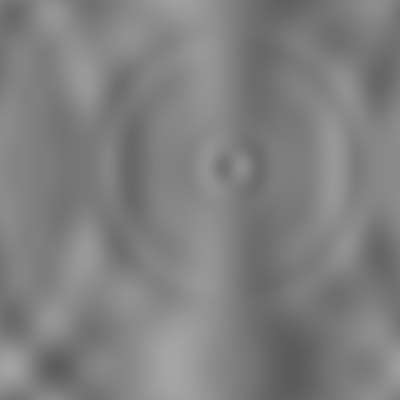}
\includegraphics[width=0.24\linewidth]{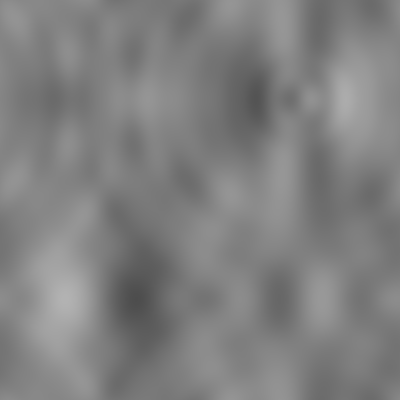} \\
\includegraphics[width=0.24\linewidth]{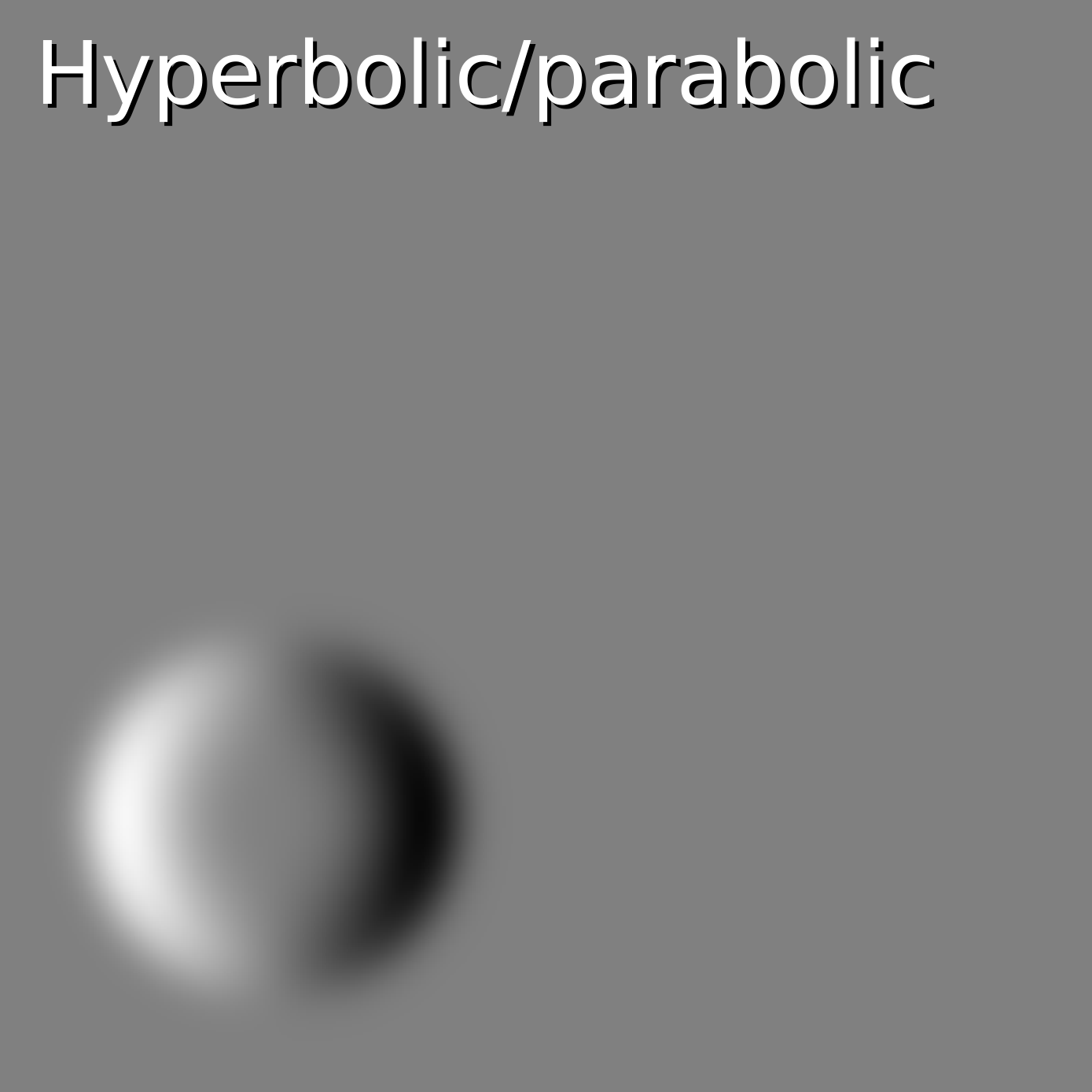}
\includegraphics[width=0.24\linewidth]{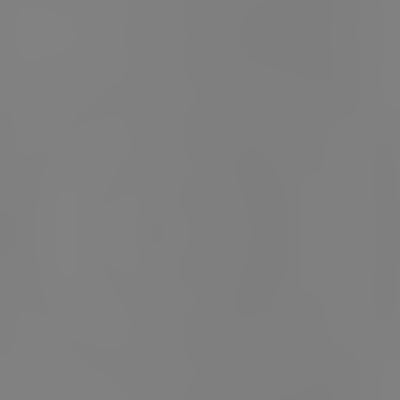}
\includegraphics[width=0.24\linewidth]{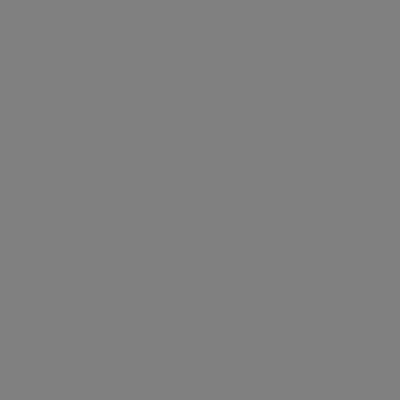}
\includegraphics[width=0.24\linewidth]{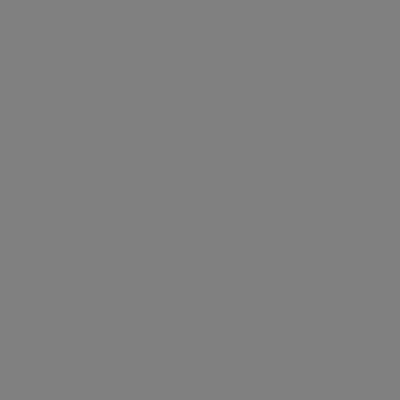} 
\end{minipage}\hspace{-2mm}
\begin{minipage}{0.1\textwidth}
\includegraphics[height=6.3\textwidth]{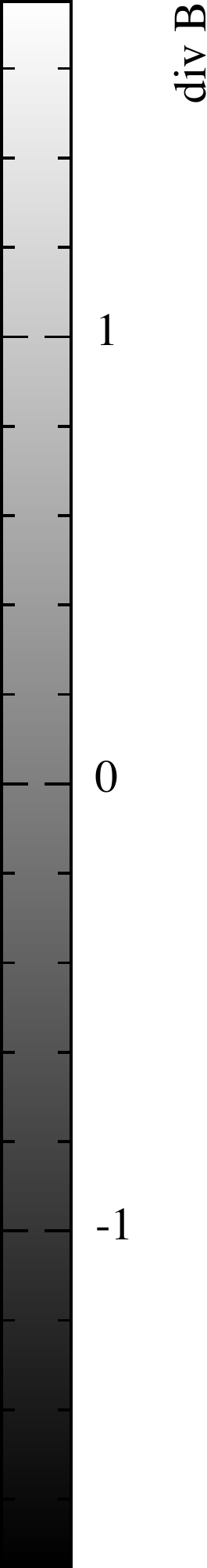}
\end{minipage}
\caption{Our fiducial model, used in the series of idealised tests, where fluid flowing towards the top right has divergence error artificially introduced in the initial, otherwise uniform, magnetic field. The renderings show the divergence of the magnetic field at $t=0, 0.33, 0.66, 1.0$ (left to right). If no divergence cleaning is applied (top row), the error passively advects with the fluid flow. Using purely hyperbolic divergence cleaning (middle row), the divergence error is spread throughout the domain. With mixed hyperbolic/parabolic divergence cleaning (bottom row), the divergence error is quickly removed producing a clean field.}
\label{fig:fiducial}
\end{figure}

Fig.~\ref{fig:fiducial} shows the fiducial model performed without divergence cleaning, with purely hyperbolic divergence cleaning ($\sigma = 0$) and with mixed hyperbolic/parabolic divergence cleaning ($\sigma = 0.3$). Without divergence cleaning, the divergence error is passively advected with the flow. With hyperbolic divergence cleaning, the error is spread throughout the domain as a series of waves, reducing the maximum value of divergence error. Coupling parabolic diffusion with hyperbolic cleaning rapidly removes the error, reducing the average error in the simulation by $\sim 5$ orders of magnitude.

\subsection{Time-varying wave cleaning speed}
\label{sec:tests-timevary}

%% color bar options
%% 8.0 x 8.0 inch page, character height 1.55

\begin{figure}
\centering
\begin{minipage}{0.87\textwidth}
\includegraphics[width=0.24\linewidth]{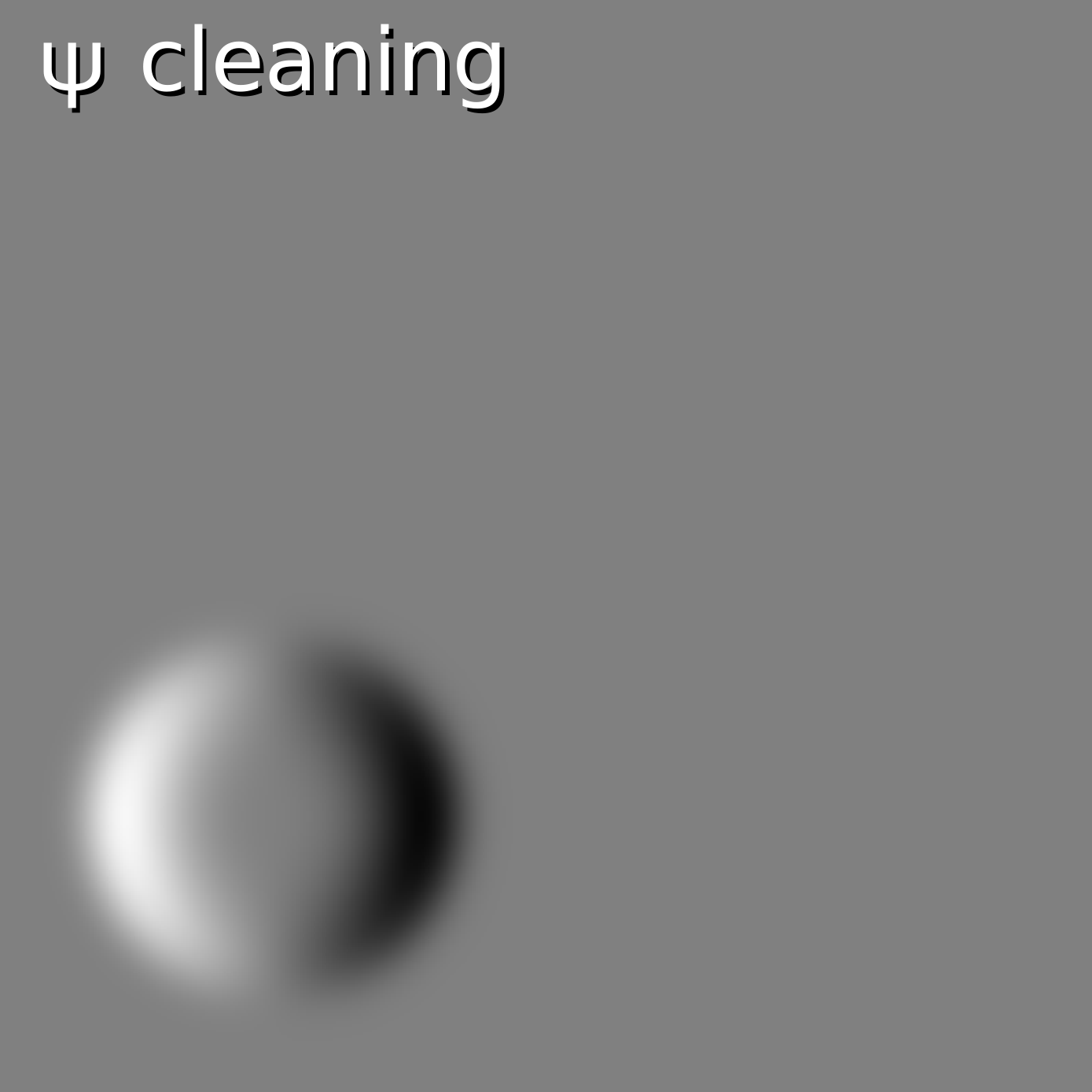}
\includegraphics[width=0.24\linewidth]{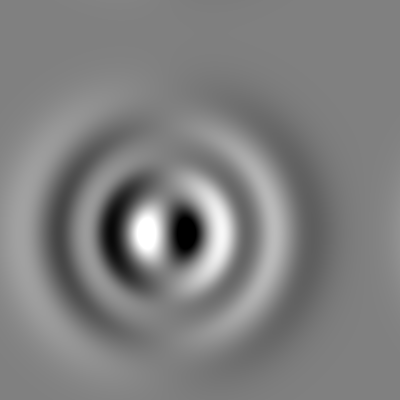}
\includegraphics[width=0.24\linewidth]{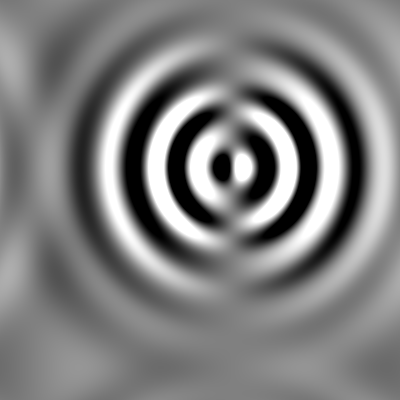}
\includegraphics[width=0.24\linewidth]{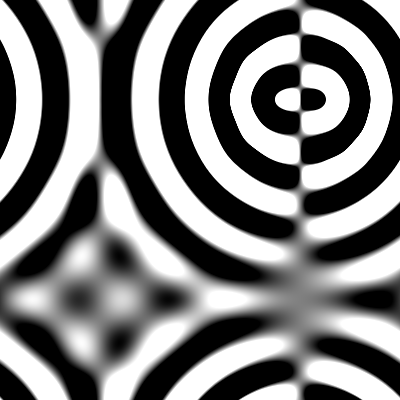} \\
\includegraphics[width=0.24\linewidth]{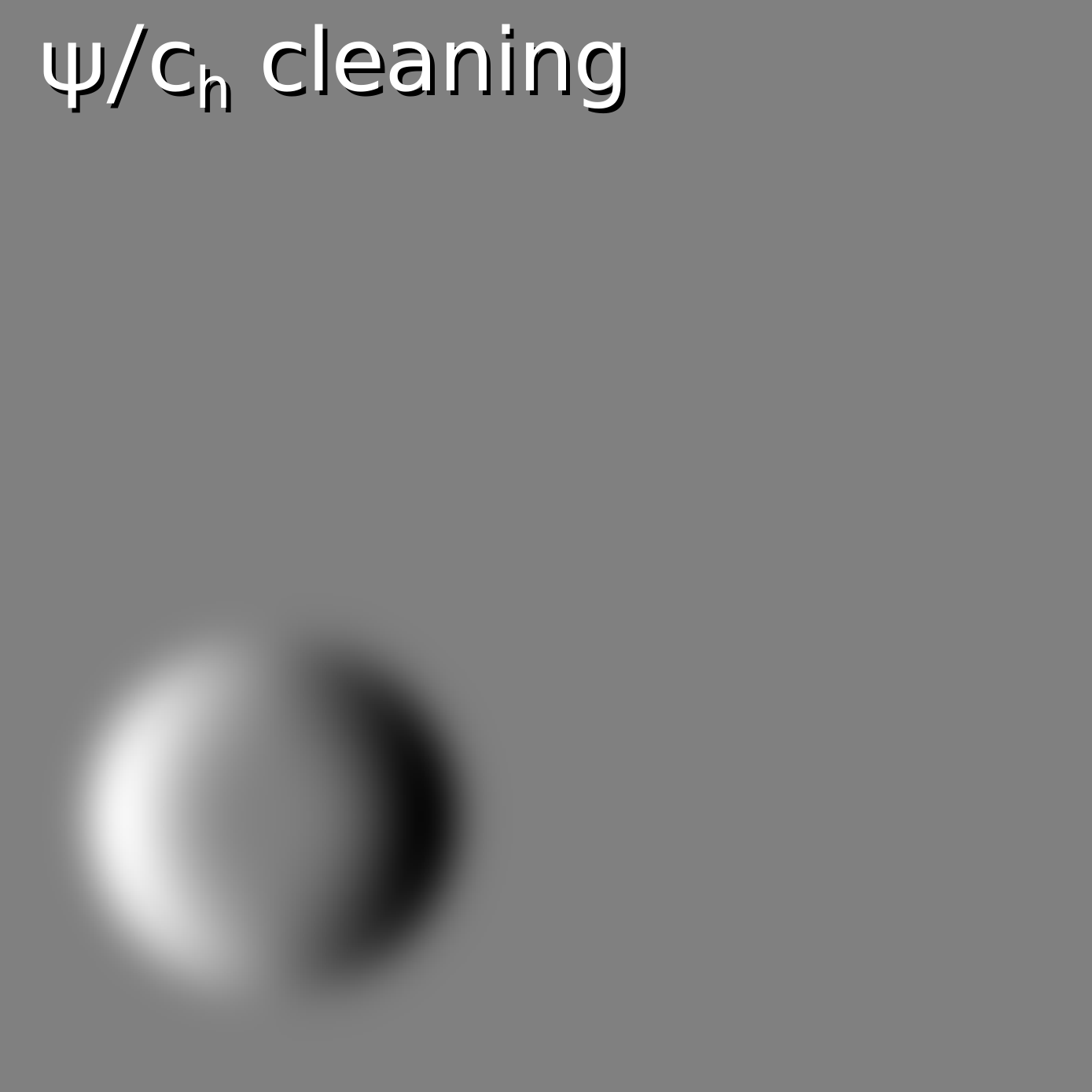}
\includegraphics[width=0.24\linewidth]{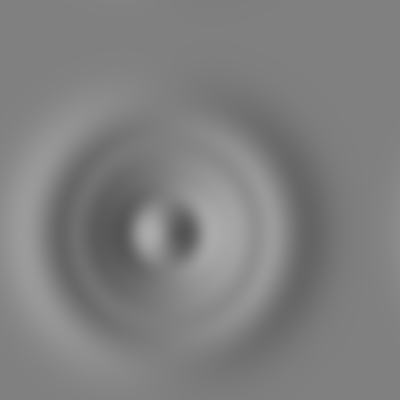}
\includegraphics[width=0.24\linewidth]{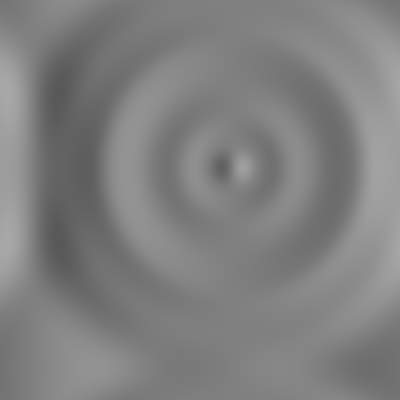}
\includegraphics[width=0.24\linewidth]{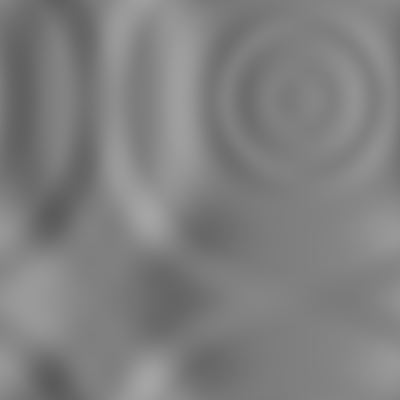} 
\end{minipage}\hspace{-2mm}
\begin{minipage}{0.1\textwidth}
\includegraphics[height=4.21\textwidth]{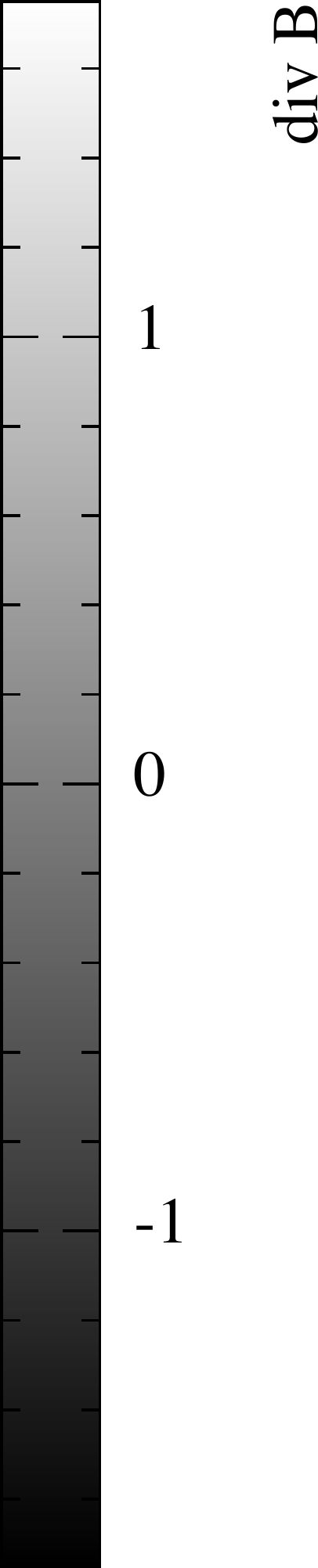}
\end{minipage}
\caption{Advection of a divergence blob using purely hyperbolic cleaning ($\sigma = 0$) where the divergence cleaning wave speed, globally for all particles, alternates between $c_{\rm h} = 1$ and $c_{\rm h} = 2$ every $t = 0.05$. Renderings are shown at $t=0, 0.33, 0.66, 1$ (left to right). The top row uses the original divergence cleaning approach, which does not account for this time variation. This leads to spurious energy generation causing runaway growth of divergence error in the magnetic field. The bottom row uses the updated divergence cleaning approach to evolve $\psi / c_{\rm h}$, naturally accounting for changes in the wave cleaning speed. For this case, energy is conserved, and no growth in divergence error occurs.}
\label{fig:timevary-render}
\end{figure}

\begin{figure}
\centering
\includegraphics[width=0.49\linewidth]{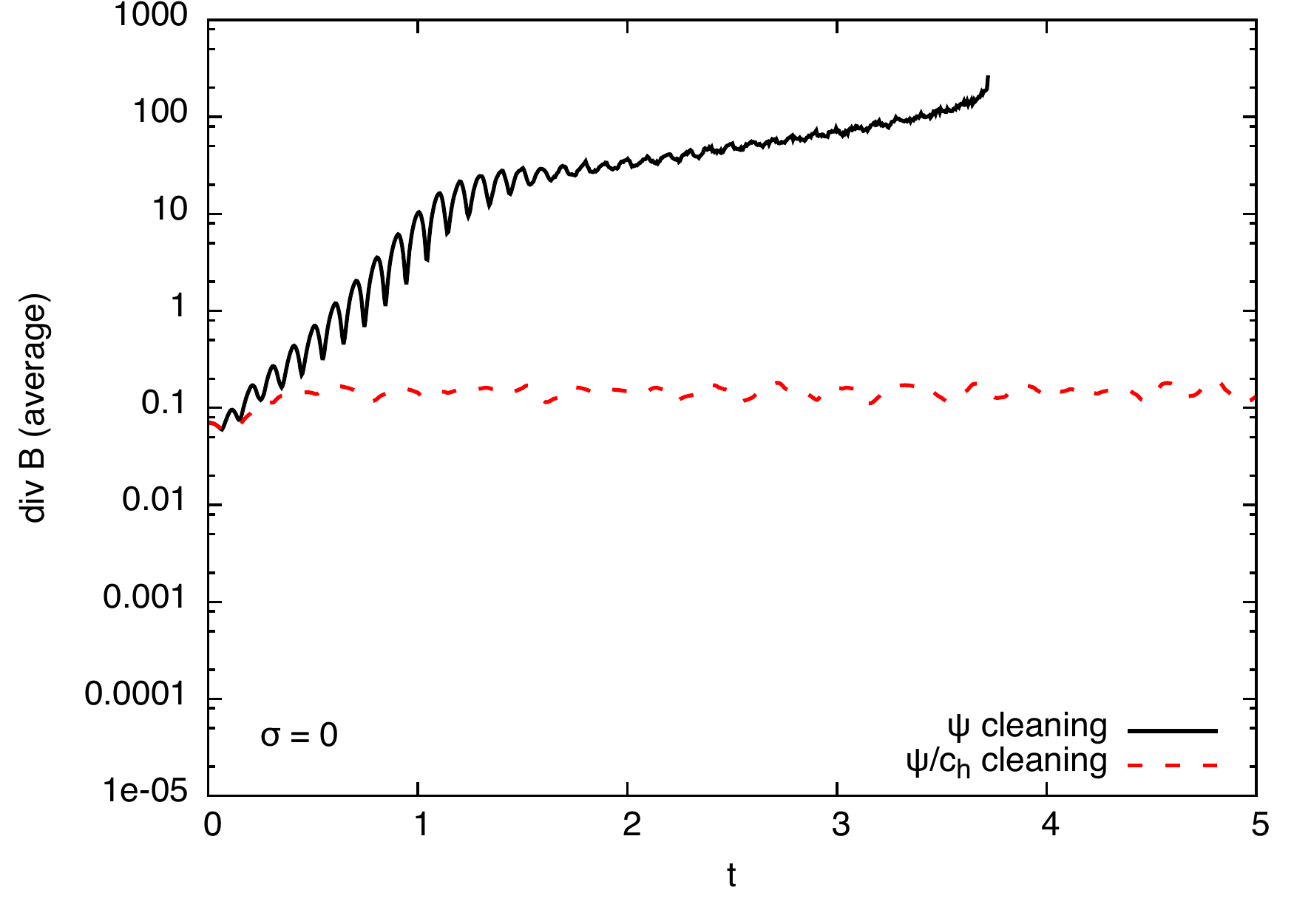}
\includegraphics[width=0.49\linewidth]{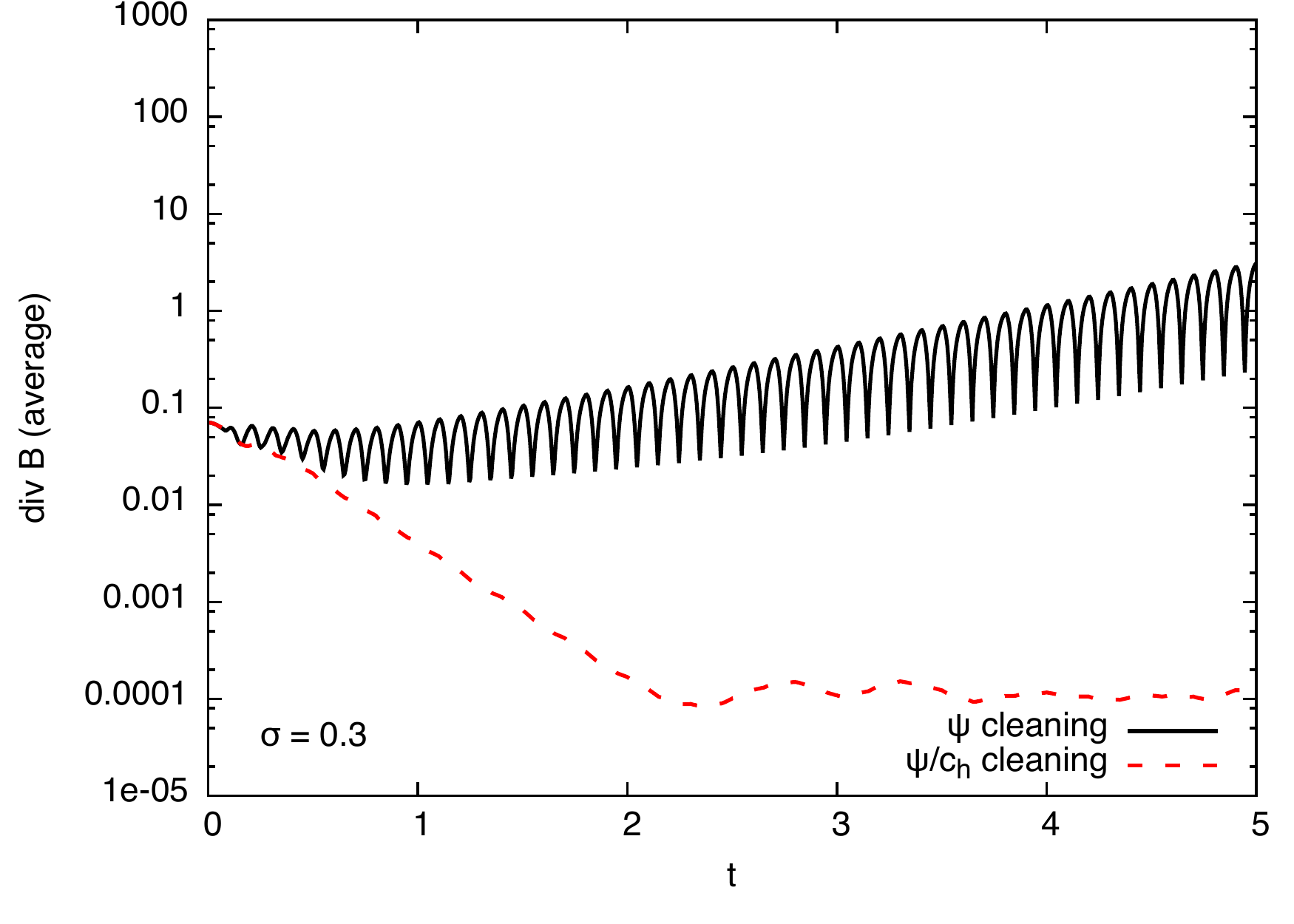}
\caption{Average divergence error as a function of time for the divergence advection test with a time-varying wave cleaning speed. The left panel is for purely hyperbolic cleaning ($\sigma = 0$) for the original divergence cleaning method (solid black line) and the new cleaning method evolving $\psi / c_{\rm h}$ (red dashed line). The original approach does not conserve energy in the presence of time variations of the wave cleaning speed, causing an increase in divergence error. At $t \sim 3.7$, the error is too large and the simulation crashes. By contrast, our new approach is conservative, and maintains divergence error at a constant level. We show results (right panel) for mixed hyperbolic/parabolic cleaning ($\sigma = 0.3$). In this case, the divergence error decays exponentially for the new method, in stark contrast to the original approach where the errors from non-conservation overpower the damping, causing the divergence error to increase.}
\label{fig:timevary-divb}
\end{figure}

Our primary goal is to show that our new divergence cleaning method addresses and fixes an issue related to a wave cleaning speed, $c_{\rm h}$, that varies in time. To test this, we use the fiducial model where the wave cleaning speed alternates between $c_{\rm h} = 1$ and $c_{\rm h} = 2$, changing every $t = 0.05$. The change in wave cleaning speed is globally applied to all particles, thus for any given timestep there is no spatial variation in $c_{\rm h}$ (this is tested separately in Section~\ref{sec:tests-xvary}).

Fig.~\ref{fig:timevary-render} shows renderings of the divergence of the magnetic field for the two divergence cleaning methods. When using the original method (top row), the divergence of the magnetic field propagates radially outwards from the initial divergence blob, but after $c_{\rm h}$ has undergone several variations, the divergence error is increased beyond the initial value. When the new cleaning method is used (bottom row), no increase in divergence error occurs and the propagation of waves proceeds in similar fashion to the fiducial model. The key difference is that, when using the original method, modifications to $c_{\rm h}$ result in a change of $e_\psi$ that is unaccounted for. In the new approach, when $c_{\rm h}$ is modified, it is balanced by a change to $\psi$ such that $e_\psi$ remains constant.

Fig.~\ref{fig:timevary-divb} shows the average divergence error as a function of time for purely hyperbolic and mixed hyperbolic/parabolic ($\sigma = 0.3$) divergence cleaning. For purely hyperbolic cleaning, the old method causes an exponential increase in divergence error which eventually destabilises the simulation. The new method keeps the average error to a near constant level. With parabolic damping included, the old cleaning approach still shows a long-term increase of divergence error, whereas the new cleaning method yields exponential decay of average divergence error, reproducing the behaviour of the fiducial model.

\subsection{Spatial discontinuities in the wave cleaning speed}
\label{sec:tests-xvary}

\begin{figure}
\centering
\includegraphics[width=0.49\linewidth]{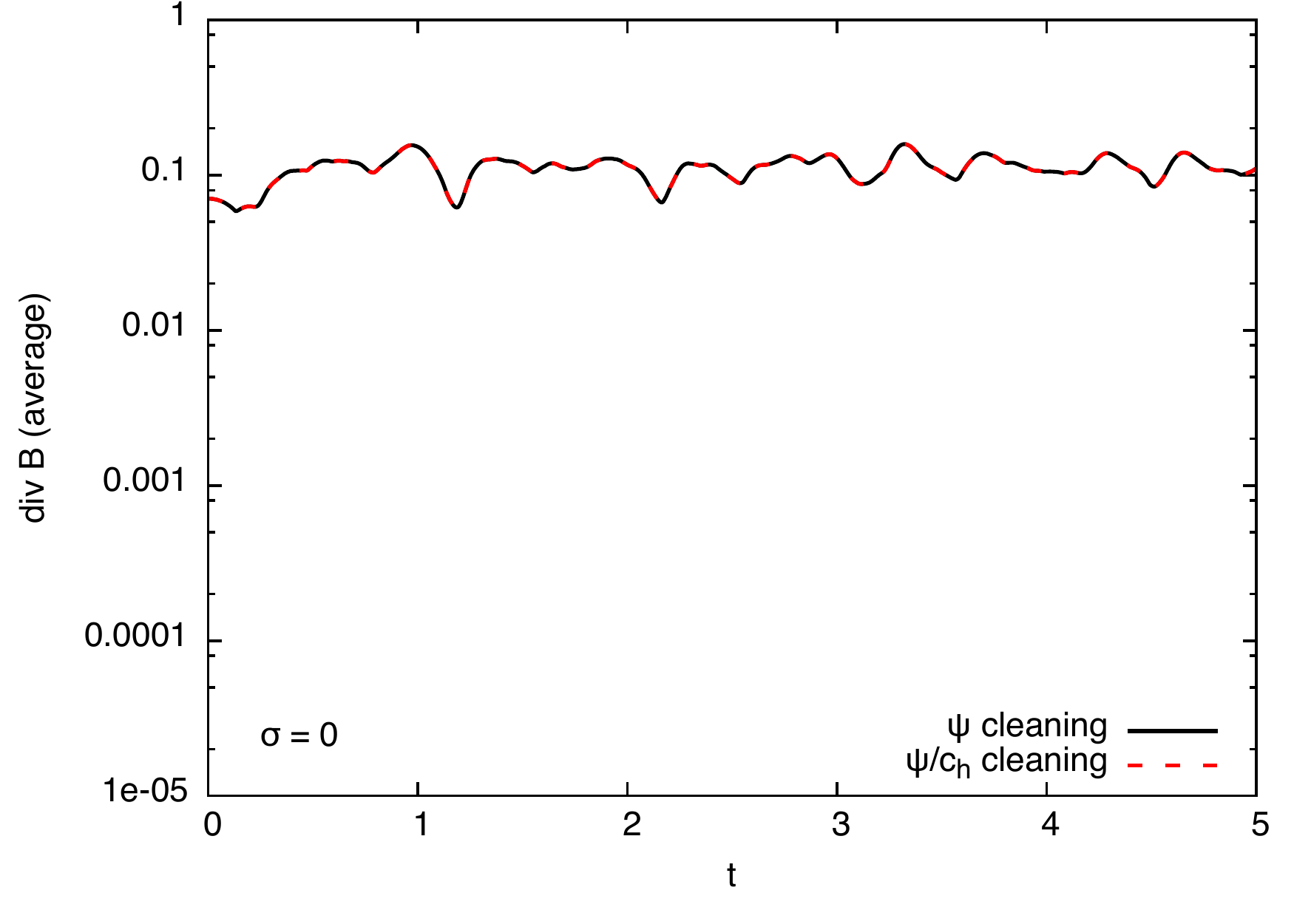}
\includegraphics[width=0.49\linewidth]{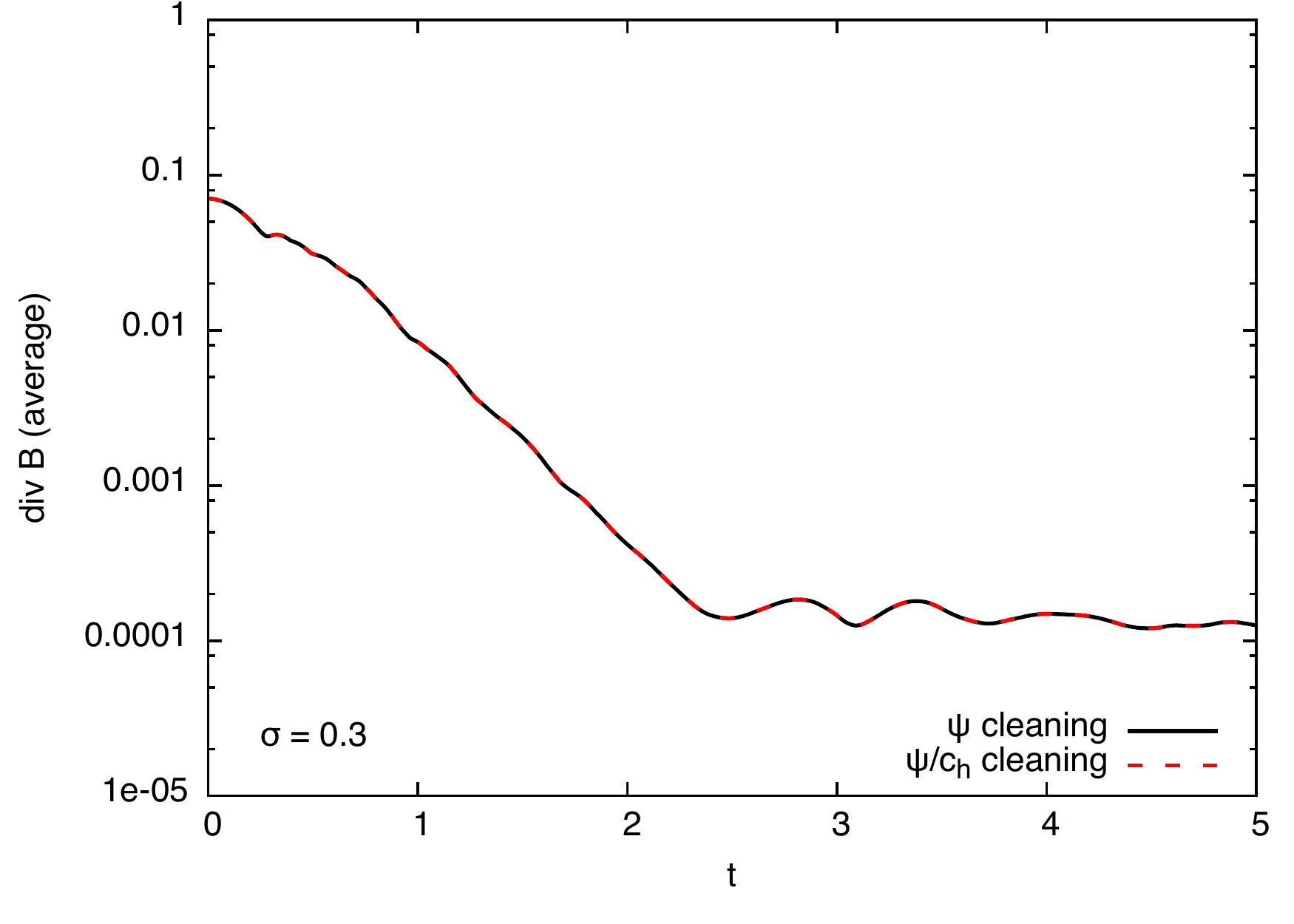}
\caption{Average divergence error for the divergence advection test with a spatially-varying wave cleaning speed. The left panel is for purely hyperbolic cleaning ($\sigma = 0$) and the right panel for mixed hyperbolic/parabolic cleaning ($\sigma = 0.3$). The solid black line is the original cleaning method, and the red dashed line the new cleaning method. Both cleaning methods yield identical results. We thus conclude that spatial variations in the wave cleaning speed do not introduce any errors into the magnetic field.}
\label{fig:xvary-divb}
\end{figure}

Now we introduce a spatial discontinuity into $c_{\rm h}$ instead of a time-variable global wave cleaning speed. This is an important case to investigate so that it can be determined if any problems occur when divergence waves cross between regions of differing wave cleaning speed, and furthermore, whether errors arise as a result of the communication between neighbouring particles which have differing wave cleaning speeds. 

The spatial variation is introduced by assigning $c_{\rm h} = 1$ to particles which have initial position $y < 0.5$, otherwise, they are assigned $c_{\rm h} = 2$. During the course of the simulation, each particle holds its assigned value fixed even though they move, thus, ${\rm d}c_{\rm h}/{\rm d}t = 0$ for each particle.

Fig.~\ref{fig:xvary-divb} shows the average divergence error for this test using both the old divergence cleaning approach evolving $\psi$, and the new approach evolving $\psi / c_{\rm h}$.  The two methods produce identical results, yielding a steady level of average divergence error for purely hyperbolic cleaning and exponential decrease in error for mixed hyperbolic/parabolic cleaning. We conclude that spatial variations in $c_{\rm h}$ do not affect the effectiveness or robustness of divergence cleaning.

\subsection{Discontinuities in $\tau$}
\label{sec:tests-disctau}

\begin{figure}
\centering
\includegraphics[width=0.49\linewidth]{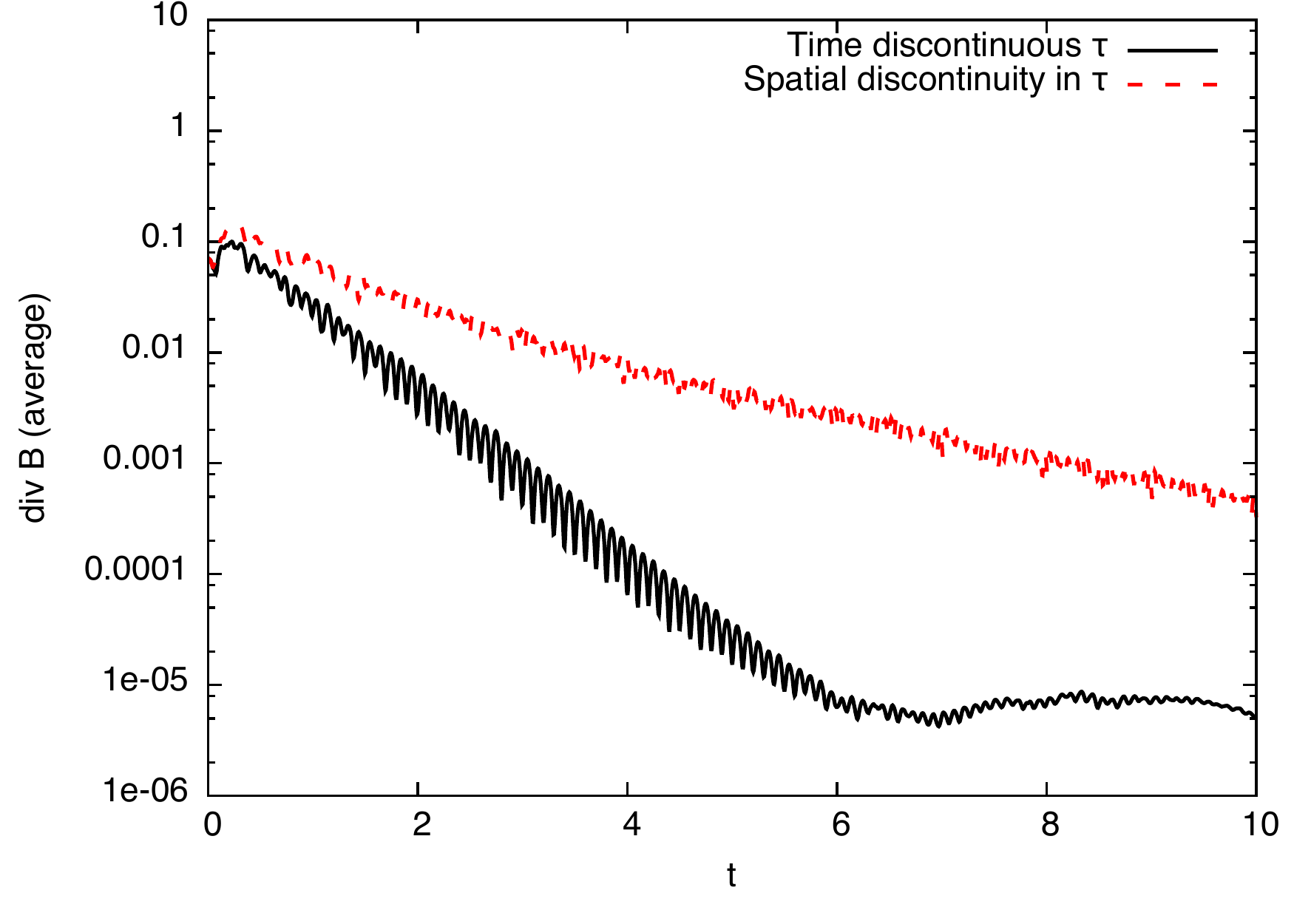}
\includegraphics[width=0.49\linewidth]{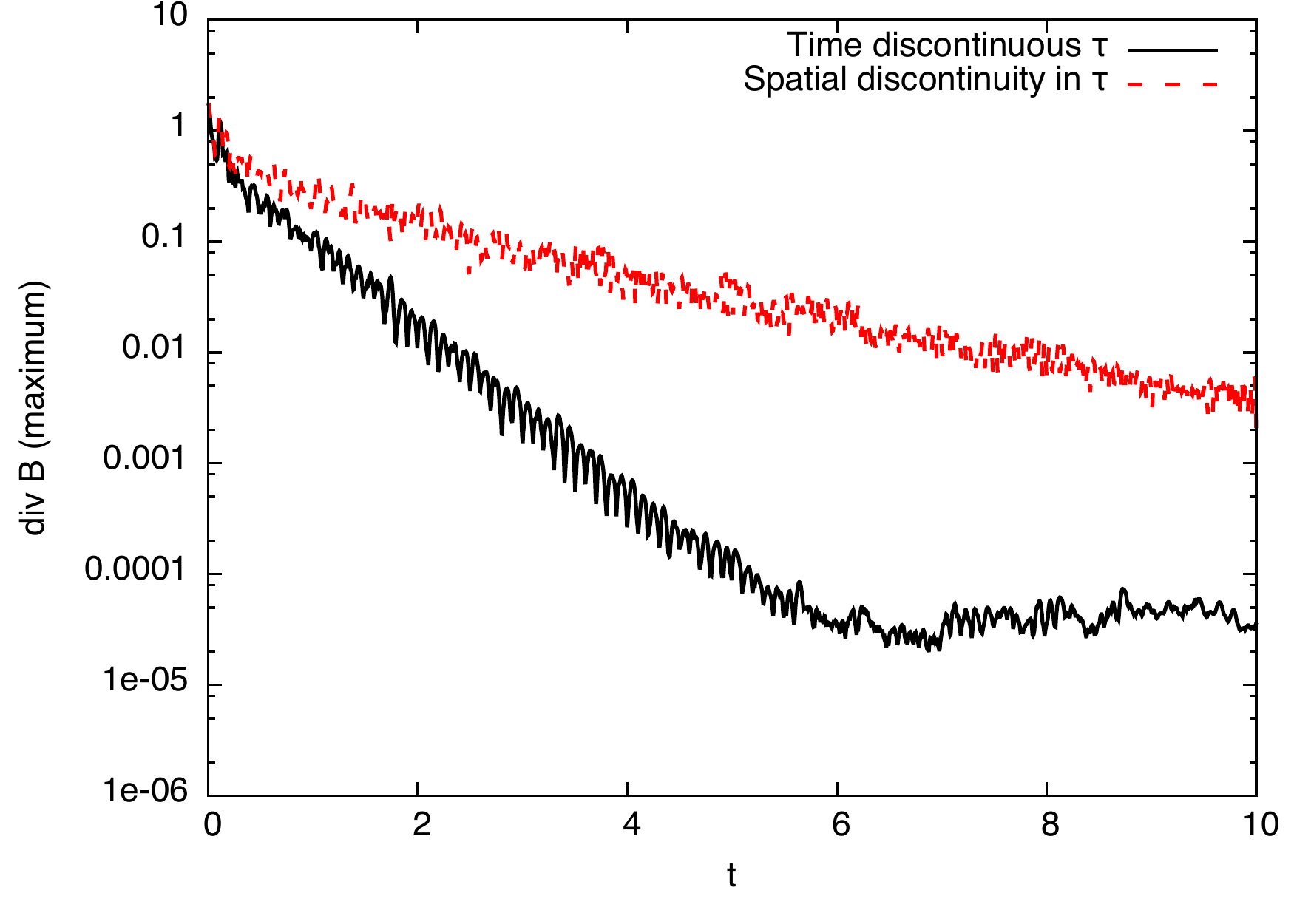}
\caption{Average (left) and maximum (right) divergence error for the divergence advection test where $\tau$ changes discontinuously in time (black solid line) and has a spatial discontinuity (red dashed line). The discontinuity in both cases is introduced by a $10$:$1$ ratio in $\sigma$. Both cases yield exponential decay of average and maximum divergence error, showing no evidence that errors are introduced by variations in $\tau$.}
\label{fig:tau-divb}
\end{figure}

Having tested discontinuities in $c_{\rm h}$, now we consider whether discontinuities in $\tau$ may lead to numerical error. Two variable $\tau$ cases are explored --- time variations and spatial variations, mirroring the tests performed for $c_{\rm h}$. To isolate any errors encountered as the result of variations in $\tau$ alone (as a reminder, $\tau = h / \sigma c_{\rm h}$), these tests use a fixed $c_{\rm h} = c_{\rm s}$ while $\sigma$ is varied. For the test with time variations, $\sigma$ is set globally for all particles, alternating between $0.1$ and $0.01$ every $t = 0.05$. For the spatial variations, half the particles are initially assigned $\sigma = 0.01$ if $y < 0.5$, otherwise $\sigma = 0.1$. In this case each particle holds fixed its assigned value of $\sigma$ so that no there is no time change. Both cases represent a $10$:$1$ discontinuity in $\tau$, larger than the ratio used in the tests of $c_{\rm h}$. The values of $\sigma = 0.1$ and $\sigma = 0.01$ are intentionally chosen to be weaker than the damping typically employed ($\sigma = 0.3$) so that the decay of $e_\psi$ happens only slowly and any errors which may be introduced are not rapidly removed.

Fig.~\ref{fig:tau-divb} shows the average and maximum divergence error for the two test cases. For both cases, the divergence error undergoes exponential decay, showing no evidence that variations in $\tau$ have any detriment on the effectiveness of the divergence cleaning. The time-varying calculation has a faster decay rate of divergence error than the spatially-varying calculation, even though both use values of $\sigma=0.1$ and $\sigma=0.01$. This is due to the spatially-varying calculation having a persistent low $\sigma$ region, since particles are split half and half between the low and high values. By contrast, for the time-varying calculation every particle will use both values of $\sigma$ throughout the calculation. Overall, it does not appear that there is any adverse effect by discontinuities in $\tau$.

\subsection{Eulerian vs. Lagrangian derivatives}
\label{sec:tests-advection}

\begin{figure}
\centering
\begin{minipage}{0.87\textwidth}
\includegraphics[width=0.24\linewidth]{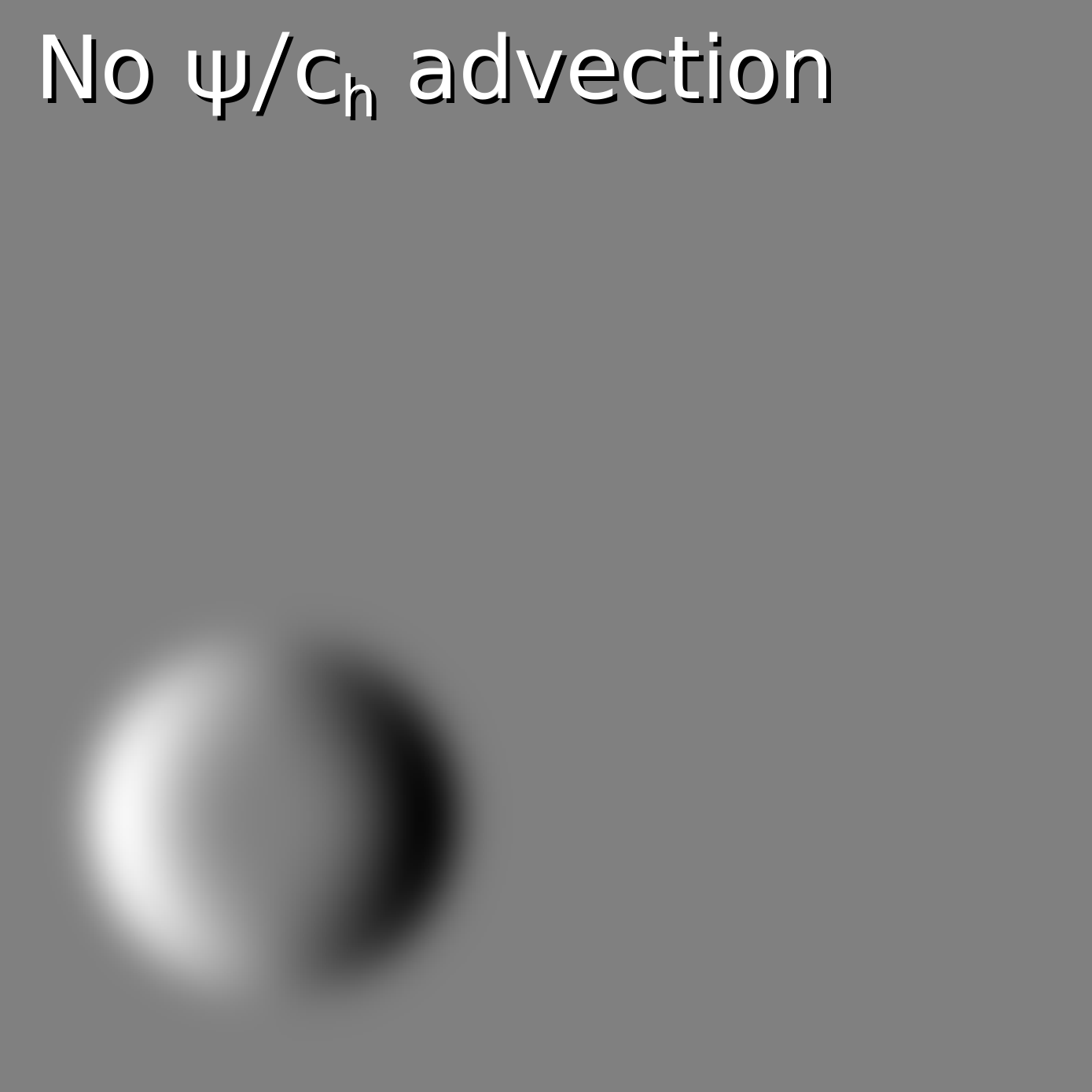}
\includegraphics[width=0.24\linewidth]{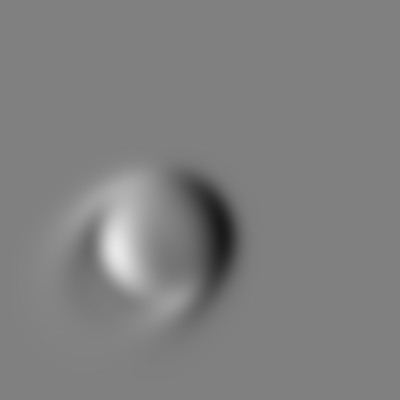}
\includegraphics[width=0.24\linewidth]{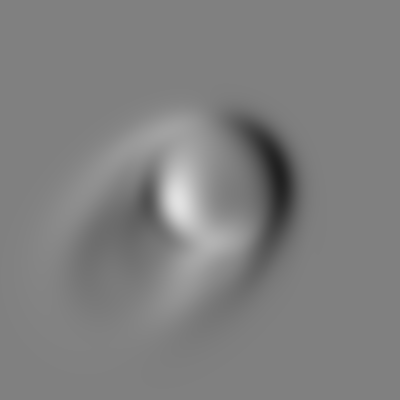}
\includegraphics[width=0.24\linewidth]{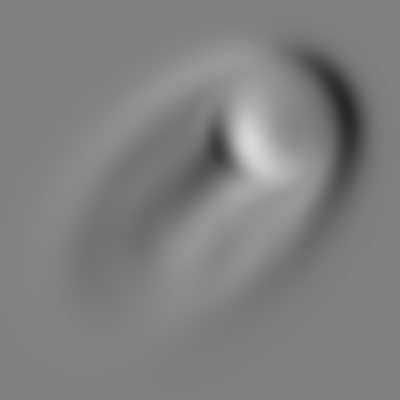} \\
\includegraphics[width=0.24\linewidth]{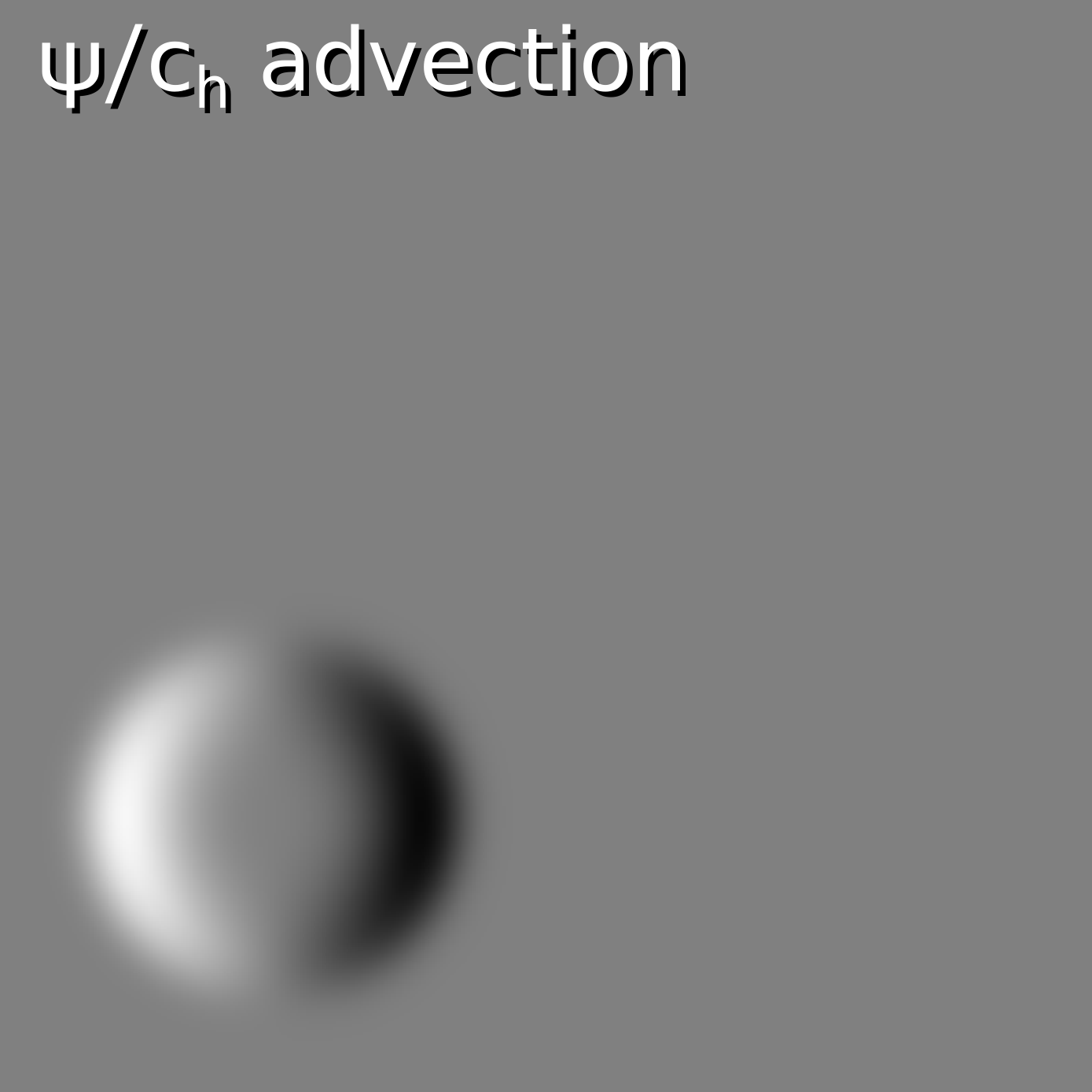}
\includegraphics[width=0.24\linewidth]{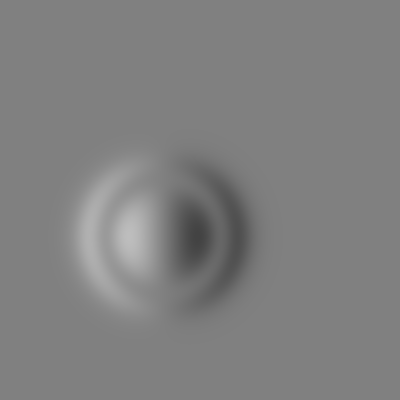}
\includegraphics[width=0.24\linewidth]{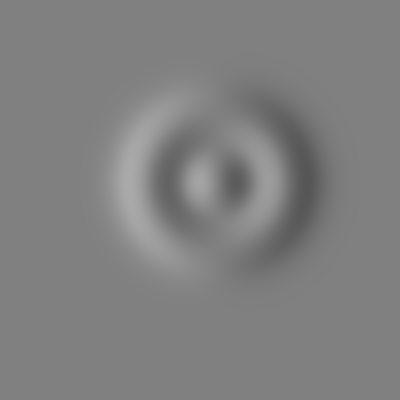}
\includegraphics[width=0.24\linewidth]{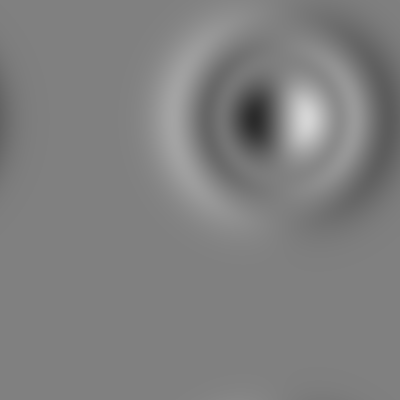} 
\end{minipage}\hspace{-2mm}
\begin{minipage}{0.1\textwidth}
\includegraphics[height=4.21\textwidth]{psiwall-colorbar2.pdf}
\end{minipage}
\caption{Advection of a divergence blob where the cleaning equations have been implemented using Eulerian derivatives (no advection of $\psi / c_{\rm h}$; top row) and Lagrangian derivatives (standard approach; bottom row) for fluid velocity at $\mathcal{M}=4$. The renderings are shown for $t = 0, 0.033, 0.066$ and $0.1$ (left to right, respectively). Using Eulerian derivatives leads to streaking of divergence error across the box due to $\psi / c_{\rm h}$ attempting to remain in the spatial location it was generated, rather than being advected with the fluid motion.}
\label{fig:eulerian-render}
\end{figure}

\begin{figure}
\centering
\includegraphics[width=0.49\linewidth]{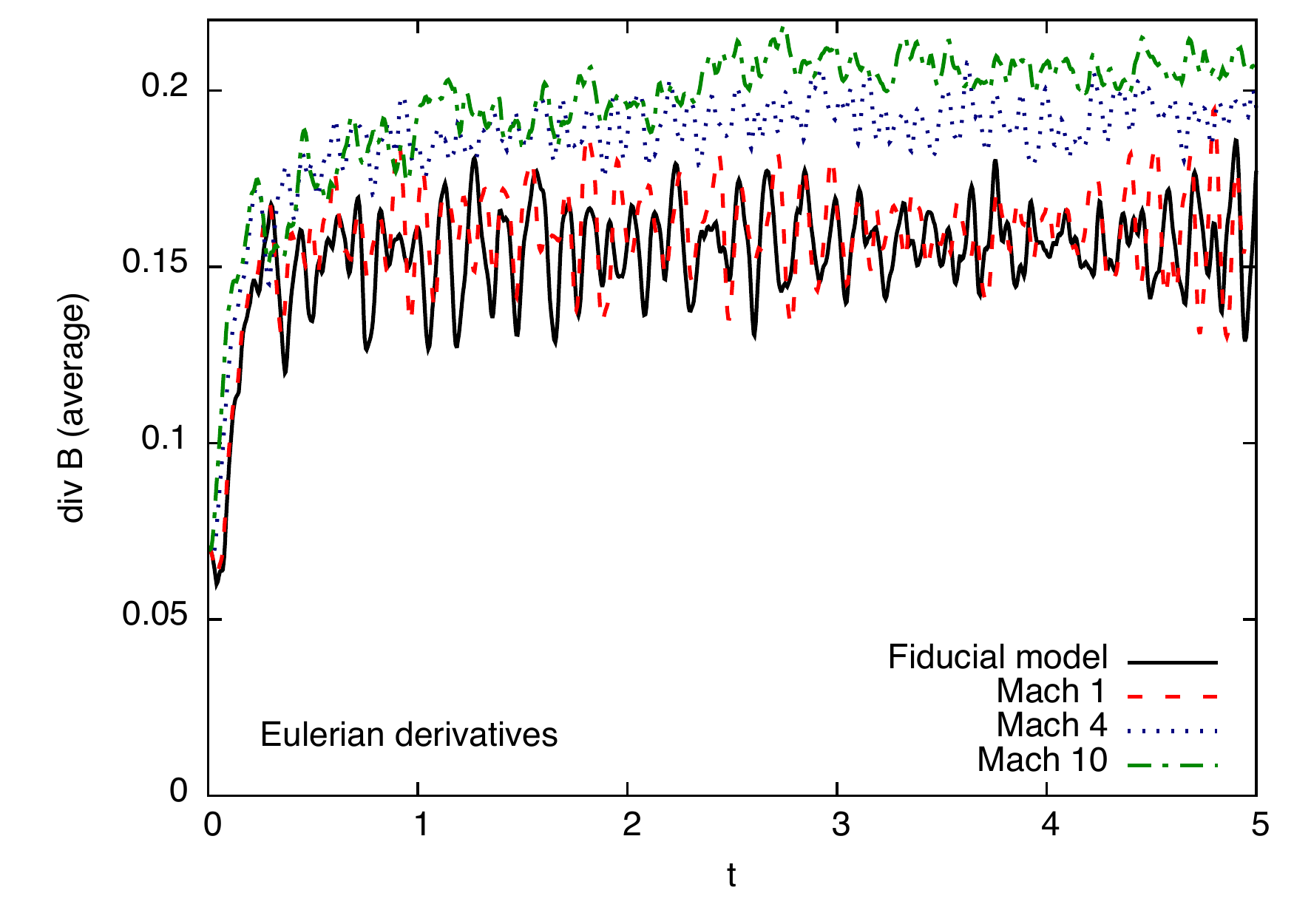}
\includegraphics[width=0.49\linewidth]{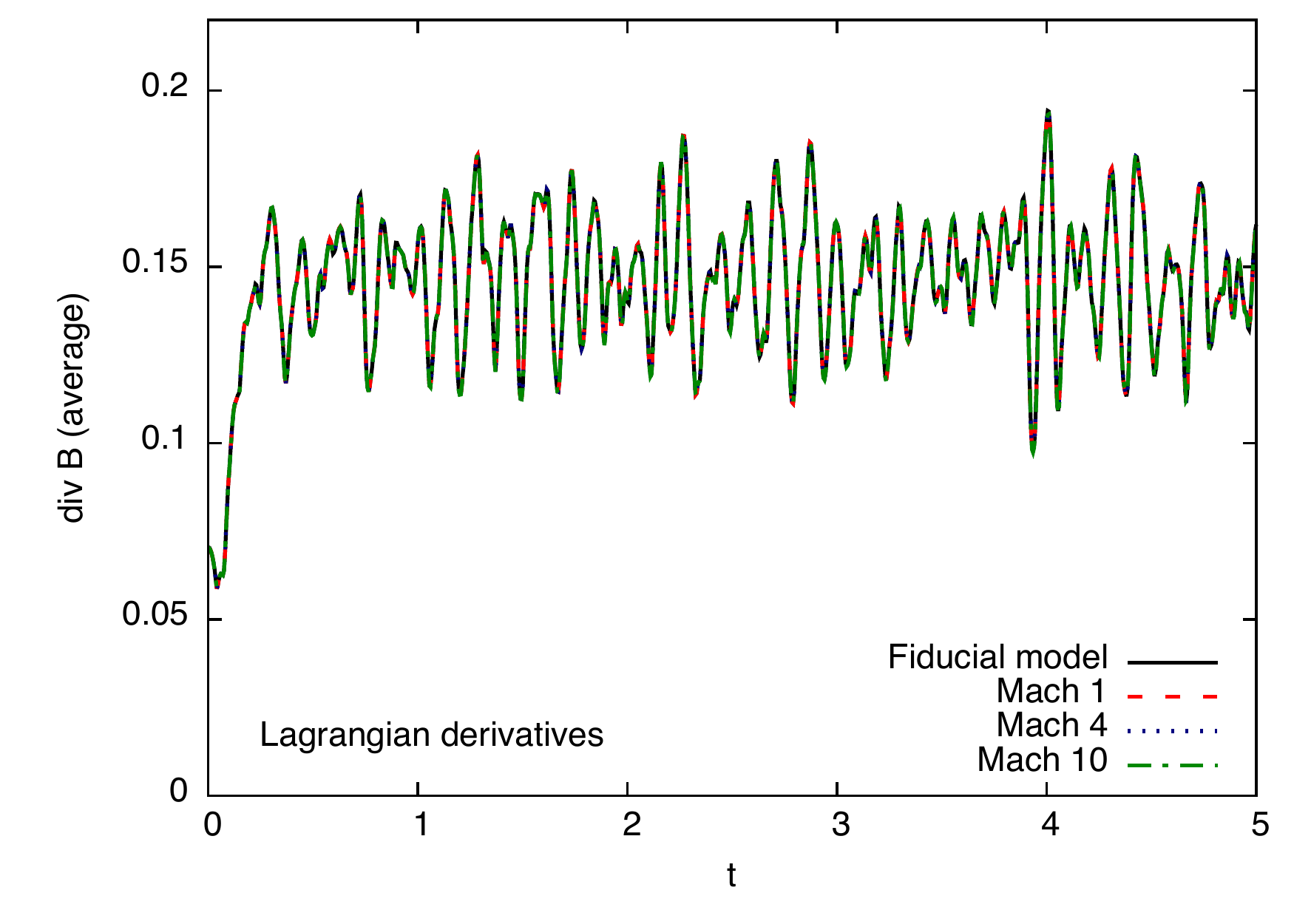}
\caption{Average divergence error for the cleaning equations implemented with Eulerian derivatives (left panel) and Lagrangian derivatives (standard approach; right panel), with the velocity of the fluid increased from the fiducial model ($\mathcal{M}=0.45$) up to $\mathcal{M}=10$. There is a $\sim30\%$ increase in average divergence error going from $\mathcal{M} = 0.45$ to $\mathcal{M} = 10$ when using Eulerian derivatives. With Lagrangian derivatives, the divergence cleaning is agnostic to the fluid velocity and produces identical results for all Mach numbers.}
\label{fig:eulerian-divb}
\end{figure}

The SPMHD cleaning equations in our method are implemented using Lagrangian derivatives (${\rm d} / {\rm d}t \equiv \partial / \partial t + {\bf v} \cdot \nabla$), which means that $\psi / c_{\rm h}$ is advected with the fluid flow. However, the original paper by \citet{dedneretal02} used Eulerian derivatives, so such advection was not part of their scheme. Here, we investigate the effect of advecting $\psi / c_{\rm h}$ by comparing our standard implementation using Lagrangian derivatives with an implementation using Eulerian derivatives. The latter is implemented by adding a `reverse advection' term to our standard implementation, in essence counteracting the Lagrangian nature of SPMHD.

To test this, we compare results of the fiducial model, where the `blob' moves subsonically at Mach number $\mathcal{M} = 0.45$, to simulations where the velocity of the fluid has been increased to $\mathcal{M} = 1, 2, 4$ and $10$. The high value of $\mathcal{M} = 10$ is motivated by our desire to simulate molecular clouds.

Fig.~\ref{fig:eulerian-render} shows renderings of the divergence error in the magnetic field for the $\mathcal{M}=4$ calculations. For the implementation using Eulerian derivatives, the divergence error is smeared behind the initial divergence blob. As energy is transferred from ${\bf B}$ to $e_\psi$, $\psi / c_{\rm h}$ remains in the spatial location it was generated, rather than remain with the fluid. For our default implementation using Lagrangian derivatives, the divergence waves can be seen to propagate symmetrically outwards from the central divergence blob since ${\bf B}$ and $\psi / c_{\rm h}$ are co-moving with the fluid. This demonstrates the hyperbolic propagation in the co-moving frame described by Equation~(\ref{eq:genwave}).

Fig.~\ref{fig:eulerian-divb} shows the average divergence error for both implementations. When Eulerian derivatives are used, the average error increases with the background velocity of the fluid, with the average error of the $\mathcal{M}=10$ calculation $\sim30\%$ larger than the subsonic fiducial model. While it does appear that there are larger variations in the average error for the calculations using Lagrangian derivatives, they all yield identical results and do not show any increase in average error due to the fluid velocity.

An important argument against using Eulerian derivatives for hyperbolic divergence cleaning in SPMHD is that it introduces a velocity dependence into the Courant timestep criterion, as is the case for grid codes. This loses one of the advantages SPMHD has over grid-based methods, in that there is no timestep restriction from the local fluid velocity since it inherently handles advection as part of the method. Adding a `reverse advection' term disables this advantage (for no benefit).

\subsection{Compression and rarefaction of $\psi / c_{\rm h}$}
\label{sec:tests-divv}

\begin{figure}
\centering
\includegraphics[width=0.49\linewidth]{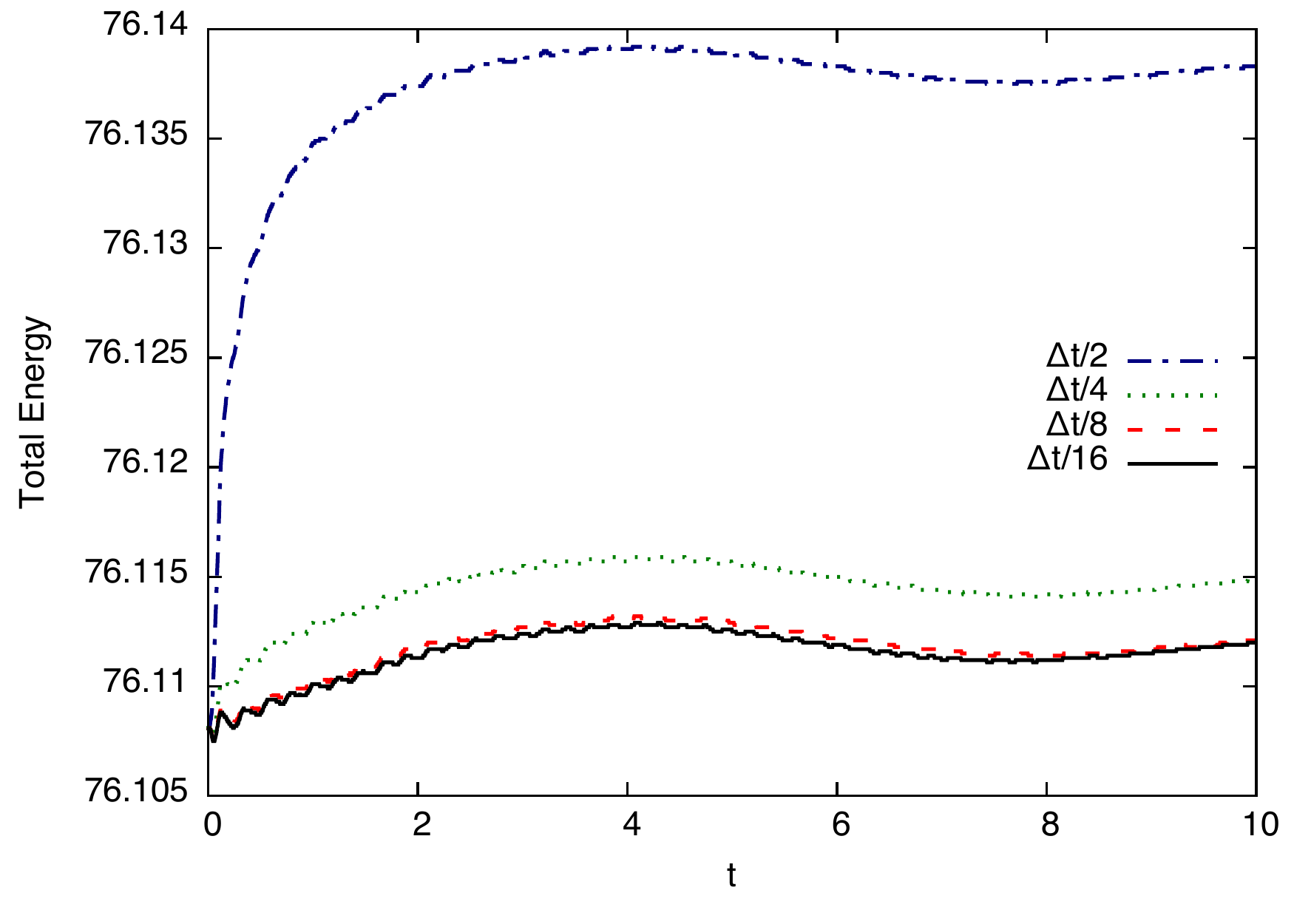}
\includegraphics[width=0.49\linewidth]{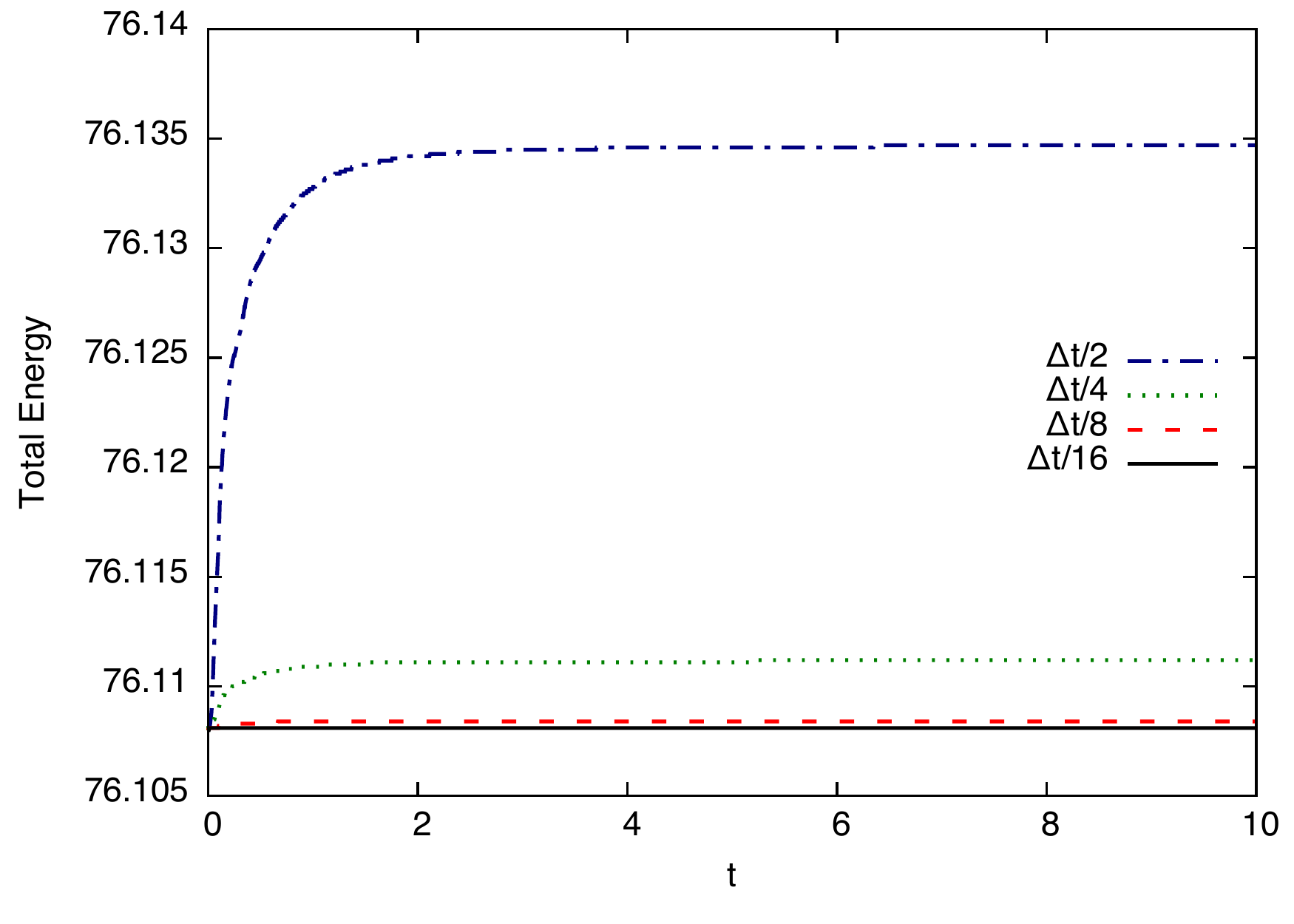}
\caption{Total energy (kinetic + thermal + magnetic + $e_\psi$) for the divergence advection test involving compression and rarefaction repeated for decreasing timestep sizes. The left panel shows results of a simulation which does not include the $\tfrac{1}{2} (\psi / c_{\rm h}) (\nabla\cdot{\bf v})$ term in the evolution equation for $\psi / c_{\rm h}$, thus changes in $\psi / c_{\rm h}$ when undergoing compression or rarefaction are not compensated for. As the errors from timestepping are reduced, the energy converges to a non-constant value, thus there exists a source of non-conservation of energy. The right panel includes the compression term, and the total energy converges to a constant value. Therefore, the compression term is indeed required to exactly conserve energy, though the non-conservation introduced by its absence is well below the level of errors from timestepping in a normal simulation.}
\label{fig:divv-nodivv}
\end{figure}

Our final algorithmic test is to confirm that the $\tfrac{1}{2} (\psi / c_{\rm h}) (\nabla \cdot {\bf v})$ term added to the evolution equation for $\psi / c_{\rm h}$ is indeed necessary to conserve energy in the presence of compression and rarefaction. To accomplish this, the velocity field is initialised to $v_x = 2 c_{\rm s} \sin(2 \pi x)$, such that the gas undergoes oscillating compression and rarefaction with initially supersonic velocities. The test is simulated with and without the compression term as part of the cleaning equations, using timesteps with Courant factors $C / 2^n$, where $C=0.3$ is the factor in the Courant condition and $n = 0, 1, 2, 3, 4$. The calculations use a second-order Runge-Kutta integrator.

Fig.~\ref{fig:divv-nodivv} shows the total energy (including $e_\psi$) for the calculations. The maximum density reaches $\rho \approx 3.5$ during the initial compression, after which the velocity becomes subsonic and the compressions give only $\sim 20 \%$ enhancements for the remainder of the calculation. The initial compression is reflected by a spike in total energy, caused by errors from the timestepping algorithm. For both sets of calculations, this error is reduced quadratically with decreasing timestep, as expected for a second-order integrator. 

For the calculations which do not include the compression term, the total energy does not converge to a constant value. For timesteps of size $C/8$ and $C/16$, the total energy exhibits a slow, long-term variation with an increase in energy over the lifetimes of the calculations. Additionally, there are short wavelength variations initially in the total energy. This implies that, for these timestep sizes, a source of error exists which is greater than that introduced by the timestepping. By comparison, the calculations including the compression term converge the total energy to a constant value as the errors from timestepping are decreased and show no initial short wavelength variations. Thus, the non-conservation of energy is resolved by the addition of the compression term. We conclude that the compression term is indeed technically required to conserve total energy in the presence of compression and rarefaction, however we note that the errors introduced by its absence are, generally, well below those due to the timestepping method in normal simulations.

\section{Practical tests}
\label{sec:practical-tests}

Now we turn attention to more standard MHD test problems. The tests chosen are the MHD blast wave \citep{bs99, ldz00}, Orszag-Tang vortex \citep{ot79} and the MHD rotor \citep{bs99}. All of these tests have been studied with SPMHD in previous works \citep{pm05, bot06, ds09}, and we report similar results here. Since constrained hyperbolic divergence cleaning was extensively tested in our previous paper \citep{tp12}, which included the blast wave and Orszag-Tang tests, our analysis is focused on the improvement, if any, the modified method has over the previous scheme. For all tests, the \citet{mm97} artificial viscosity switch (with $\alpha = [0.1, 1]$) and the \citet{tp13} artificial resistivity switch (with $\alpha_{\rm B} = [0, 1]$) have been used. We measure the divergence error, as usual, using the dimensionless quantity $h \vert \nabla \cdot {\bf B} \vert / \vert {\bf B} \vert$.

\subsection{Blast wave in a strongly magnetised medium}
\label{sec:tests-mhdblast}

\begin{figure}
\centering
\includegraphics[height=0.295\linewidth]{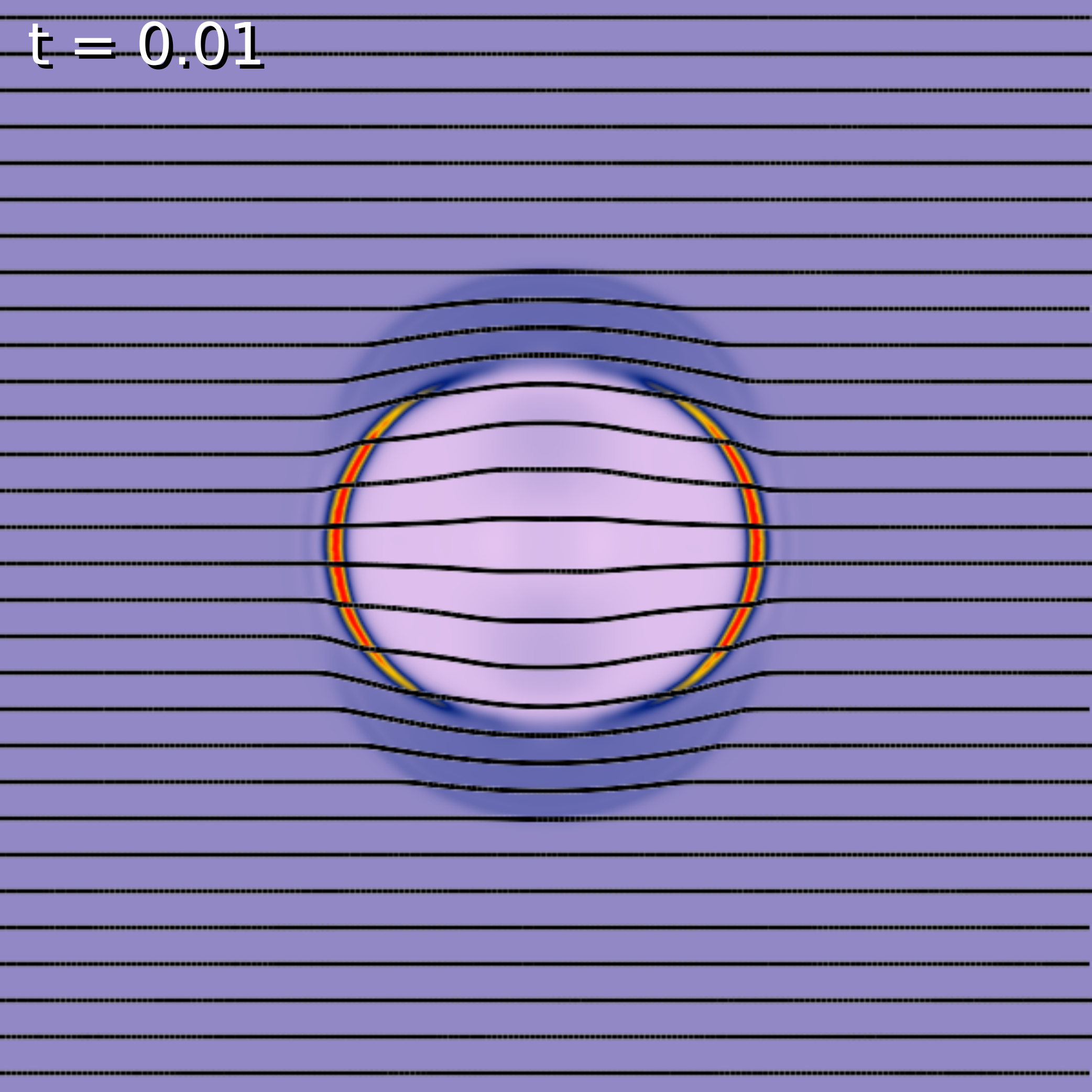}
\includegraphics[height=0.295\linewidth]{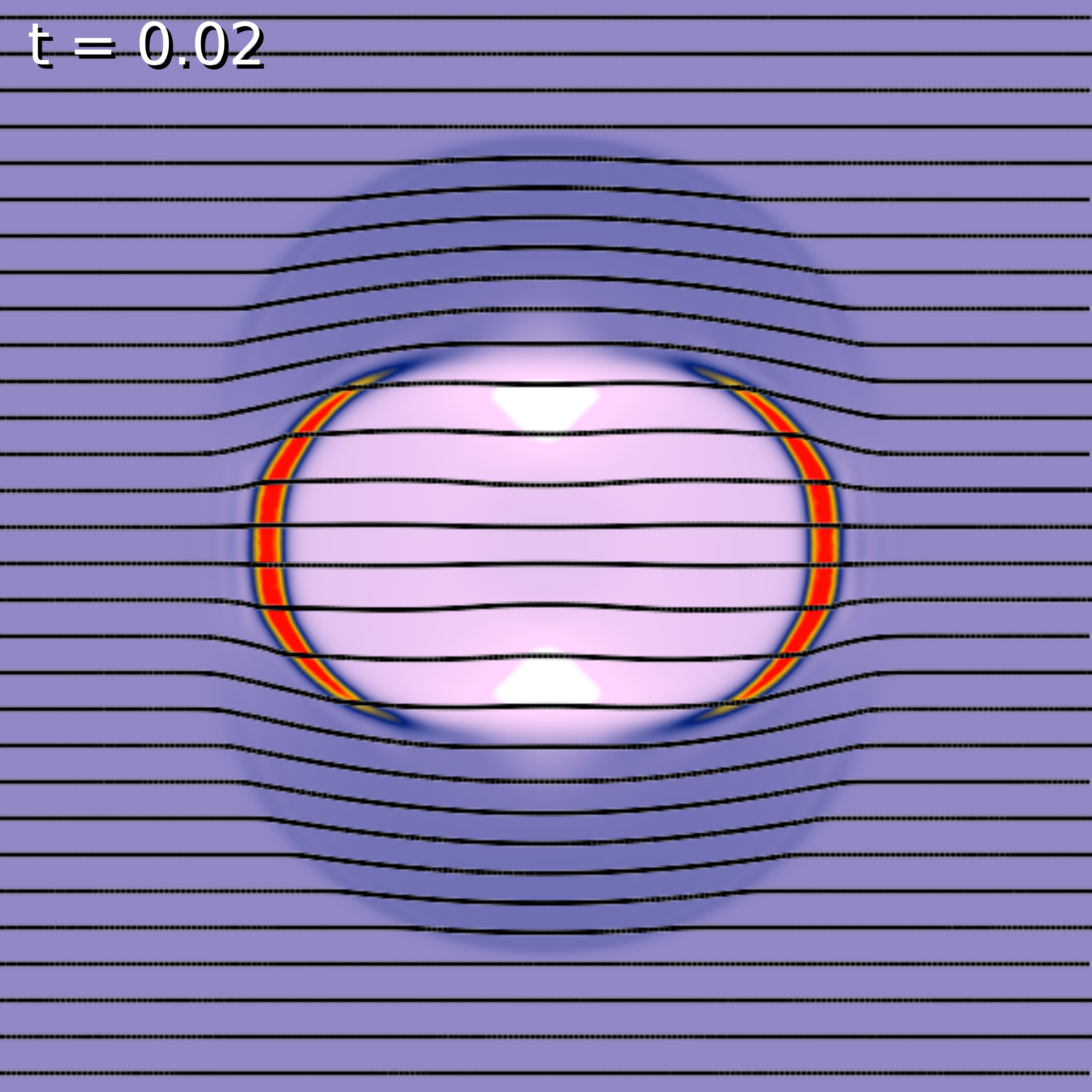}
\includegraphics[height=0.295\linewidth]{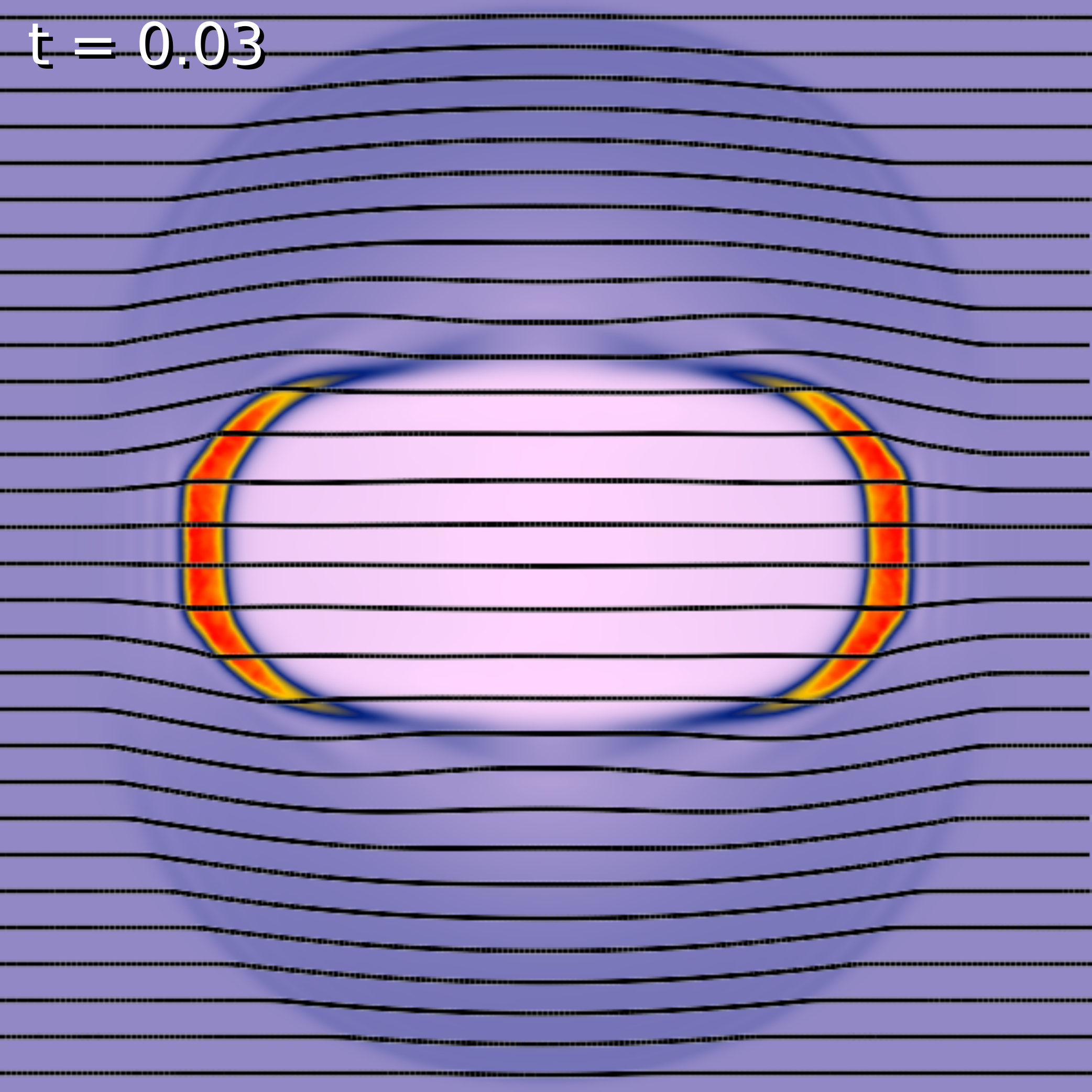}
\includegraphics[height=0.295\linewidth]{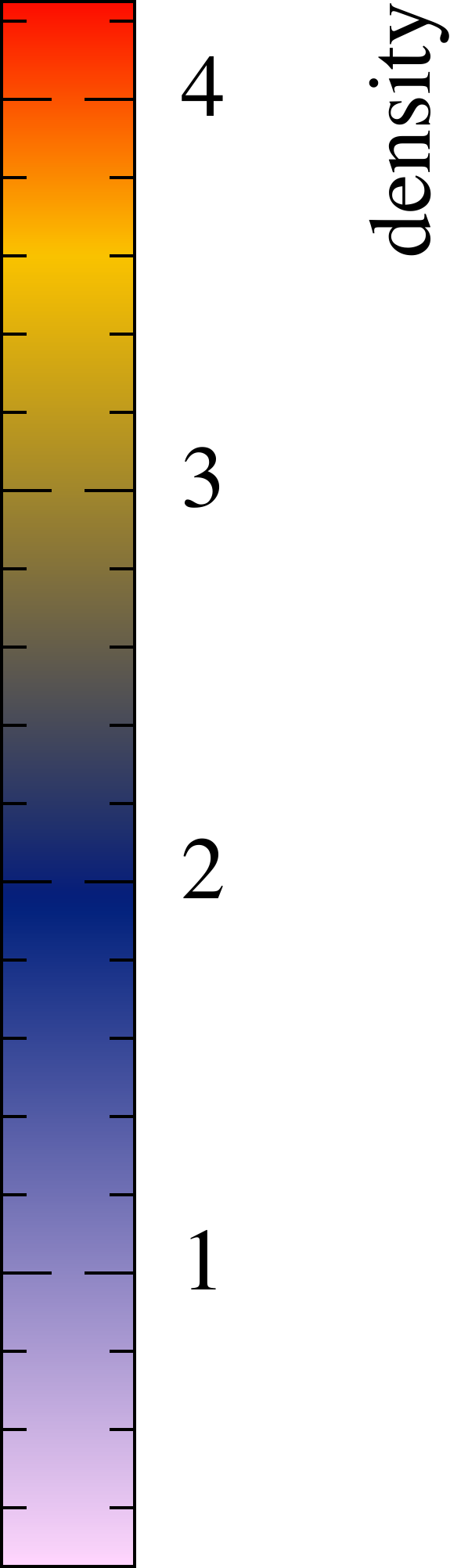}
\caption{Renderings of the density, with overlaid magnetic field lines, of the MHD blast wave test at $t = 0.01, 0.02, 0.03$ (left to right). The tension in the magnetic field causes the expansion of the blast wave to be preferentially directed along the magnetic field lines.}
\label{fig:mhdblast-render}
\end{figure}

First, we investigate a blast wave in the presence of a strong magnetic field \citep{bs99}. We follow the initial conditions of \citet{ldz00}. The domain is $x,y = [-0.5, 0.5]$, with $\rho = 1$, $B_x = 10$, ${\bf v} = 0$ and $P = 1$ with $\gamma = 1.4$ except for a central region of radius $R = 0.125$ that has its pressure increased to $P = 100$. These conditions yield initial plasma $\beta = 2$ in the disc and $\beta = 0.02$ in the surrounding medium. The simulation is initialised using $256 \times 296$ particles arranged on a triangular lattice. Fig.~\ref{fig:mhdblast-render} shows the evolution of the density of the simulation, with overlaid magnetic field lines. The initially circular blast region preferentially expands along the magnetic field lines due to the magnetic tension. This test was performed in our original paper \citep{tp12}, so here we are mainly interested in differences compared to our \citet{tp12} divergence cleaning scheme. For a detailed comparison on the effectiveness of hyperbolic divergence cleaning for this test, we refer the reader to our earlier paper. 

% color bar range: [0.25, 4.25]
% page 8 x 8 inch, character height 2.1

\begin{figure}
\centering
\includegraphics[width=0.49\linewidth]{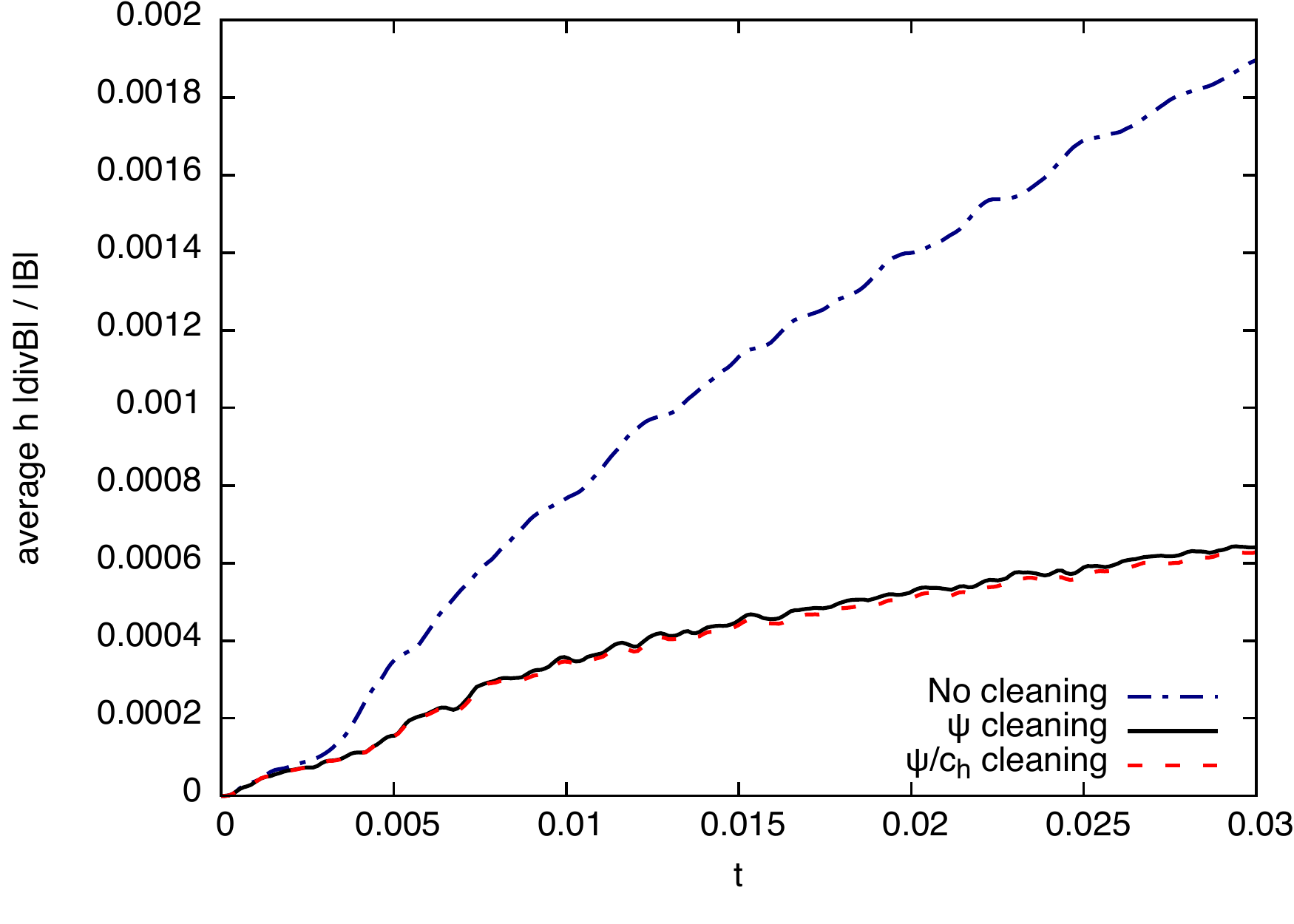}
\includegraphics[width=0.49\linewidth]{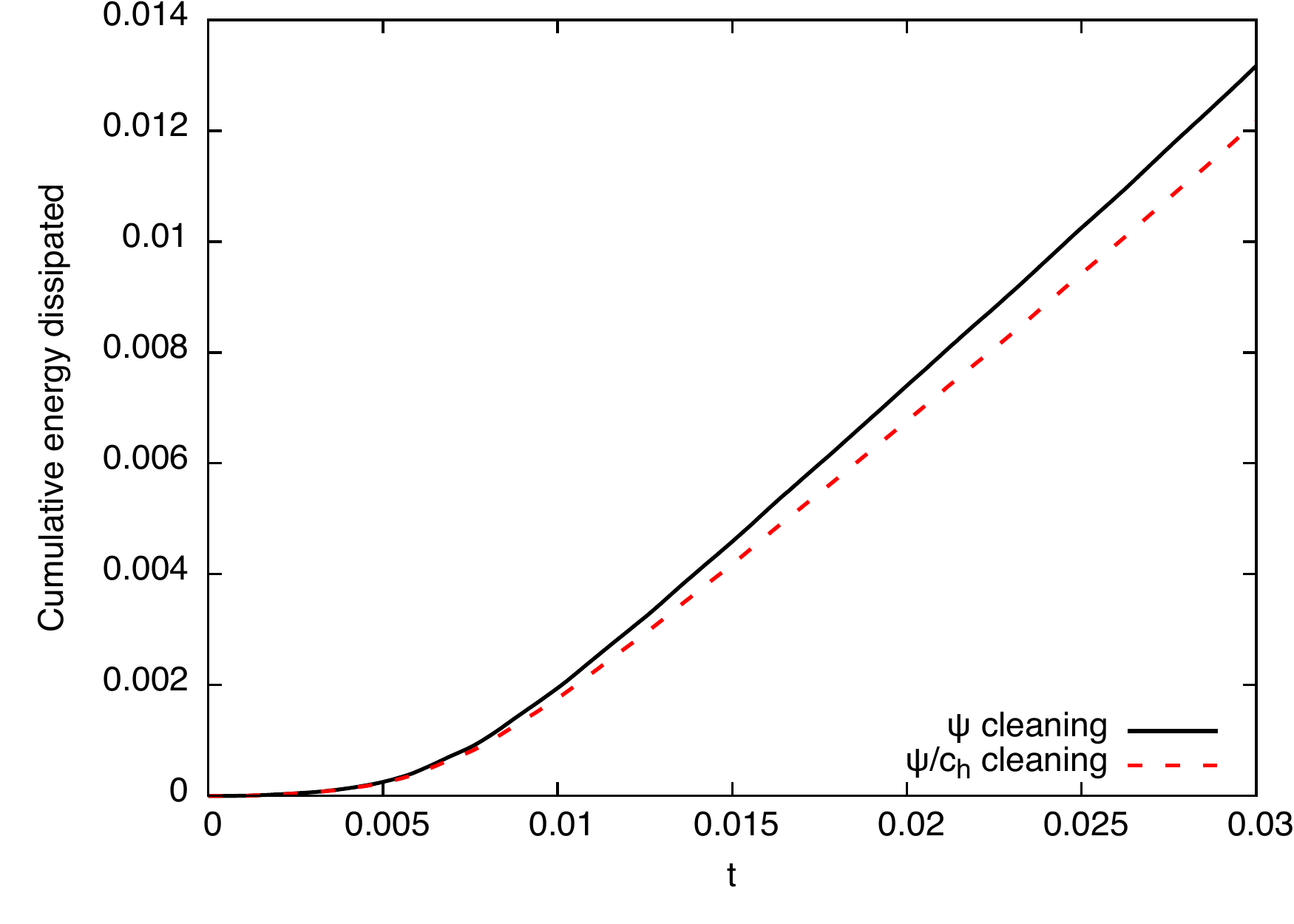}
\caption{Results for the blast wave in a strongly magnetised medium. {\it Left panel:} Average divergence error, $h \vert \nabla \cdot {\bf B} \vert / \vert {\bf B} \vert$, as a function of time for the old cleaning method (black solid line), new $\psi / c_{\rm h}$ cleaning method (red dashed line), and without divergence cleaning (blue dot-dashed line). The new method yields lower average divergence error at all times, with the average error $\sim 3\%$ lower by the end of the simulation. {\it Right panel:} Cumulative magnetic energy dissipated by the two cleaning schemes (Equation~(\ref{eq:psiheat})). The new cleaning approach dissipates magnetic energy at a rate $5\%$ less than the original method. For reference, the magnetic energy is $\sim 50$, implying that dissipation associated with divergence cleaning is insignificant. Though marginal, the new method is more effective at reducing divergence errors for this test and is less dissipative, without requiring additional computational expense.}
\label{fig:mhdblast}
\end{figure}

The left panel of Fig.~\ref{fig:mhdblast} shows the average divergence error for the previous and updated scheme. It is found that there is a slight reduction in average error when adopting the new $\psi / c_{\rm h}$ cleaning method, though the difference is marginal with at most a $3\%$ reduction in absolute error. Still, the average divergence error is lower at all times for no additional computational expense. Compared to a simulation with no divergence control, the average divergence error is reduced by a factor of $3$.

The right panel of Fig.~\ref{fig:mhdblast} shows the cumulative energy dissipated by the cleaning scheme, measured by storing the $e_\psi$ lost per particle according to Equation~(\ref{eq:psiheat}). The new $\psi / c_{\rm h}$ cleaning method dissipates magnetic energy at a rate $5\%$ less than the original method, measured by fitting a straight line to $t > 0.01$. At the end of the simulation, the cumulative energy dissipated is $\sim 0.012$, which, given that the magnetic energy is $\sim~50$, amounts to $< 0.03\%$ of the magnetic energy. By contrast, the shock capturing method dissipates $10 \times$ more magnetic energy, meaning divergence cleaning does not represent a significant source of dissipation. Overall, adopting the new divergence cleaning method shows two advantages: it has lower average divergence error and reduces numerical dissipation of magnetic energy, at no additional computational expense.

\subsection{Orszag-Tang vortex}
\label{sec:tests-orszag}

% color bar range: [0.04, 0.5]
% page size 8 x 8 inch, character height 2.1
\begin{figure}
\centering
\includegraphics[height=0.295\linewidth]{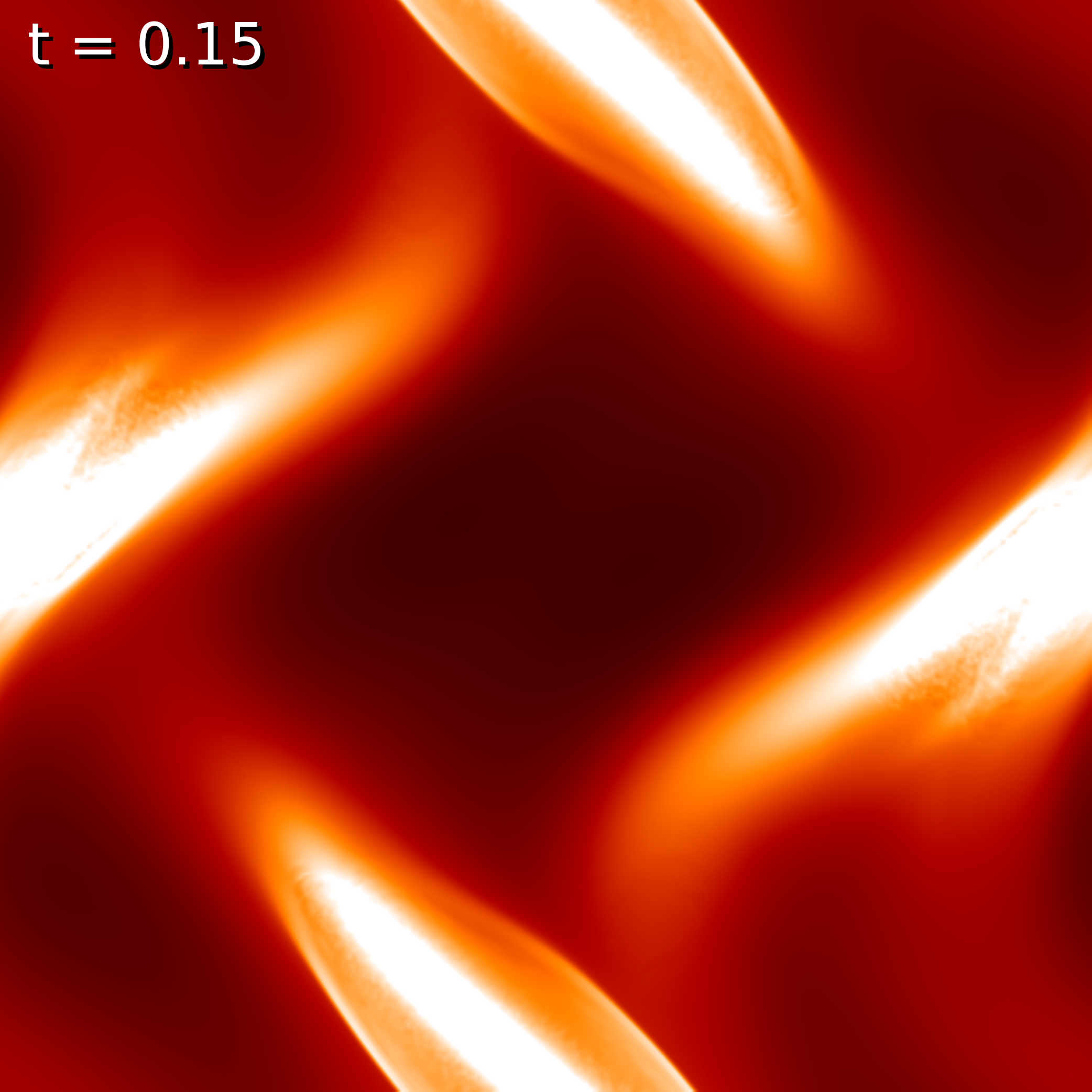}
\includegraphics[height=0.295\linewidth]{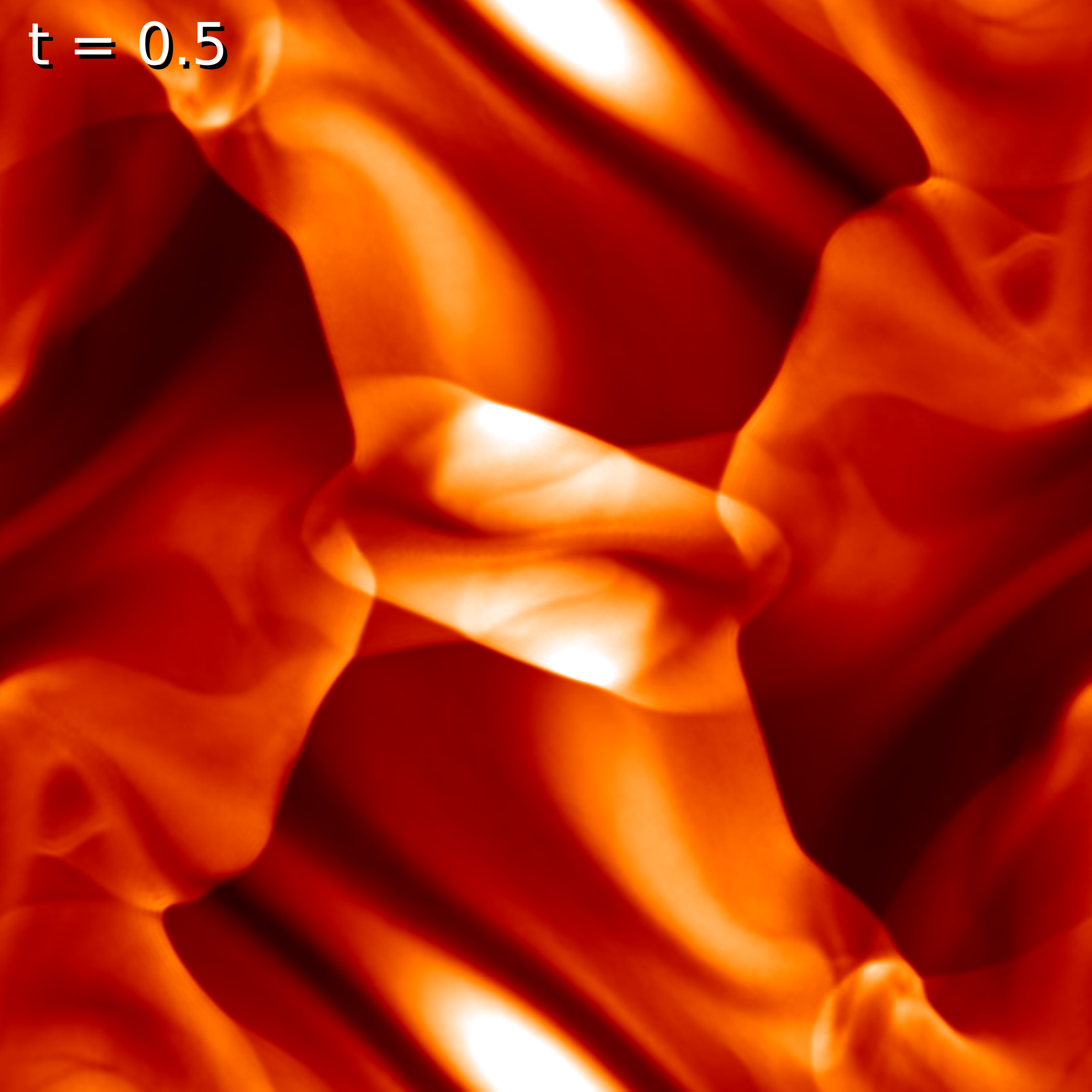}
\includegraphics[height=0.295\linewidth]{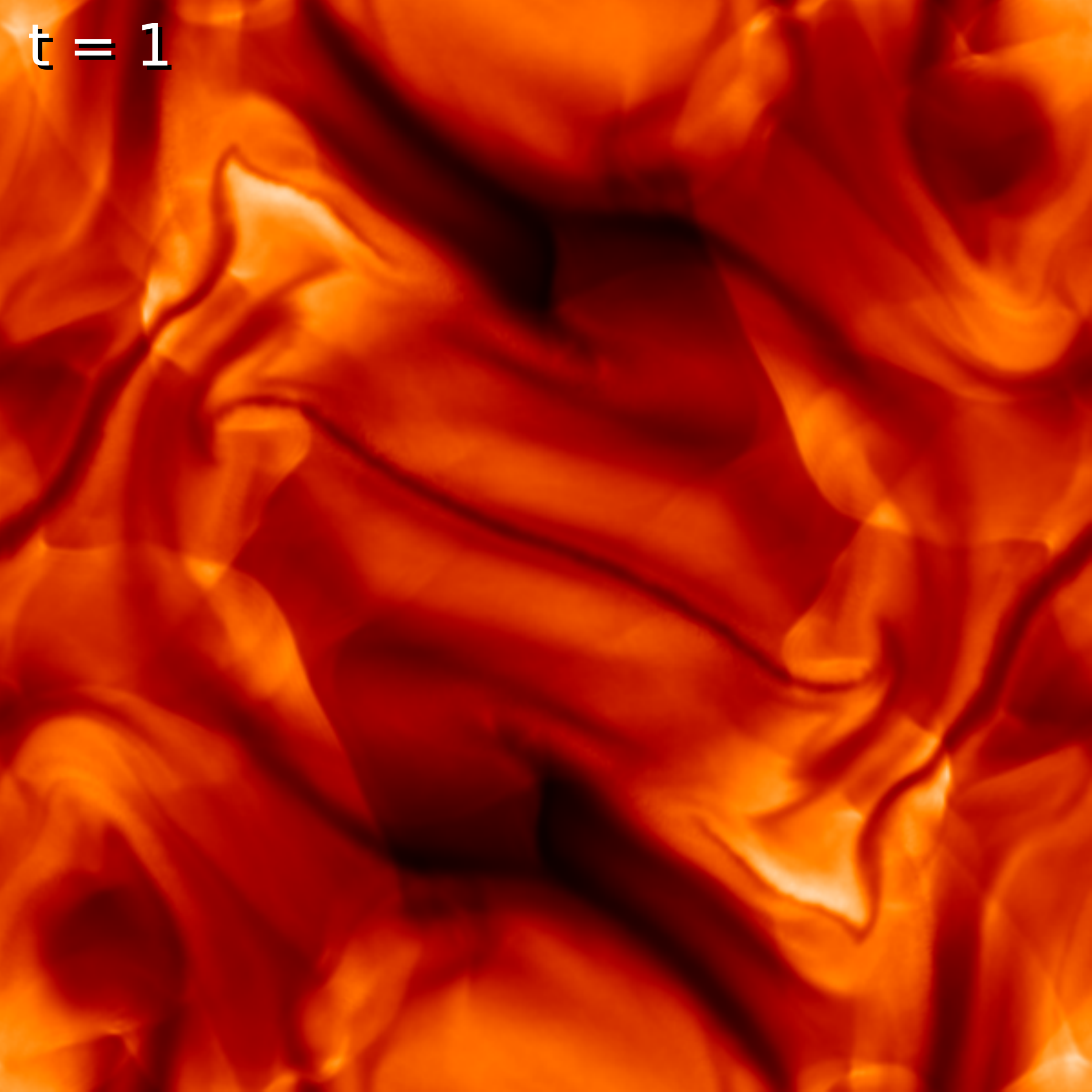}
\includegraphics[height=0.295\linewidth]{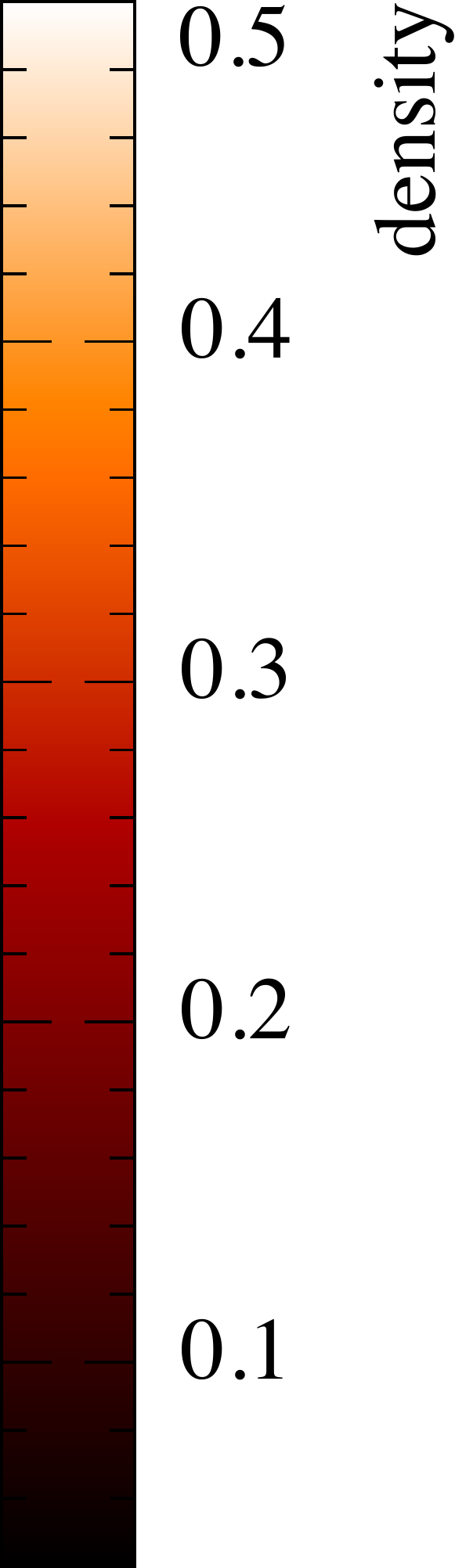}
\caption{Renderings of the density of the Orszag-Tang vortex at $t = 0.15, 0.5$ and $1$ (left to right). The initial vortex structures ($t=0.15$) produce shock waves that collide and interact ($t=0.5$), forming complex structures which begin the early stages of turbulence ($t=1$).}
\label{fig:orszag-render}
\end{figure}

\begin{figure}
\centering
\includegraphics[width=0.49\linewidth]{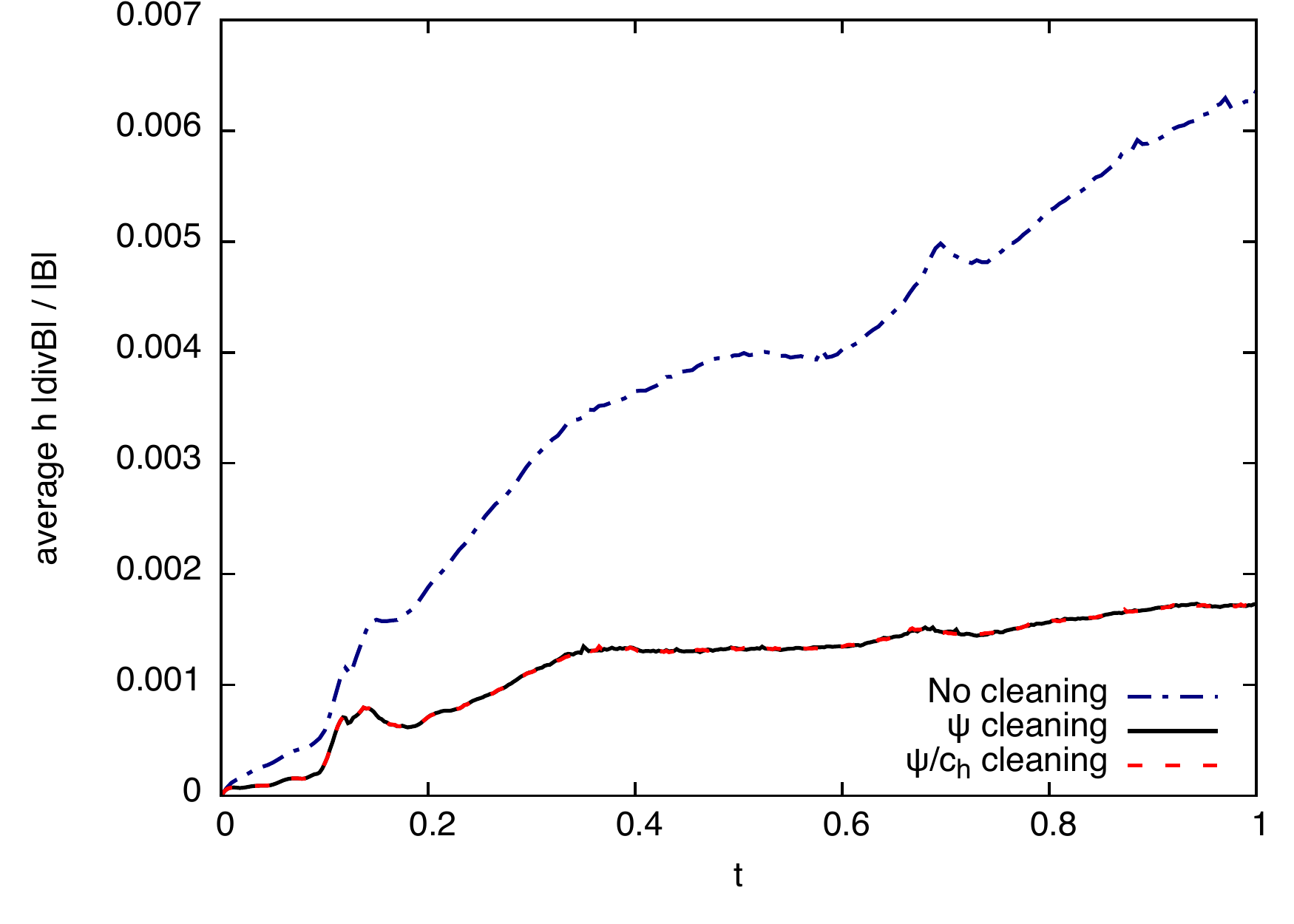}
\caption{Average divergence error, $h \vert \nabla \cdot {\bf B} \vert / \vert {\bf B} \vert$, in the Orszag-Tang vortex test, showing comparison between original divergence cleaning method (solid black line) and the new $\psi / c_{\rm h}$ cleaning method (dashed red line), with a reference simulation without divergence cleaning (blue dot-dashed line). Both produce similar levels of average divergence error, exhibiting small deviations but no long-term systematic difference.}
\label{fig:orszag}
\end{figure}

Next we consider the Orszag-Tang vortex \citep{ot79}, a two-dimensional problem where initial velocity and magnetic field vortices interact to produce turbulence. The test involves supersonic motion with several classes of interacting shockwaves, relevant for astrophysical applications. The Orszag-Tang vortex was extensively studied in our original method paper, thus we restrict our analysis to differences between the original and new, updated method. For a detailed comparison of results between hyperbolic divergence cleaning with alternative divergence control measures, along with optimal $\sigma$ values for damping and a resolution study, we refer the reader to our earlier paper.

The initial conditions are ${\bf v} = [- \sin(2 \pi y),$ $\sin(2 \pi x)]$, ${\bf B} = [-\sin(2 \pi y), \sin(4 \pi x)]$, $\rho = 25/(36\pi)$ and $P=5/(12\pi)$ with $\gamma=5/3$. The problem is set up using $512 \times 512$ particles initially arranged on a square lattice, set up by creating one quadrant of the lattice then mirroring the particles so that symmetry of the initial conditions is exactly preserved. Renderings of the density evolution are shown in Fig.~\ref{fig:orszag-render}, showing representative times of the early vortex structure ($t=0.15$), formation and interaction of shocks ($t=0.5$), and onset of turbulence ($t=1$).

Fig.~\ref{fig:orszag} shows the mean $h \vert \nabla \cdot {\bf B} \vert / \vert {\bf B} \vert$ as a function of time, comparing results between the old and new divergence cleaning methods along with a simulation without divergence cleaning for reference. The average divergence error as a function of time shows short-term variations between the two methods on the order of $\sim 1$--$3\%$, but no long-term deviation exists, similar to our findings in the blast wave test.  %The cumulative magnetic energy dissipated by the divergence cleaning (right panel) is indistinguishable between the two methods. For either case, divergence cleaning represents a tiny source of dissipation, with the cumulative dissipated energy less than $0.5\%$ of the magnetic energy by the end of the simulation.

\subsection{MHD Rotor}
\label{sec:tests-rotor}

% color bar range: [0.4, 14]
% page size 8 x 8 inch, character height 2.1
\begin{figure}
\centering
\includegraphics[height=0.295\linewidth]{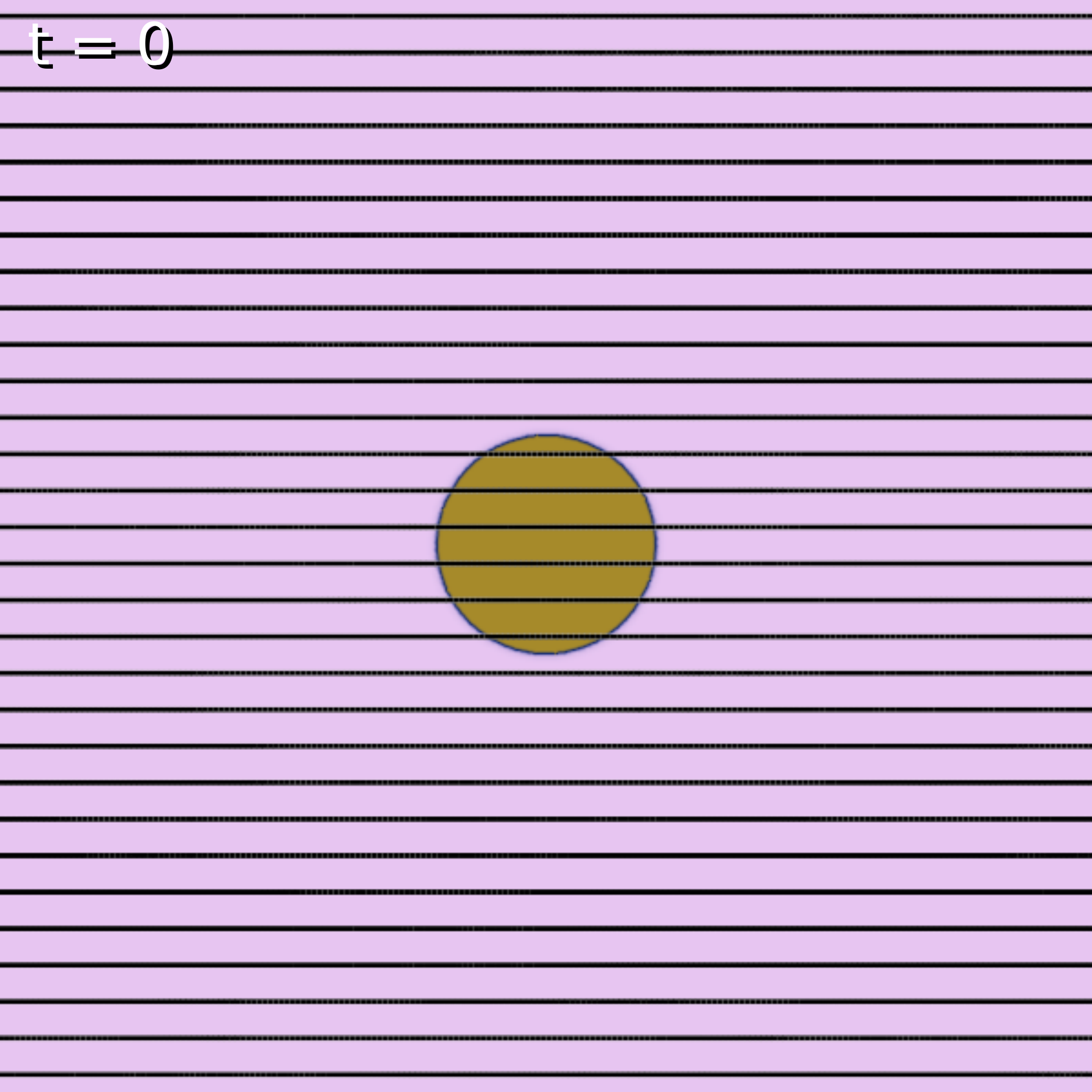}
\includegraphics[height=0.295\linewidth]{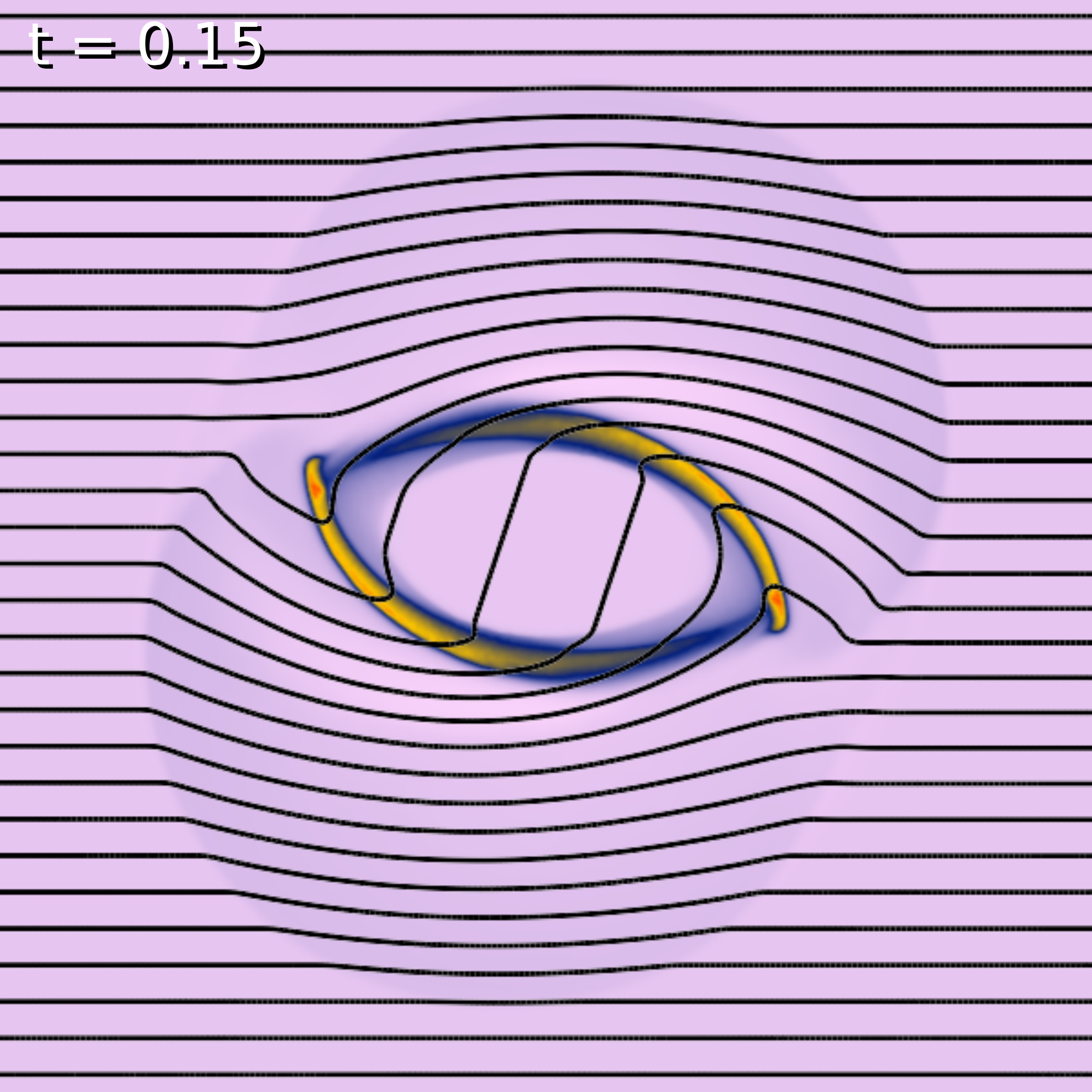}
\includegraphics[height=0.295\linewidth]{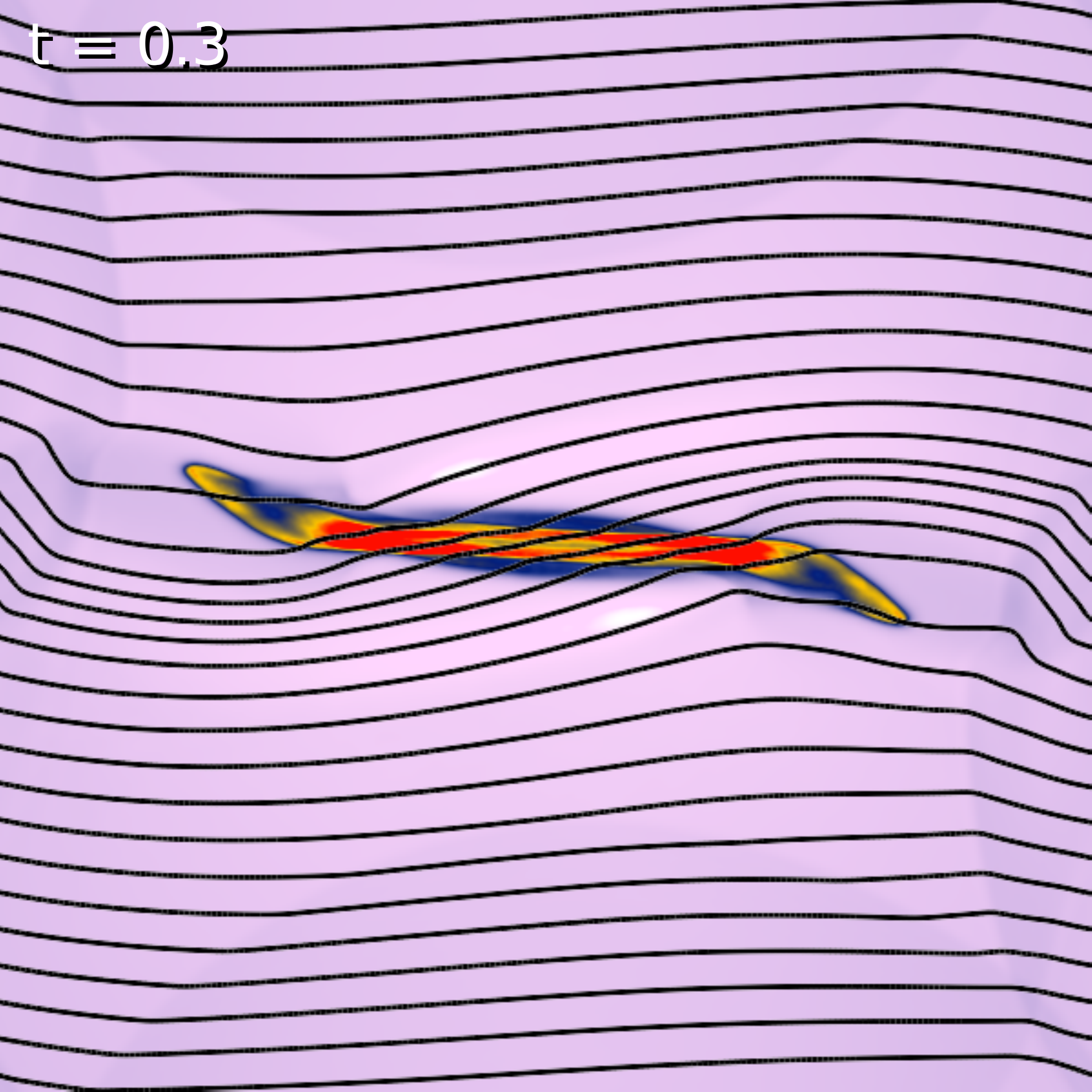}
\includegraphics[height=0.295\linewidth]{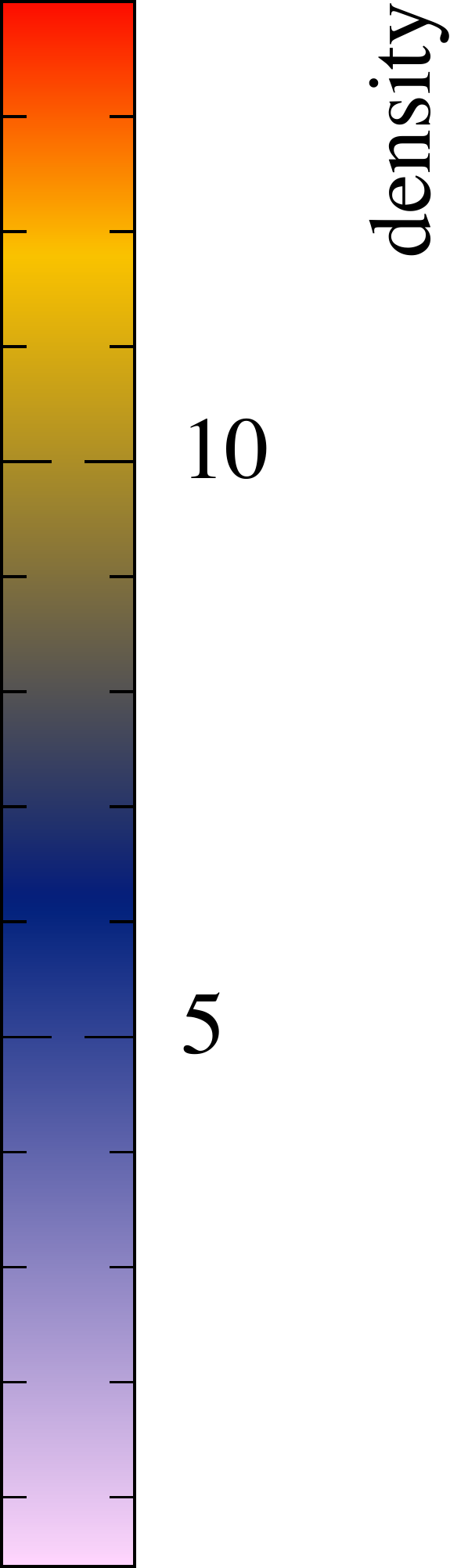}
\caption{Renderings of the density and magnetic field lines for the MHD rotor test at the initial time ($t = 0$, left panel) and evolved time slices $t = 0.15, 0.3$ (centre and right, respectively). The central dense disc is initially rotating, launching strong torsional Alfv\'en waves. As the magnetic field lines twist, the rotor is compressed in the $y$-direction, leading to the formation of a single dense filament by the end of the calculation.}
\label{fig:rotor-render}
\end{figure}

\begin{figure}
\centering
\includegraphics[width=0.49\linewidth]{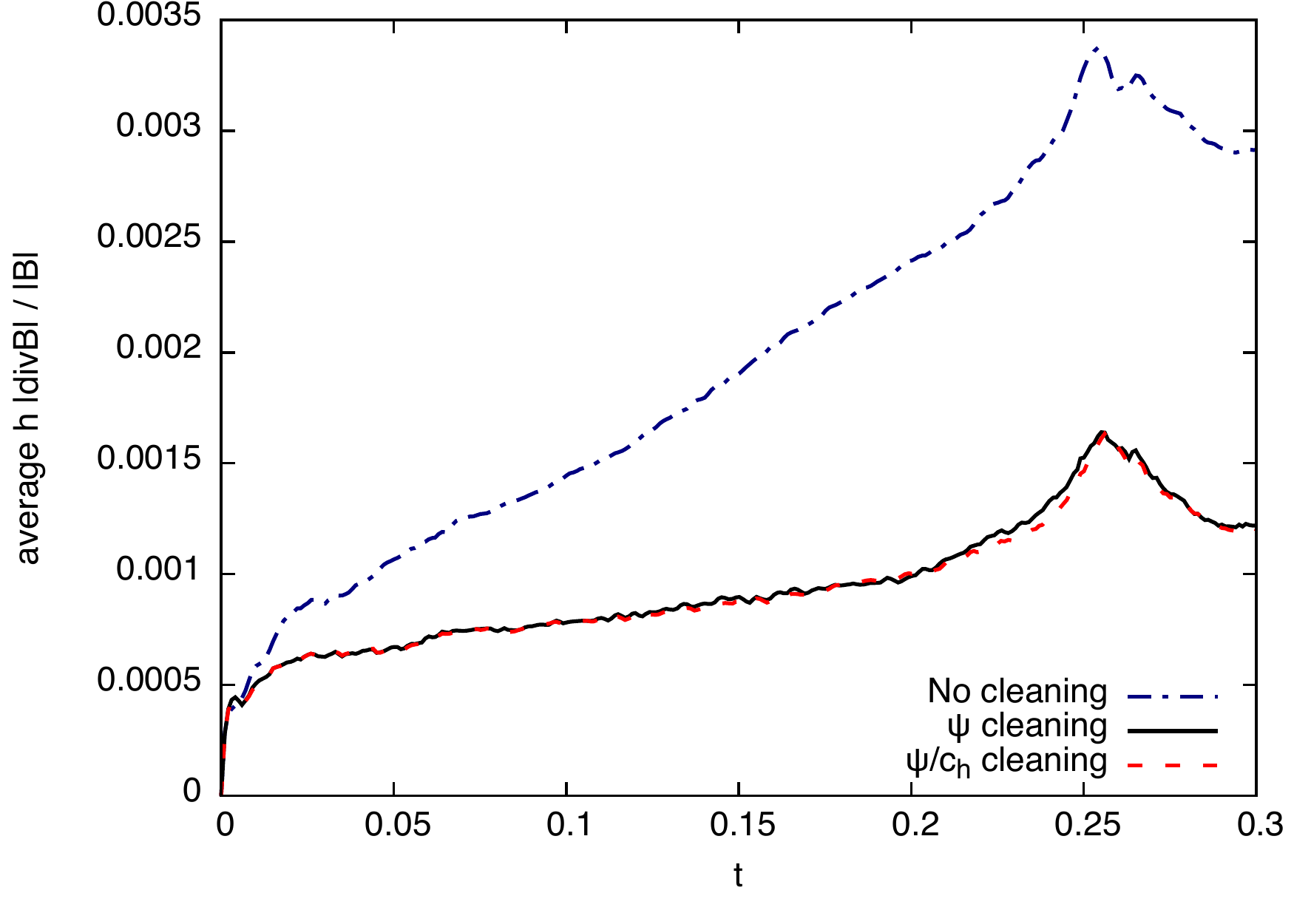}
\caption{Average divergence error as a function of time for the MHD rotor test. The previous cleaning method (solid black line) and new $\psi / c_{\rm h}$ cleaning method (red dashed line) yield similar levels of average divergence error, showing minimal difference with the new scheme having  average errors which are lower by $\sim 1$--$2\%$. At $t \sim 0.25$, the rotor merges in on itself and we see a corresponding dip in average divergence error where the new scheme is $5\%$ lower than the previous scheme.}
\label{fig:rotor}
\end{figure}

The MHD rotor \citep{bs99} consists of a rotating dense disc embedded in a lower density medium. As the disc turns, it twists the magnetic field lines launching strong torsional Alfv\'en waves. The domain is $x, y = [-0.5, 0.5]$ with initial conditions $P = 1$, $\gamma = 1.4$ and $B_x = 5 / \sqrt{4 \pi}$. The disc, of radius $R = 0.1$ located in the centre of the domain, has density $\rho = 10$ and angular velocity $\omega = 20$. The surrounding medium has density $\rho = 1$ and is at rest (${\bf v} = 0$). The outer region is formed from a triangular lattice of $256 \times 296$ particles with the $R = 0.1$ central region excised, with the inner disc trimmed from a triangular lattice of $161 \times 186$ particles scaled to $1/5${\it th} the size of the outer lattice. The total number of particles is $96 914$. The simulation is performed until $t = 0.3$. Fig.~\ref{fig:rotor-render} shows renderings of the density evolution of the simulation with overlaid magnetic field lines. Due to the presence of the strong magnetic field, the disc becomes oblate as it rotates, eventually merging in on itself becoming a single dense filament at the end of the calculation.

Fig.~\ref{fig:rotor} shows the average divergence error as a function of time for the two cleaning methods, with a calculation without divergence cleaning for reference. Both methods yield similar results, differing by $1$--$2\%$ in average $h \vert \nabla \cdot {\bf B} \vert / \vert {\bf B} \vert$ throughout the duration of the simulations, except around $t\sim0.25$ where the new $\psi / c_{\rm h}$ divergence cleaning method exhibits average divergence error which is $\sim 5\%$ lower. At this time, the dense edges of the disc are beginning to merge to make the final, single dense filament, causing the cleaning wave speed of low density region to rapidly increase as it is compressed. This rapid change in cleaning wave speed is accounted for by the $\psi / c_{\rm h}$ cleaning approach, yielding a small, but measurable, improvement in the effectiveness of the divergence cleaning. %The cumulative magnetic energy dissipated by divergence cleaning is shown in the right panel of Fig.~\ref{fig:rotor}. The new approach has less dissipation by $1$--$3\%$ at all times. Note that the magnetic energy of the calculation is $\sim 1$, thus divergence cleaning is a negligible source of dissipation, dissipating less than $0.05\%$ of the magnetic energy over the lifetime of the calculation.

\section{Achieving $\nabla \cdot {\bf B} = 0$ to machine precision}
\label{sec:divbzero}

An important aspect of any divergence cleaning algorithm is whether ``$\nabla \cdot {\bf B} = 0$'' is well defined in terms of the numerical operator used to evaluate the divergence of the magnetic field. For example, in a projection method (e.g. \citealt{bb80}), one solves the two equations
\begin{equation}
\nabla^2 \phi = \nabla \cdot {\bf B}^* \label{eq:poisson}
\end{equation}
and
\begin{equation}
{\bf B} = {\bf B}^* - \nabla \phi,  \label{eq:projection}
\end{equation}
where ${\bf B}^*$ is a magnetic field with non-zero $\nabla \cdot {\bf B}^*$, and ${\bf B}$ is the resultant clean magnetic field. As discussed by \citet{toth00}, this will only result in $\nabla \cdot {\bf B} = 0$ to machine precision for the chosen discrete operator if the numerical operators used to evaluate $\nabla \cdot {\bf B}^*$ and $\nabla \phi$ in Equations~(\ref{eq:poisson}) and (\ref{eq:projection}) are the same as those used to evaluate $\nabla^2$ in Equation~(\ref{eq:poisson}).

Here we demonstrate that this consistency is satisfied by our divergence cleaning method, that is, it is possible to achieve $\nabla \cdot {\bf B}=0$ to machine precision. We will demonstrate that in the limit $t \to \infty$ (or equivalently, $c_{\rm h} \to \infty$), our discretised cleaning equations (\ref{eq:cleaning-spmhd1})--(\ref{eq:cleaning-spmhd2}) result in $\nabla\cdot{\bf B} = 0$ to machine precision when measured with the numerical operator used on the right hand side of Equation~(\ref{eq:cleaning-spmhd2}).

Our approach is to sub-cycle the divergence cleaning equations between timesteps, updating only the magnetic field via the cleaning equations (Equations~(\ref{eq:cleaning-spmhd1})--(\ref{eq:cleaning-spmhd2})) with the position and velocity of each particle held fixed. In effect, this iterates toward the solution of the Poisson equation for $\nabla \cdot {\bf B}$ (e.g., \citealt{toth00}). \citet{yalimetal11} have used a similar technique in an Eulerian code, except they iterate only the hyperbolic equations with no parabolic damping. 

\subsection{Sub-cycling the divergence cleaning equations}

To begin, we examine the degree to which the divergence error of the magnetic field can be reduced through divergence cleaning. To test this, we perform a `static' test, similar to those performed by \citet{tp12}, whereby the magnetic field evolves only by sub-cycling the divergence cleaning equations (Equations~(\ref{eq:cleaning-spmhd1})--(\ref{eq:cleaning-spmhd2})) with the position and velocity of each particle held constant. In order for results to be applicable to a `real' application, rather than an artificial setup, we use the particle and magnetic field structure taken from the $t=1$ evolved state of the Orszag-Tang vortex calculation in Section~\ref{sec:tests-orszag}. 

\begin{figure}
\centering
\includegraphics[width=0.49\linewidth]{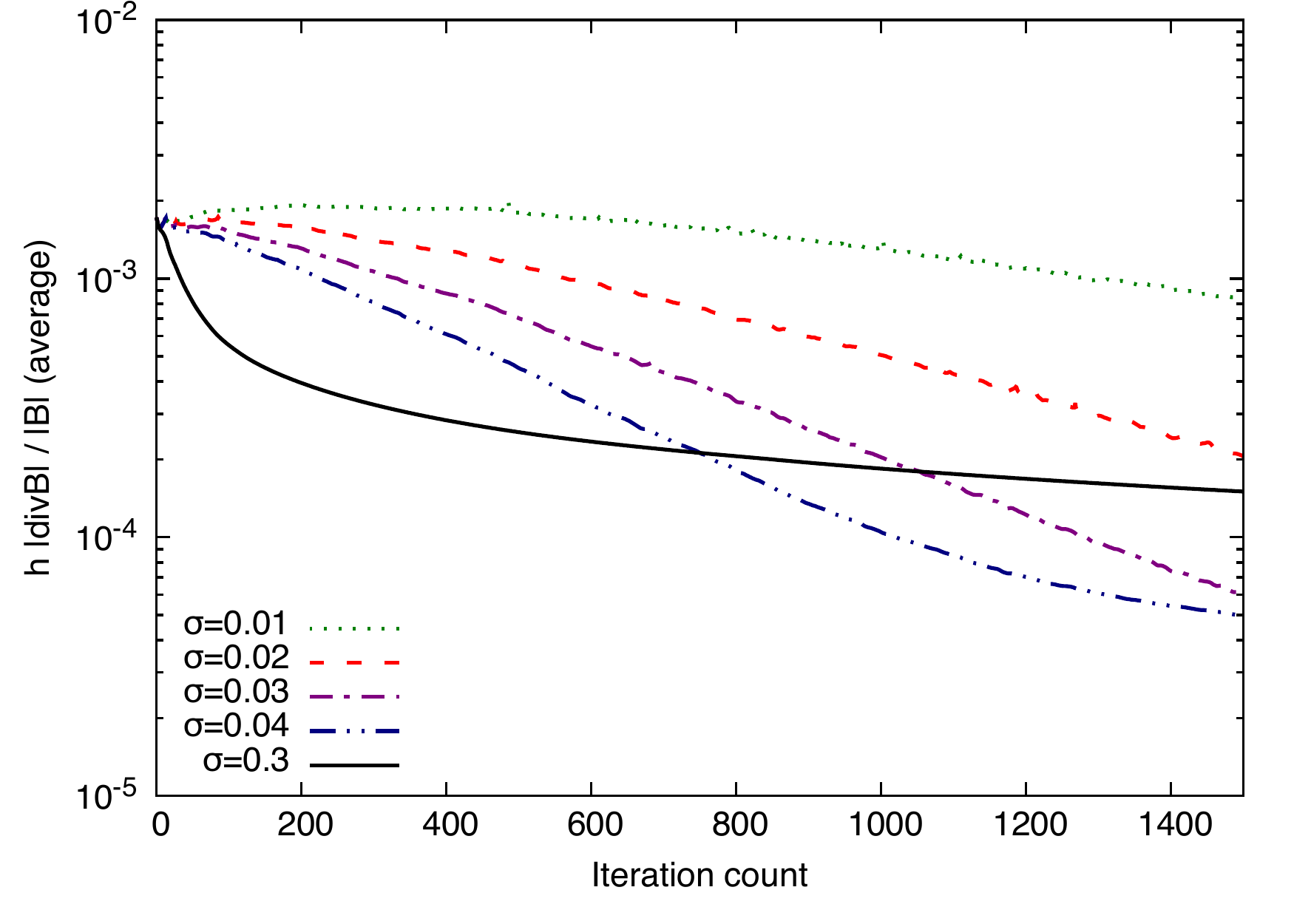} 
\includegraphics[width=0.49\linewidth]{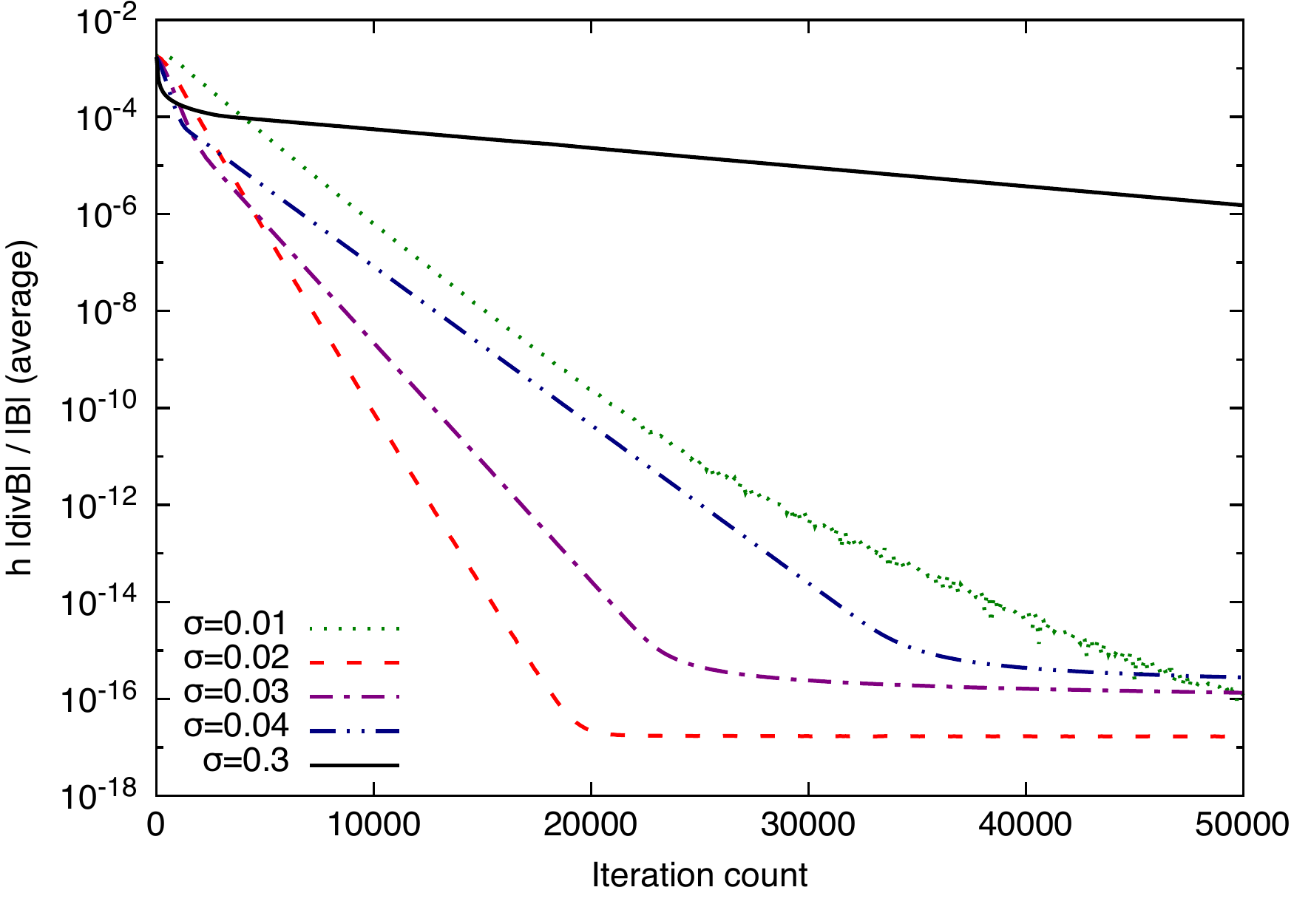}
\caption{Comparing values of $\sigma$ in the damping parameter to obtain an optimal value for sub-cycling, with the left panel the first $1500$ iterations and right panel $50~000$ iterations. Short wavelength errors are quickly removed using the default value of $\sigma = 0.3$ (left panel), though this value performs poorly at removing long wavelength modes (right panel). Using $\sigma = 0.02$--$0.03$, though initially worse at reducing divergence error, is found to remove long wavelength errors in the shortest number of iterations.}
\label{fig:cleantozero}
\end{figure}

\begin{figure}
\includegraphics[width=\columnwidth]{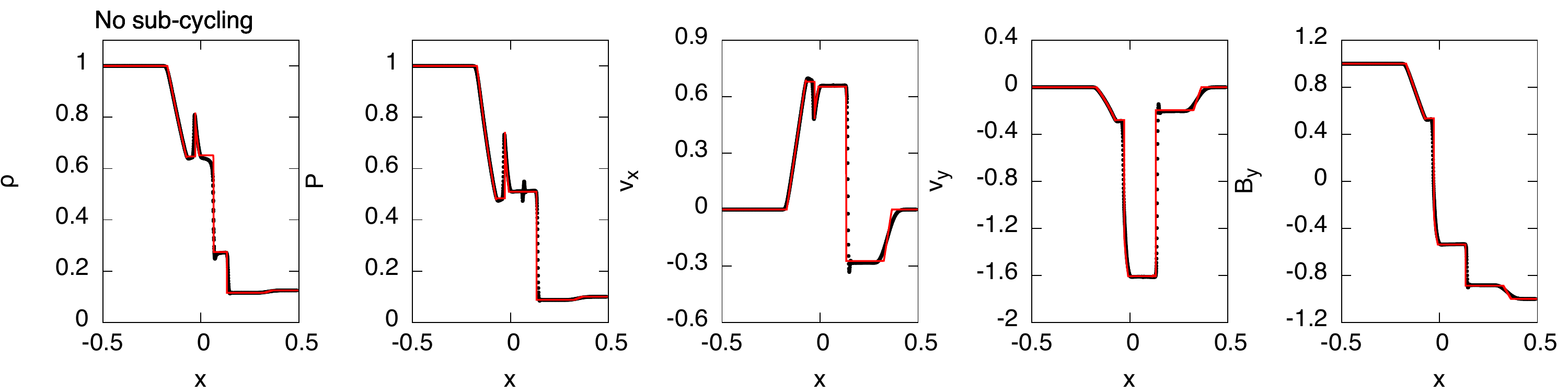} \\
\includegraphics[width=\columnwidth]{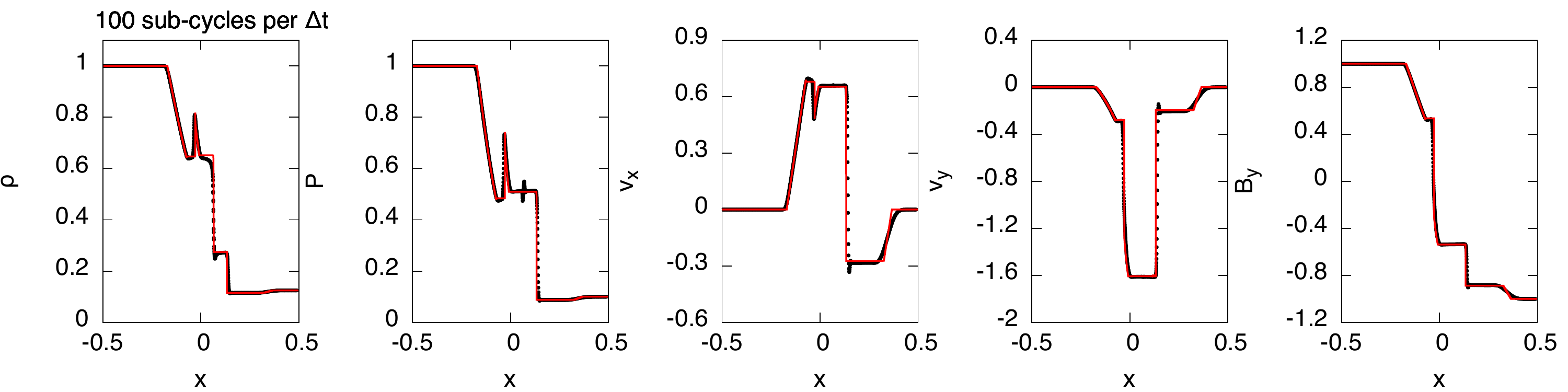} 
\caption{Profile for the Brio-Wu shocktube test at $t=0.1$ using the standard SPMHD implementation (top row) and where the divergence cleaning equations are sub-cycled $100\times$ per timestep (bottom row). Black circles are the particle data, which is in agreement with a reference solution obtained using {\sc Athena} with $10^4$ grid cells. No detrimental effect is found in the quality of the shock solution by sub-cycling the divergence cleaning equations.}
\label{fig:subcycling-accuracy}
\end{figure}

% Do \sigma = 0.01, 0.02, 0.04, 0.3

Fig.~\ref{fig:cleantozero} shows the average of $h \vert \nabla \cdot {\bf B} \vert / \vert {\bf B} \vert$ on the particles as a function of the number of iterations of the cleaning equations. We tested a series of values of the parabolic damping parameter, $\sigma$. The initial decay of divergence error is most rapid for $\sigma=0.3$, but with a turnover in decay rate occurring once the average error has been reduced by around an order of magnitude. This turnover may be understood due to the differing rates of removal of short and long wavelength divergence errors. Divergence errors are introduced into simulations at short wavelengths, e.g.\ from shocks, which this level of damping is most effective at removing. Hence, this value of $\sigma$ is optimal when the simulation is evolving and continually injecting divergence error, as found by \citet{tp12}. However, once short wavelength errors have been removed, the decay rate slows significantly because only long wavelength modes remain which decay slowly.

The most effective value of $\sigma$ for removing long wavelength modes is $0.02$--$0.03$. Since these values are less effective at removing short wavelength errors than $\sigma=0.3$, they are initially slower at reducing the average $h \vert \nabla \cdot {\bf B} \vert / \vert {\bf B} \vert$. However, the smaller $\sigma$ value allows the hyperbolic waves to propagate more effectively, spreading the divergence waves throughout the simulation, in turn allowing the diffusion term to become more effective at reducing the long wavelength modes. Thus, over a large number of iterations ($\sim10^4$), using $\sigma = 0.02$--$0.03$ will reduce the average $h \vert \nabla \cdot {\bf B} \vert/ \vert {\bf B} \vert$ to zero in the shortest number of iterations. For this simulation, it took $20~000$ iterations for $\sigma=0.02$ to reduce to the average divergence error to zero, compared to $\sigma=0.3$ which still has average error of $\sim10^{-6}$ after $50~000$ iterations.

%For example, Fig.~\ref{fig:cleantozero} shows that $\sigma=0.3$ reduces the initial average $h \vert \nabla \cdot {\bf B} \vert / \vert {\bf B} \vert$ by half in $\sim10$ iterations, whereas it takes $\sim90$ iterations for $\sigma=0.03$ to achieve the same level of reduction. 

For every value of $\sigma$ that we tested, the average $h \vert \nabla \cdot {\bf B} \vert / \vert {\bf B} \vert$ in the simulation could be reduced to zero within machine precision ($\sim10^{-16}$) given enough iterations, demonstrating that it is possible in principle to achieve $\nabla \cdot {\bf B}=0$ with divergence cleaning. Further work to reduce the number of iterations required to achieve $\nabla\cdot{\bf B} = 0$ would be highly valuable as a step towards achieving true tolerance-based control of magnetic divergence errors in SPMHD simulations.

\subsection{Accuracy analysis}
\label{sec:divbzero-accuracy}

Sub-cycling the divergence cleaning equations can reduce divergence errors, but it is crucial that this process not degrade the quality of the solution. To investigate this, we simulate the Brio-Wu shocktube \citep{bw88} using calculations without sub-cycling (divergence cleaning in its usual form) and calculations where $10$ and $100$ sub-cycles are performed each timestep. The shocktube has initial left state $[\rho, P, v_x, v_y, B_x, B_y] = [1, 1, 0, 0, 0.75, 1]$ and right state $[\rho, P, v_x, v_y, B_x, B_y] = [0.125, 0.1, 0, 0, 0.75, -1]$, using $\gamma = 5/3$. The particles are arranged on triangular lattices, with the left side composed of $800 \times 30$ particles and the right side $300 \times 10$ particles. Results are compared against those from a high-resolution {\sc Athena} \citep{athena} calculation using $10^4$ grid cells.

%  no substeps: L1 error: 0.0094343235106205     L2 error: 0.049109565990399
%  10 substeps: L1 error: 0.0093828954655377     L2 error: 0.04855037076824
% 100 substeps: L1 error: 0.009348155458488     L2 error: 0.048228700994865
% 1000 substeps: L1 error: 0.0093615515659785     L2 error: 0.048283363759621

% 1000 substeps max h divb / |B| error: 2.841 e-7

Fig.~\ref{fig:subcycling-accuracy} shows the profile of the shocktube at $t=0.1$ for the calculation with no sub-cycles (default SPMHD) and when $100$ sub-cycles are performed per timestep. The $100$ sub-cycle calculation yields the correct shock profile, with results which are indistinguishable by eye compared to the default calculation. There is no evidence that sub-cycling the divergence cleaning equations is detrimental to the behaviour of the shock. This may be quantified by measuring the L$_2$ error of $B_y$. For the default case, L$_2$=$4.911 \times 10^{-2}$, and for the simulations using $10$ and $100$ sub-cycles per timestep, L$_2$=$4.855 \times 10^{-2}$ and L$_2$=$4.823 \times 10^{-2}$. Note that the L$_2$ error is primarily dominated by the accuracy of the shock capturing method, so the differences are marginal. However, there is evidence that the L$_2$ error decreases with increasing number of sub-cycles. The maximum $h \vert \nabla \cdot {\bf B} \vert / \vert {\bf B} \vert$ error at $t=0.1$ is $9.62 \times 10^{-5}$ for the default calculation, and $1.98 \times 10^{-5}$, and $4.45 \times 10^{-6}$ for the $10$ sub-cycle and $100$ sub-cycle calculations, respectively (an $80\%$ and $95\%$ reduction). We conclude that sub-cycling the divergence cleaning equations can provide reduced divergence error without negatively affecting the quality of the solution obtained.

\section{Summary}
\label{sec:summary}

We have developed a new formulation of hyperbolic/parabolic divergence cleaning for SPMHD which takes account of the variability in the wave cleaning speed. This is accomplished by evolving $\psi / c_{\rm h}$ instead of $\psi$ as the primary variable. In Section~\ref{sec:cleaningv2}, cleaning equations were derived in terms of this quantity. Using this set of equations ensures that divergence cleaning cannot lead to increases in magnetic energy, as the parabolic damping can only remove magnetic energy and the hyperbolic terms are guaranteed to exactly conserve $e_\psi$ and magnetic energy. The new cleaning equations remain similar to the previous equations, differing only by factors of $c_{\rm h}$, but permit the wave cleaning speed to evolve in time without needing an explicit expression for the time derivative of $c_{\rm h}$. In Section~\ref{sec:sourceterms}, the generalised wave equation was derived demonstrating that the propagation of divergence errors remains hyperbolic/parabolic, but occurs in the co-moving frame and naturally accounts for changes in the density, wave speed and parabolic damping term. The new method was tested using a series of idealised tests (Section~\ref{sec:idealised-tests}) and standard MHD test problems (Section~\ref{sec:practical-tests}).

The issue related to variable wave cleaning speeds was demonstrated in Section~\ref{sec:tests-timevary} using a simplified test of the advection of a divergence blob. When the wave cleaning speed was varied in time, it led to exponential increases of magnetic energy in the form of increased divergence error. This occurred both for purely hyperbolic cleaning ($\sigma = 0$) and mixed hyperbolic/parabolic cleaning. No such errors were found when the test was repeated for wave cleaning speeds that were constant in time but which had spatial variations (Section~\ref{sec:tests-xvary}), nor for time or spatial discontinuities in the parabolic damping parameter (Section~\ref{sec:tests-disctau}). 

In Section~\ref{sec:tests-advection}, the effect of advecting $\psi/c_{\rm h}$ was tested. The motivation for this test was that the original \citet{dedneretal02} formulation used Eulerian derivatives, i.e.\ no advection of $\psi$, however, the constraint of energy conservation requires the use of Lagrangian derivatives, adding advection of $\psi/c_{\rm h}$ to our scheme. Using the divergence advection test, we found that if the cleaning equations are implemented using Eulerian derivatives, the average divergence error increased by $30\%$ when the background velocity of the fluid increased from $\mathcal{M}=0.45$ to $\mathcal{M}=10$. By contrast, our Lagrangian implementation produced equivalent results for all flow velocities. Furthermore, computing Eulerian derivatives require `reverse advection' terms be added to counteract the Lagrangian nature of SPMHD, adding a velocity dependence into the Courant timestep constraint. For these reasons, we conclude that the cleaning equations should be implemented with Lagrangian derivatives.

Our final idealised test was to confirm that the $\tfrac{1}{2} (\psi / c_{\rm h}) (\nabla \cdot {\bf v})$ term added to account for compression and rarefaction is indeed necessary to exactly conserve energy (Section~\ref{sec:tests-divv}). To investigate this, supersonic compressional motions were added to the divergence advection test. As the errors due to time-stepping were reduced through reductions of the Courant factor, the total energy of the simulations with the compression term converged to a constant value in time, whereas the simulations without the term did not. Thus, the compression term resolves a source of non-conservation of energy, and we conclude that this term is strictly required to exactly conserve energy, though we note that the errors introduced by its absence are smaller than those from the time-stepping algorithm in general simulations.

In Section~\ref{sec:practical-tests}, the new cleaning method was applied to simulations of a blast wave in a magnetised medium (Section~\ref{sec:tests-mhdblast}), the Orszag-Tang vortex (Section~\ref{sec:tests-orszag}), and the MHD rotor problem (Section~\ref{sec:tests-rotor}). In general, using the new cleaning method provided reductions of average divergence error of $1$--$2\%$, to a maximum of $5\%$ occurring when the wave cleaning speed underwent its most rapid changes. For the blast wave test, using the new cleaning equations led to less overall dissipation of magnetic energy, dissipating magnetic energy at a rate $5\%$ less than the original method. We note that the dissipation of magnetic energy from divergence cleaning is $\lesssim 10$\% of that from artificial resistivity, meaning that it is only a minor contribution to the total dissipation.

Finally, in Section~\ref{sec:divbzero}, we demonstrated that it is possible to clean a magnetic field to arbitrarily small values of $\nabla \cdot {\bf B}$ in SPMHD, albeit with a large number of iterations. We found that using a lower value for the damping parameter ($\sigma=0.02$--$0.03$ in 2D) was optimal for reducing long wavelength divergence modes, though higher values ($\sigma=0.3$ in 2D) remained optimal for removing short wavelength errors and therefore for general use in simulations. Sub-cycling the divergence cleaning equations between timesteps was not found to have any adverse effect on the quality of the solution of the Brio-Wu shocktube test (Section~\ref{sec:divbzero-accuracy}), indeed only leading to further reductions in divergence error.

In summary, we recommend that our new divergence cleaning method be universally adopted over the previous method. The previous method had a numerical issue which could cause, in certain circumstances, an increase in magnetic energy and divergence error that would reduce the effectiveness of divergence cleaning. Though this effect is likely small in practical simulations, adopting the new method removes this source of energy growth, potentially yielding improvements in the reduction of average divergence error with lower associated numerical dissipation. It is trivial to adapt existing codes to evolve $\psi / c_{\rm h}$, and doing so provides a more robust, numerically stable method at no additional computational expense.

\section*{Acknowledgments}

The authors thank the three anonymous referees whose critiques have improved the quality of this paper. TST thanks G\'{a}bor T\'{o}th for useful discussions at Astronum 2015 in Avignon, France which helped to motivate this work. DJP thanks G\'{a}bor T\'{o}th for further useful discussions at Astronum 2016 in Monterey, California. TST is supported by a CITA Postdoctoral Research Fellowship. TST and MRB acknowledge support by the European Research Council under the European Community's Seventh Framework Programme (FP7/2007--2013 grant agreement no.\ 339248). DJP is supported by a Future Fellowship (FT130010034) from the Australian Research Council (ARC). This work, and MRB's visit to Australia in 2014, were part-funded by ARC Discovery Project DP130102078. This research has made use of NASA's Astrophysics Data System.

\bibliographystyle{elsarticle-harv}
\bibliography{bib}

\end{document}